\documentclass{article}
\usepackage[left=3cm,right=3cm,top=3cm,bottom=3cm]{geometry} 
\usepackage{authblk}
\pdfoutput=1
\usepackage{graphicx,psfrag,epsf}
\usepackage{graphicx,hyperref}
\usepackage{amssymb}
\usepackage{amsmath}
\usepackage{multirow}
\usepackage{float}
\usepackage{mathtools}
\usepackage{xcolor}
\usepackage{lmodern}
\mathtoolsset{showonlyrefs=true}
\usepackage{natbib}
\usepackage{subfig}
\usepackage{lipsum}
\usepackage{graphicx}
\usepackage{epsfig}
\usepackage[normalem]{ulem}

\usepackage{xr}
\usepackage{fancyhdr,lipsum}

\usepackage[title]{appendix}

\usepackage{amssymb}

\begin{document}


\title{Modeling animal movement with directional persistence and attractive points}

\author{{Gianluca} {Mastrantonio}}
\affil{Politecnico di Torino, Dipartimento di Scienze Matematiche}

	\date{}

	\maketitle

			\begin{abstract}

GPS technology is currently easily accessible to researchers,  and many  animal movement datasets are   available.
Two of the main features  that a model which describes an animal's path can possess are
directional persistence and  attraction to a point in space. In this work, we propose a new approach that can have both characteristics.

Our proposal is a hidden Markov model with a new emission distribution. The emission distribution models the two aforementioned characteristics, while the latent state of the hidden Markov model  is  needed  to account for   the behavioral modes. We  show that the model is easy to implement in a Bayesian framework.

We estimate our proposal on the motivating data that represent  GPS locations of a Maremma Sheepdog  recorded in Australia. The  obtained results are easily interpretable and   we show that our proposal outperforms the main  competitive model.

		\end{abstract}

		\maketitle

\section{Introduction}


The use of statistical models to understand the movement  of animals has become increasingly popular. The  data of such models are generally in the form of a time series of 2-dimensional spatial locations,
recorded using   a GPS device attached to the animal, with a time-interval between observations programmed by the researcher \citep{Cagnacci2010}.
These data are  often called ''trajectory tracking data'', and they  allow    features of an animal's movement, such as
habitat selection \citep{Hebblewhite}, spatio-temporal patterns  \citep{morales2004,Fryxell2008,Nathan2008,Frair2010} and animal behavior
\citep{MERRILL2000,Anderson2003}, to be investigated; for a detailed review,  the reader may refer to \cite{hooten2017animal}. These  approaches   can be grouped into three main categories:
 point processes  \citep{Johnson2013,Brost2015}, continuous-time   dynamic models (CTM) \citep{BLACKWELL199787,Johnson2008a,fleming2014non,hanks2015,  Buderman2018a, Buderman2018} and  discrete-time  dynamic models (DTM) \citep{morales2004,Jonsen2005,McClintock2012}.
 Animal movement is a continuous process, but it is generally easier to interpret and analyze a DTM  if  the time-interval between observations is fixed \citep{CODLING2005573,Patterson2017}. For a discussion on the differences between and similarities of these  approaches, the reader may refer  to \cite{McClintock2014}.


Most of the proposed models, in both continuous- and discrete-time frameworks,  are based on  biased random walks (BRWs) and  correlated random walks (CRWs). In a BRW, the animal is attracted to a specific spatial location, called \emph{center-of-attraction} \citep[see, for example,][]{BLACKWELL199787,dunn77}, which can be used to model a  tendency to move toward a patch of space \citep{McClintock2012} or the home range \citep{Christ2008}. On the other hand, in a CRW, the movement direction, at a given time, depends on the previous one;  this property is called directional persistence.   A CRW is often expressed
using  \emph{movement-metrics}, e.g.,  \emph{step-length} and  \emph{turning-angle} or  \emph{step-length} and  \emph{bearing-angle}, which are proxies of the speed and the direction measured between consecutive locations. These models   are often referred to as \emph{step and turn}  models \citep{Parton2017}.
  Independence  between the movement-metrics is usually assumed, see, for example, \cite{morales2002} and
\cite{Patterson2017}, and they  are rarely  dependent \citep{MASTRANTONIO2018,mastrantonio2019}. {The correlation between locations is a necessary  but not  sufficient condition for a model to be a CRW,  e.g., a two-dimensional AR(1) model is a BRW, but not a CRW, since there is no correlation between consecutive turning-angles or bearing-angles.   }
Most of the proposed  models have only a BRW or CRW characteristic, but there are some approaches,   called biased and correlated random walks (BCRWs), which have both  \citep{Schultz2001,Fortin2005,codling2008,McClintock2012}.
In order to be able to detect  changes in the  behavior of an animal (behavioral modes), the data  are generally modeled conditionally to a latent discrete state that describes the  behavior  assumed by the animal at a given time \citep{Patterson2222222}. Switching between behavioral modes is often assumed to be temporally structured,
 and if it  follows a Markov process the model is said to be a hidden Markov model (HMM)  \citep{Michelot2016,Langrock2012}. In the HMM context, the distribution of the data is   called  emission distribution \citep{Volant2014}.

In this work, we propose an HMM with a new  emission distribution { which  is called   \emph{step and turn with an attractive point (STAP)}; the STAP  belongs to the  BCRW family.}
We define the STAP by combining a CRW and a BRW.
For the CRW part, we envision the step-length and turning-angle as the polar-coordinate representation of  Cartesian coordinates. The Cartesian coordinates are assumed to be normally distributed, which induces a conditional normal distribution on the spatial locations.
 We use  an AR(1) model with a normal conditional distribution for the BRW  part.
We can thus combine the two   by introducing a parameter $\rho \in[0,1]$, which defines a conditional  normal  distribution with  a mean and covariance matrix that reduces to those of the BRW when $\rho=0$ and to those of the CRW   if $\rho=1$, with   a bias toward the center-of-attraction and a directional persistence for any value in between.
The distribution we define for $\rho$ allows us  to detect whether a behavior follows BRW, CRW or   BCRW dynamics.
Our HMM is formalized in a Bayesian setting, using the  sticky hierarchical Dirichlet process HMM  (sHDP-HMM) of \cite{fox2011}. {The sHDP-HMM  is a general framework that is used to define an HMM with any emission distribution,   and it is based on the Dirichlet process (DP). We indicate  as  STAP-HMM the  sHDP-HMM  with our proposal as the emission distribution.
 The use of  the DP  and of the distribution we define for $\rho$, allows us to treat the number of behaviors as a random variable, which can then be estimated during the model fitting, and to select the behavior-type without the need to resort to information criteria or a trans-dimensional Markov chain Monte Carlo algorithm (MCMC), such as the reversible jump MCMC (RJMCMC),  which involves challenges  in its implementation \citep{Hastie2012}.}


Although other  BCRWs  model the circular mean of the  turning- or bearing-angle     to induce attraction and directional persistence \citep{Fortin2005,Barton2009,rivest2016}, we approach the problem from a different point of view by  directly modeling the movement path with a normal distribution, which allows not only a  straightforward inference but,  as a by-product,  a projected-normal distribution to be induced over the bearing- and turning-angle, which is, to-date, the most flexible distribution for circular data \citep{mastrantonio2015}. On the other hand,   competitive models, {i.e., HMMs with a  BCRW   emission distribution,}   use unimodal and  symmetrical circular distributions for interpretability and easiness of implementation (e.g., the  von-Mises, wrapped Normal or wrapped Cauchy), thereby limiting the flexibility of the model.
Most  BCRW models can be formalized in the  general framework proposed by \cite{McClintock2012}, which we  consider as our main competitive model.

We use our proposal to model the trajectory tracking data of a Maremma Sheepdog.
 These dogs  have been used for centuries in Europe and Asia to protect livestock from predators, but only recently in Australia  \citep{Gehring,bommel2}. They generally work  with the shepherd to keep the stock together but, due to the   size of the properties,  this is not always possible in Australia. Since  an owner is often unaware of his/her dog's behavior, it is of interest to understand and characterize such behavior.
We analyze a dataset  taken from the movebank repository (\url{www.movebank.org}) in which the spatial locations of a dog, used to protect livestock in a property in Australia, are recorded using a GPS device  \citep{SheepdogRep,Sheepdog}.
We also show that our proposal outperforms the model of \cite{McClintock2012} in terms of ``deviance information criterion'' (DIC) \citep{celeux2006deviance} and ``integrated classification likelihood'' (ICL) \citep{Biernacki:2000}.
%
%
%

The paper is organized as follows.  We formalize the model in Section \ref{sec:Model}, by  introducing the   CRW (Section \ref{sec:crw}),  the  BRW (Section \ref{sec:brw}),  then the STAP models (Section \ref{sec:TheModel}) and finally  the STAP-HMM  (Section  \ref{sec:stap-hmm}). In Section \ref{sec:diff}, we show the similarities and differences between our proposal and the general framework of \cite{McClintock2012}. The conclusive remarks are presented in \ref{sec:final}.
{In the Appendix,    we show how to implement the  MCMC algorithm   (Appendix \ref{sec:imp}), some simulated examples (Appendix \ref{sec:sim}), a description of the  \cite{McClintock2012} proposal (Appendix \ref{sec:mc}),  how the turning-angle distribution changes if we change the time-interval between the recorded coordinates (Appendix  \ref{sec:Bimod}),  a sensitivity analysis of the sHDP-HMM hyper-parameter priors (Appendix  \ref{sec:priors}),
simulated paths obtained with  parameters estimated from the real data application (Appendix \ref{sec:simTraj}), and the results of   STAP-HMMs that are only able to estimate CRW or BRW behaviors  (Appendix \ref{sec:realOtherRes}).}



\section{Methods} \label{sec:Model}

\begin{figure}[t]
    \centering
    \includegraphics[scale=0.5]{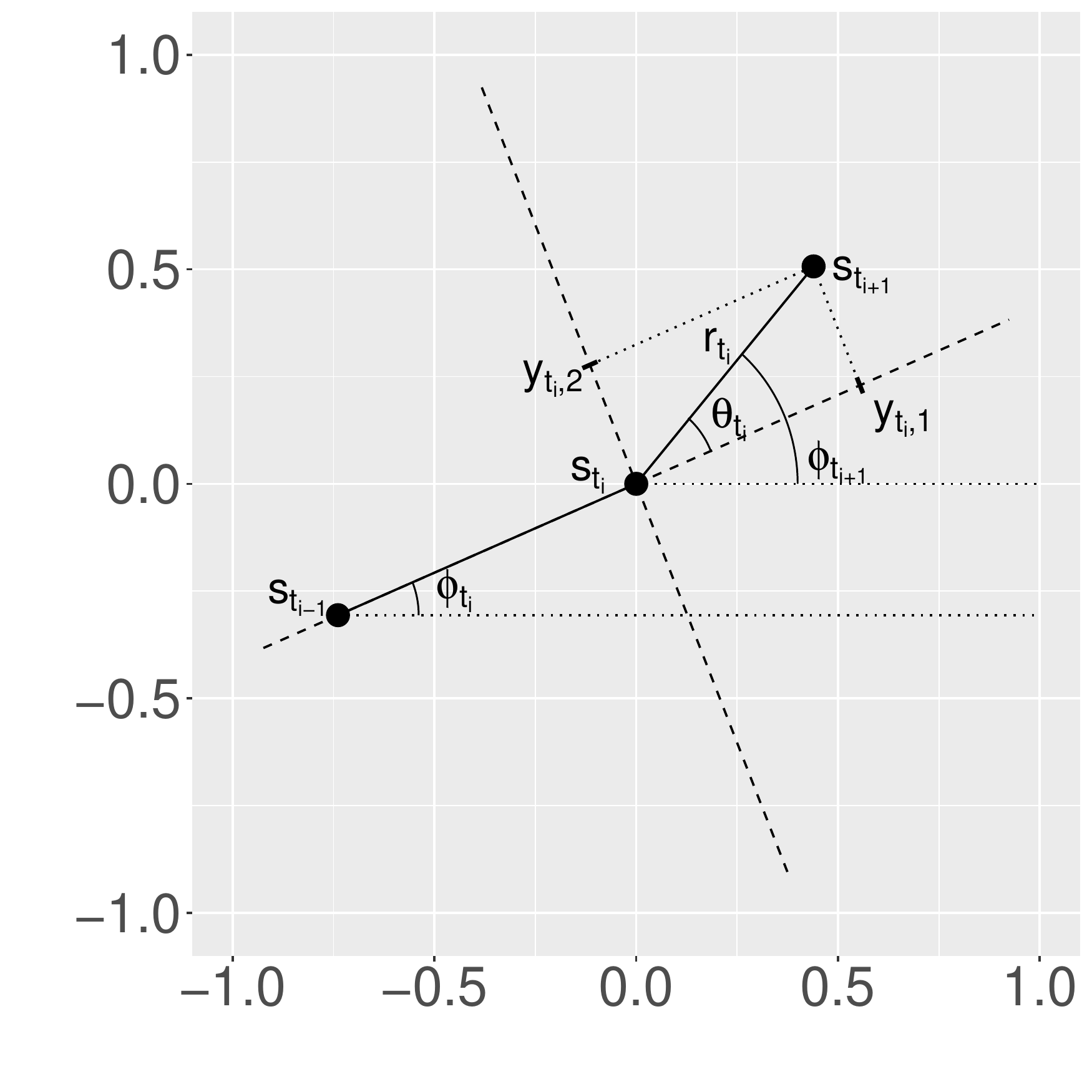}
    \caption{A graphical representation  of the relations between the spatial locations, the movement-metrics  and the displacement-coordinate.   {The arrow represents   $\protect\vec{\mathbf{F}}_{{i}}$, while the ellipse is an area containing  95\% of the probability mass of the conditional distribution of
      $\mathbf{s}_{{i+1}}$, computed using \eqref{eq:ell}}.
    } \label{fig:trans}
\end{figure}


{We assume we have a time-series of spatial coordinates $\mathbf{s} = (\mathbf{s}_{{1}},\dots, \mathbf{s}_{{T}})'$, with  $\mathbf{s}_{{i}} = (s_{{i},1},s_{{i},2}) \in \mathcal{D} \subset \mathbb{R}^2$, which represent the animal's path  in a two-dimensional space.
We  assume  there is no measurement error and
the time-difference between consecutive observations is fixed.}
We consider coordinates  $\mathbf{s}_{{i+1}}$ as the realization of a random variable with the following conditional distribution:
\begin{equation} \label{eq:scond}
    \mathbf{s}_{{i+1}} |\mathbf{s}_{{i}},\mathbf{s}_{{i-1}}, \dots , \mathbf{s}_{{1}} \sim N(\mathbf{s}_{{i}}+\mathbf{M}_{{i}},\mathbf{V}_{{i}} ), \, \, i \in \{1, \dots , T-1\},
\end{equation}
{where $\mathbf{M}_{{i}} = ({M}_{{i,1}},{M}_{{i,2}})' \in \mathbb{R}^2$ and $\mathbf{V}_{{i}}$ is a $2 \times 2 $ covariance matrix.
The vector  $\mathbf{M}_{{i}}$ and  the matrix $\mathbf{V}_{{i}}$ may depend on previous locations and other parameters, while $\mathbf{s}_{{1}}$ is known and fixed. } The conditional set  is composed of the entire past but, as we will show in Section \ref{sec:TheModel},  only the two previous   locations are needed  in our proposal,   and  $\mathbf{s}_{{0}}$ is  therefore considered  as a parameter that is able to define the conditional distribution of $\mathbf{s}_{{2}}$.
Equation \eqref{eq:scond} models the path intuitively, since the  animal moves at each time-point  following a normal distribution, with the mean coordinates given by the previous location and parameter   $\mathbf{M}_{{i}}$,  while $\mathbf{V}_{{i}}$ measures the variability.

{  In  Figure \ref{fig:trans}, it is possible to see how $\mathbf{M}_{{i}}$ and $\mathbf{V}_{{i}}$ are connected  to the observed path, since the end-point of the arrow represents the mean value of the conditional distribution of $\mathbf{s}_{i+1}$  (which is $\mathbf{s}_{{i}}+\mathbf{M}_{{i}}$), while   the ellipse    is the set of  points $\mathbf{s}^* \in\mathbb{R}^2$ that satisfy
$(\mathbf{s}^*-\mathbf{s}_{{i}}-\mathbf{M}_{{i}})'\mathbf{V}_{{i}}^{-1}(\mathbf{s}^*-\mathbf{s}_{{i}}-\mathbf{M}_{{i}}) = c$, where the value c is defined to ensure that the  area   $\mathcal{E} \in \mathbb{R}^2$ inside the ellipse  contains  95\% of the probability mass of the conditional distribution of
  $\mathbf{s}_{{i+1}}$, i.e.,
\begin{equation}\label{eq:ell}
\int_{\mathbf{s} \in \mathcal{E} } (2 \pi)^{-1}|\mathbf{V}_{{i}}|^{-\frac{1}{2}} \exp
\left(-\frac{(\mathbf{s}-\mathbf{s}_{{i}}-\mathbf{M}_{{i}})'\mathbf{V}_{{i}}^{-1}(\mathbf{s}-\mathbf{s}_{{i}}-\mathbf{M}_{{i}})}{2} \right) d \mathbf{s}= 0.95.
\end{equation}
It should be noted that the  value of the conditional density of $\mathbf{s}_{{i+1}}$ is the same for all $\mathbf{s}^*$ on the ellipse, hence  the ellipse is  a contour of the normal density, and it can be used to represent and to infer the characteristics of the covariance matrix $\mathbf{V}_{{i}}$ \citep{Friendly2012}.
For example,  let $x \in (-\pi/2, \pi/2]$ be the inclination angle  of the major axis of the ellipse, then
the correlation is >0 if $0< x< \pi/2$, <0 if $-\pi/2<x < 0 $, and there is no correlation if $x =0$ or $x = \pi/2$ , i.e., the  major axis of the ellipse is parallel to one of the two axes of the  Euclidian reference system. If we indicate   the largest values of the first and second Euclidean coordinate  in  $\mathcal{E}$ as $x_{1}^*$ and $x_{2}^*$, respectively,  the distances
$x_{1}^*-{s}_{i,1}-{M}_{i,1}$ and $x_{2}^*-{s}_{i,2}-{M}_{i,2}$ are    proxies of the two  variances of $\mathbf{V}_i$. If the ellipse is a circle, there is no correlation and the two variances have the same value.
%
}

 Parameter $\mathbf{M}_{{i}}$  can be used to  introduce such  movement features   as directional persistence and  a center-of-attraction. To describe these two features, we  introduce  the   vector $\vec{\mathbf{F}}_{{i}}$, which is a vector  with initial and terminal points  $\mathbf{s}_{{i}}$ and $\mathbf{s}_{{i}}+\mathbf{M}_{{i}}$, respectively (see Figure \ref{fig:trans}), and it represents the expected movement between time ${i}$ and ${i+1}$ {since its initial position is the previous observed location $\mathbf{s}_{{i}}$,
 while the terminal point is equal to $\mathbb{E}(\mathbf{s}_{{i+1}}|\mathbf{s}_{{i}})=\mathbf{s}_{{i}}+\mathbf{M}_{{i}}$.
  The  length of $\vec{\mathbf{F}}_{{i}}$,
 $\xi_{{i}}$, is equal to
 $$
 || (\mathbf{s}_{{i}}+\mathbf{M}_{{i}})-\mathbf{s}_{{i}} ||_2 = || \mathbf{M}_{{i}} ||_2,
 $$
while  its direction, $\lambda_{{i}}$,  is equal to
\begin{equation}
\text{atan}^*(({s}_{{i,2}}+{M}_{{i,2}})-{s}_{{i,2}},({s}_{{i,1}}+{M}_{{i,1}})-{s}_{{i,1}})= \text{atan}^*({M}_{{i,2}},{M}_{{i,1}}) \in [-\pi, \pi),
\end{equation}
where $\text{atan}^*(\cdot)$ is the two-argument tangent function \citep{Jammalamadaka2004}, i.e.,
the direction and length of $\vec{\mathbf{F}}_{{i}}$ are equal to those of $\mathbf{M}_{{i}}$.  }
{We   introduce the bearing-angle
\begin{equation}
\phi_{{i}} = \text{atan}^*({s}_{{i+1},2}-{s}_{{i} ,2}, {s}_{{i+1},1}-{s}_{{i},1}). \label{eq:Phi}
\end{equation}
which measures the direction between locations $\mathbf{s}_{{i}}$ and $\mathbf{s}_{{i+1}}$.}
{ The distribution  of
$\phi_{{i}}$, in a BRW,  is independent of  the previous bearing-angles and  $\lambda_i$ is equal to $\text{atan}^*({\mu}_{{2}}-{s}_{{i} ,2},{\mu}_{{1}}-{s}_{{i} ,1})$, which is the direction between the location in space
$\boldsymbol{\mu} = ({\mu}_{{1}},{\mu}_{{2}})' \in \mathbb{R}^2$
 and   the previous observed location  $\mathbf{s}_{{i}}$. In this case, the vector $\boldsymbol{\mu}$ is a parameter that is used  to define $\mathbf{M}_{{i}}$. Since the expected movement at time $i+1$,  measured by   $\vec{\mathbf{F}}_{{i}}$, starts from $\mathbf{s}_{{i}}$ and points to  $\boldsymbol{\mu}$, the latter is called the attractor, since  it ``attracts'' the animal. }
 {The  distribution of $\phi_{{i}}$, in a CRW, depends on the previous bearing-angles (albeit generally only on the first one, $\phi_{{i-1}}$),
 which means that the directions are correlated. This is often done by assuming that $\lambda_i$ is a function of $\phi_{{i-1}}$, see, for example, \cite{mastrantonio2019}. It should be noted that a CRW needs to be at least second-order Markovian since two previous locations are needed to compute   $\phi_{{i-1}}$. }
 {In both models,  $\xi_i$ is a measure of how far  the mean value of the conditional distribution of $\mathbf{s}_{i+1}$ is from   $\mathbf{s}_{i}$, and it can be considered a proxy of the expected bi-dimensional  movement speed. Parameter $\xi_i$ can be also used, in a BRW, to evaluate the strength of the attraction, see Section \ref{sec:brw}.  As we will show in Sections \ref{sec:crw} and \ref{sec:brw}, parameter  $\mathbf{V}_{{i}}$,  in our definition of a CRW and BRW, depends on the previous direction  in the CRW, and is independent of the previous direction and the attractor in the BRW. }




Models based on equation \eqref{eq:scond} with   only one of the two aforementioned features, {i.e., directional persistence or an attractive point, } have  already been   proposed  (see, for example, \cite{Christ2008} and \cite{mastrantonio2019}), and one of the main contributions of this paper is that it allows both properties to be considered.

\subsection{Displacement-coordinates and  movement-metrics} \label{sex:dcmv}

\begin{figure}[t]
    \centering
    \includegraphics[scale=0.31]{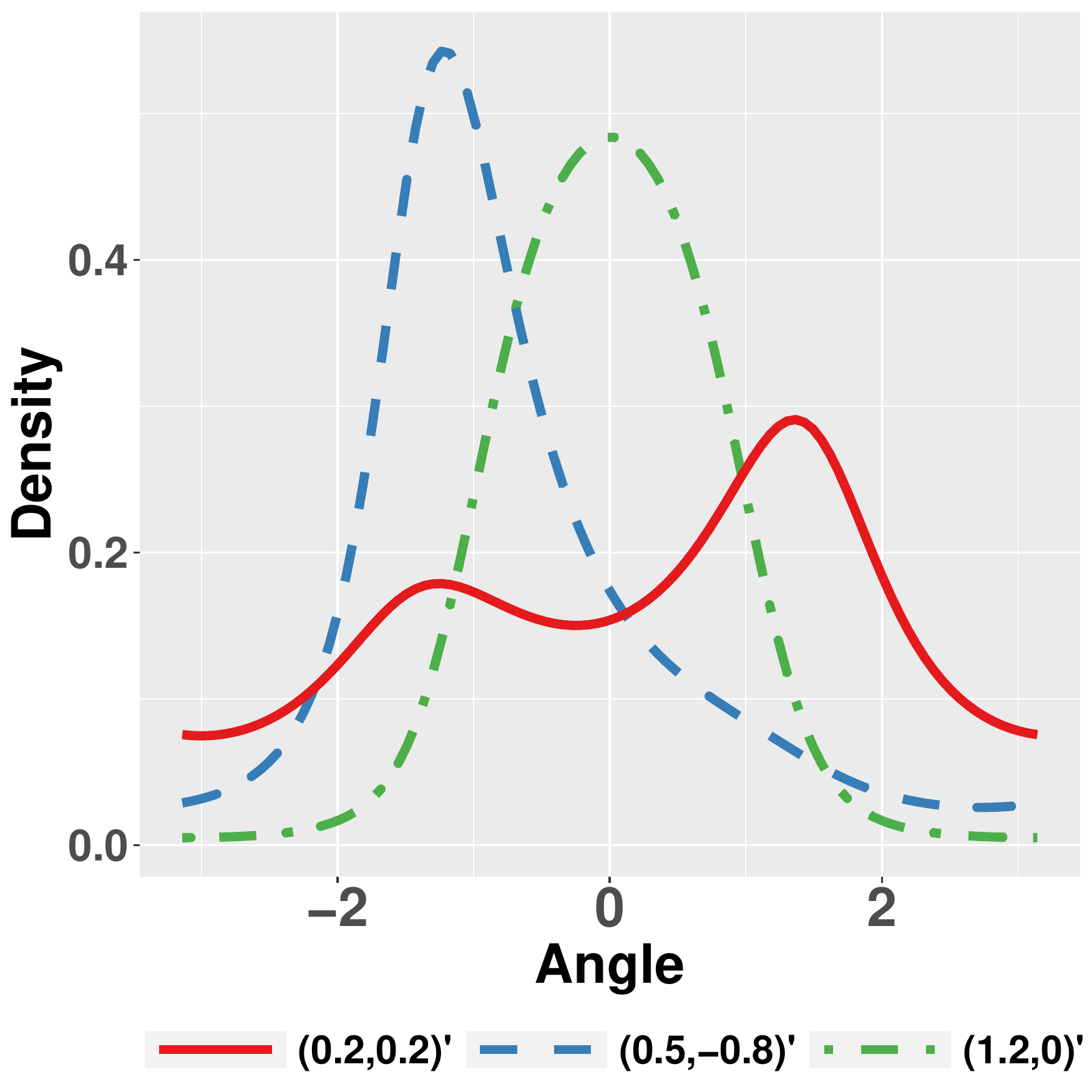}
    \caption{Representation of projected-normal densities based on a bivariate normal distribution with variances (0.5,1), 0 correlation and different   mean vectors.
    } \label{fig:ProjN}
\end{figure}

{In this section, we introduce  different ways of parametrizing the movement path} that are useful to understand and interpret our approach. The relations between these quantities are also depicted in Figure \ref{fig:trans}.

%
%
%
%
%
%
%
%
The path $\mathbf{s}$ can be described equivalently by using  $\mathbf{s}_{{1}}$ and the displacement-coordinates  $\mathbf{v} =  (\mathbf{v}_{{1}},\dots, \mathbf{v}_{{T-1}})'$, where
$\mathbf{v}_{{i}} =  \mathbf{s}_{{i+1}}- \mathbf{s}_{{i}}$,
 since
 \begin{equation} \label{eq:pathV}
 \mathbf{s}_{{i+1}} = \mathbf{s}_{{1}}+\sum_{l=1}^{i} \mathbf{v}_{{l}},  \, \, i \in \{1, \dots , T-1\}.
 \end{equation}
Variable $\mathbf{v}_{{i}}$  represents the movement between time ${i}$ and ${i+1}$
and, from equation \eqref{eq:scond}, we can easily derive that
\begin{equation} \label{eq:scondV1}
    \mathbf{v}_{{i}} |\mathbf{v}_{{i-1}}, \dots , \mathbf{v}_{{1}} \sim N(\mathbf{M}_{{i}},\mathbf{V}_{{i}} ), \, \, i \in \{1, \dots , T-1\}.
\end{equation}
The coordinates $\mathbf{v}_{{i}}$ are useful to describe an attractive point, since we can evaluate how it changes with respect to the distance from the attractor.
{Vector $\mathbf{v}_{{i}}$ can be expressed in polar-coordinates using
its magnitude   $r_{{i}} = ||\mathbf{v}_{{i}}||_2$  and the bearing-angle  $\phi_{{i}}$ which can be computed using equation \eqref{eq:Phi} or the following:}
\begin{equation}
\phi_{{i}} =  \text{atan}^*({v}_{{i},2},{v}_{{i},1}) . \label{eq:Phi22}
\end{equation}
The value of $r_{{i}}$ is the animal speed, and $\phi_{{i}}$ measures the direction with respect to the ``general''  reference-system, where  $\phi_{{i}}=0$ indicates the  East direction, and the sense of rotation  is anticlockwise.
In animal-movement literature  $r_{{i}}$ is known as the step-length.

We can define a new set of displacement-coordinates if we let the reference-system change at each time-point, and align the x-axis in the direction of the previous increment $\mathbf{s}_{{i}}- \mathbf{s}_{{i-1}}$  with origin on the  coordinate $\mathbf{s}_{{i}}$.
We  introduce the rotation matrix
\begin{equation} \label{eq:RotMat}
  \mathbf{R}(\omega)=
  \left(
\begin{array}{lr}
\cos(\omega) &  -\sin (\omega)\\
\sin(\omega) & \cos(\omega)
\end{array}
\right),
\end{equation}
which is a matrix  that can be  used to perform a rotation in a 2-dimensional space, and we define the new displacement-coordinates as
\begin{equation} \label{eq:disp1}
\mathbf{y}_{{i}} = \mathbf{R}'(\phi_{{i-1}}) \mathbf{v}_{{i}}  = \mathbf{R}'(\phi_{{i-1}})(\mathbf{s}_{{i+1}}- \mathbf{s}_{{i}}),
\end{equation}
which represent the projection of  $\mathbf{v}_{{i}}$  onto the direction of the increment $\mathbf{s}_{{i}}- \mathbf{s}_{{i-1}}$  measured by the bearing-angle $\phi_{{i-1}}$, see Figure \ref{fig:trans}.
From \eqref{eq:scondV1} and \eqref{eq:disp1}, we obtain
\begin{equation} \label{eq:scondY}
    \mathbf{y}_{{i}} |\mathbf{y}_{{i-1}}, \dots , \mathbf{y}_{{1}} \sim N(\mathbf{R}'(\phi_{{i-1}})\mathbf{M}_{{i}},\mathbf{R}'(\phi_{{i-1}})\mathbf{V}_{{i}}
    \mathbf{R}(\phi_{{i-1}}) ), \, \, i \in \{1, \dots , T-1\},
\end{equation}
where $\phi_{{0}}$ is a parameter, since it is a function of the fixed coordinate $\mathbf{s}_{{1}}$ and parameter $\mathbf{s}_{{0}}$.
The transformation \eqref{eq:disp1} preserves the vector magnitude, which then results in  $r_{{i}}= ||\mathbf{v}_{{i}}||_2= ||\mathbf{y}_{{i}}||_2$.
The polar-coordinate representation of $\mathbf{y}_{{i}}$
is the couple $(\theta_{{i}},r_{{i}})$ where
\begin{equation}\label{eq:theta_atan}
\theta_{{i}} = \text{atan}^*({y}_{{i},2},{y}_{{i},1}).
\end{equation}
The angle $\theta_{{i}} $ is called  turning-angle and it is related to the bearing angle through the  relation
\begin{equation} \label{eq:theta}
\theta_{{i}} = \phi_{{i}} -\phi_{{i-1}},
\end{equation}
which highlights how $\theta_{{i}} $ and  $\mathbf{y}_{{i}}$ represent a change in direction between two consecutive time-points and they are therefore  useful to describe and introduce directional persistence, see the next section.

The polar-coordinates   $(\phi_{{i}},r_{{i}})$ and $(\theta_{{i}},r_{{i}})$ are  called movement-metrics, and they are often modeled instead of the coordinates, especially in CRWs, but also in  BCRWs \citep{McClintock2012}. The distribution over the angular variable is generally assumed to be unimodal and symmetric  \citep{Michelot2016,Patterson2017}, with a few exceptions (see, for example, \cite{MASTRANTONIO2018}). On the other hand, since both $\mathbf{v}_{{i}}$ and $\mathbf{y}_{{i}}$ are normally distributed, the bearing- and turning-angle of our proposal are projected normal distributed
 \citep{mastrantonio2015b}, which is one of the most flexible distributions for circular data,   with a density that can be unimodal symmetric, asymmetric,  bimodal or antipodal; some examples of projected normal densities are shown in Figure \ref{fig:ProjN}, and more details can be found in \cite{Wang2013}.


\subsection{Correlated random walk} \label{sec:crw}

\begin{figure}[t]
    \centering
    {\subfloat[CRW]{\includegraphics[trim = 10 0 10 30,scale=0.3]{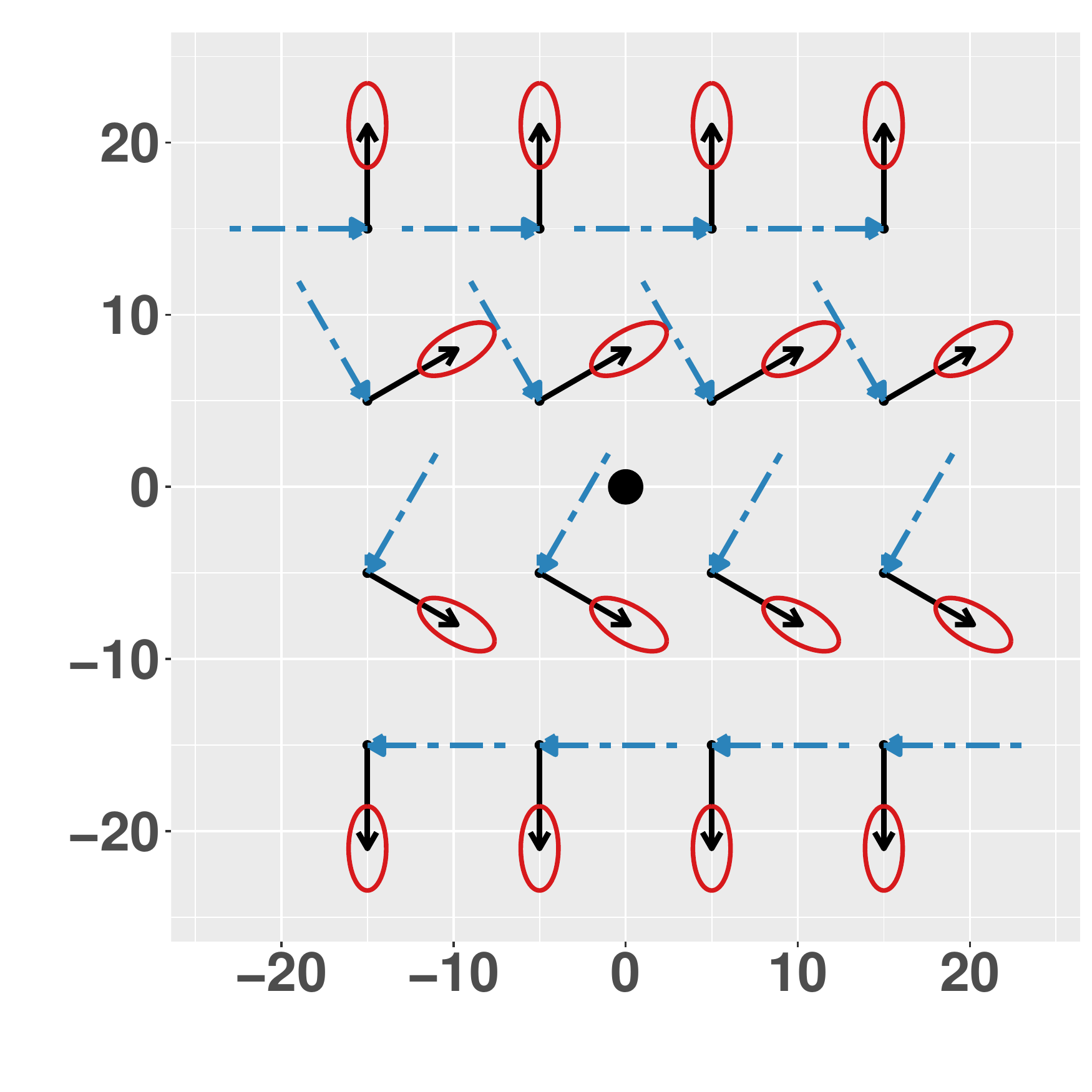}}}
    {\subfloat[BRW]{\includegraphics[trim = 10 0 10 30,scale=0.3]{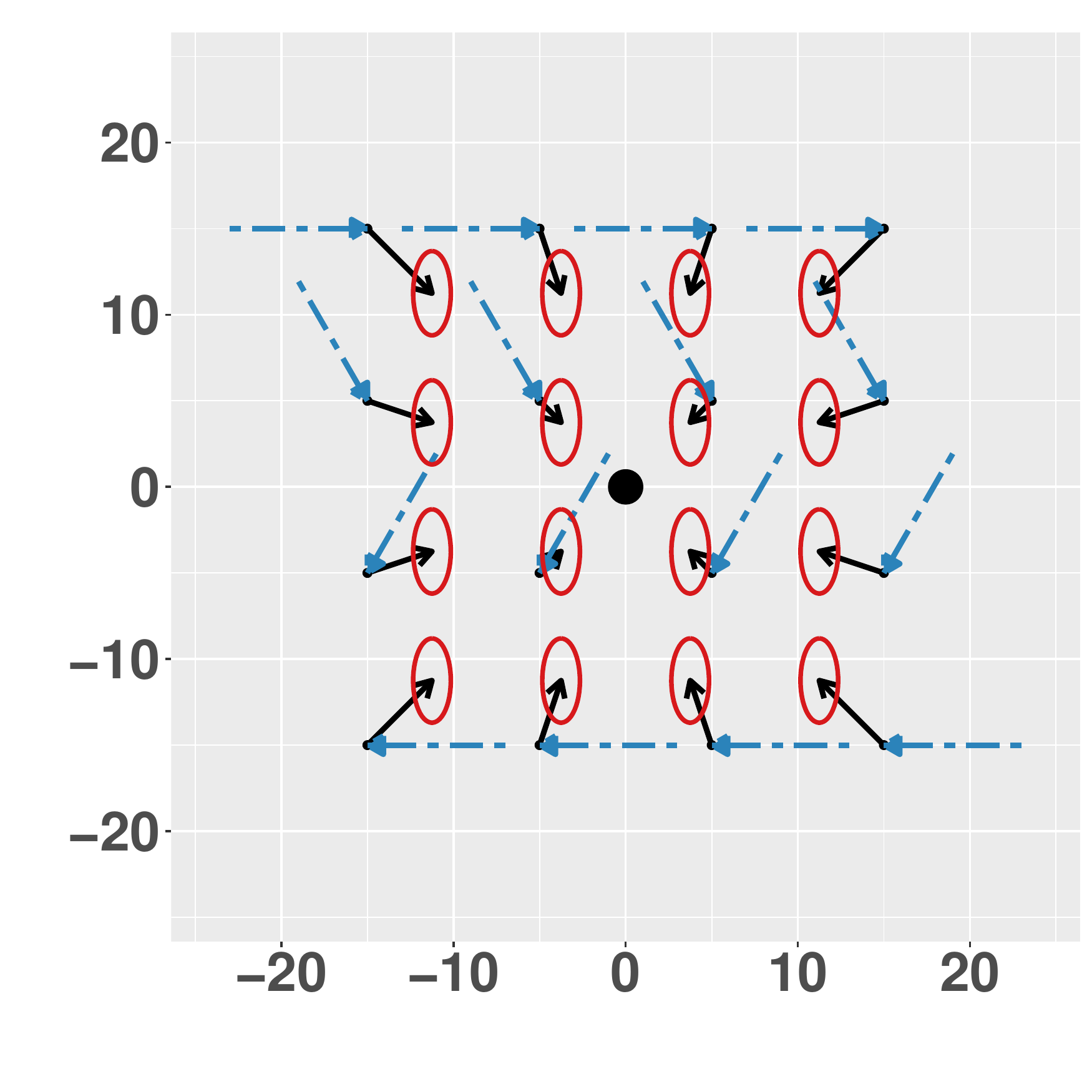}}}\\
    {\subfloat[BCRW with $\rho=2/3$]{\includegraphics[trim = 10 0 10 30,scale=0.3]{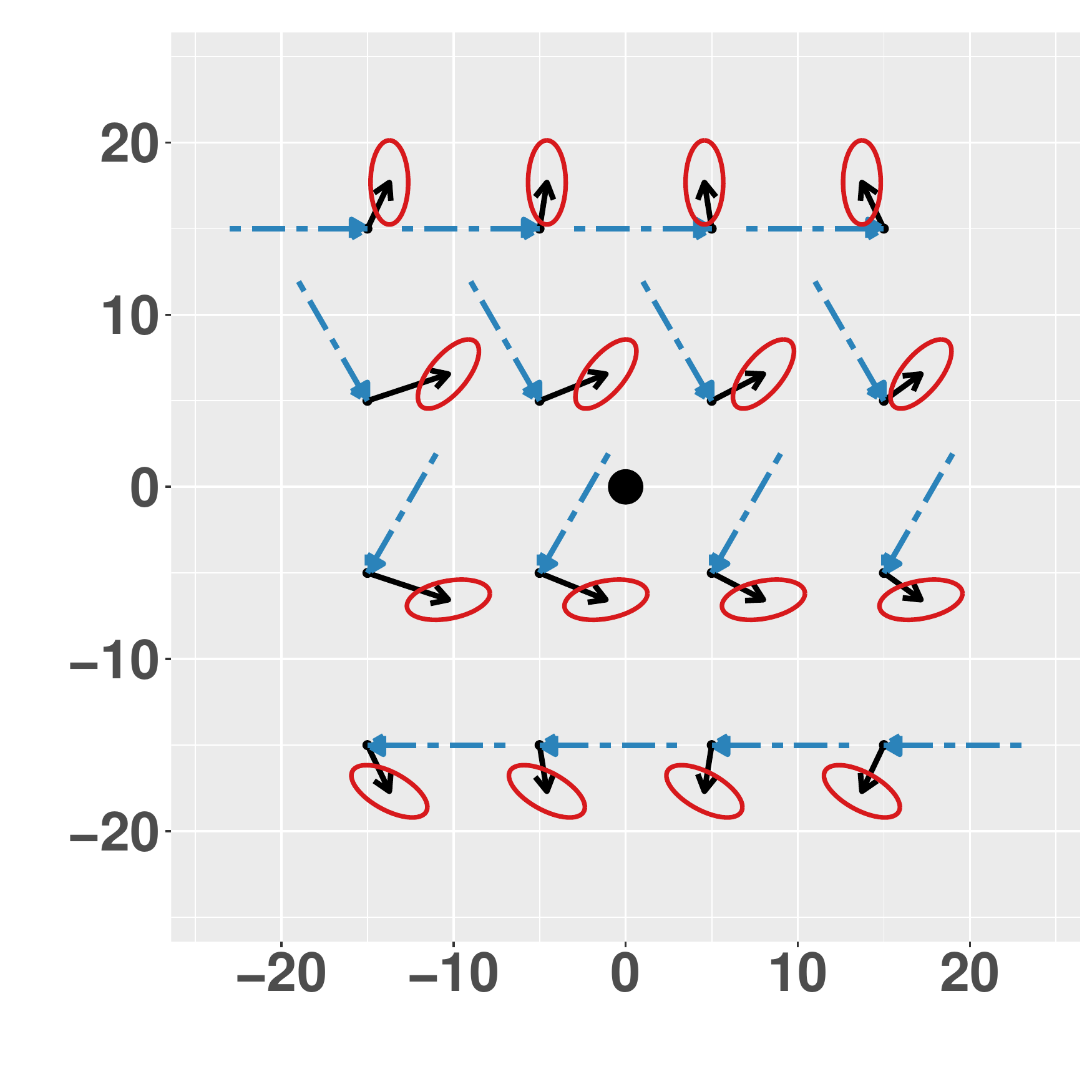}}}
    {\subfloat[BCRW with $\rho=1/3$]{\includegraphics[trim = 10 0 10 30,scale=0.3]{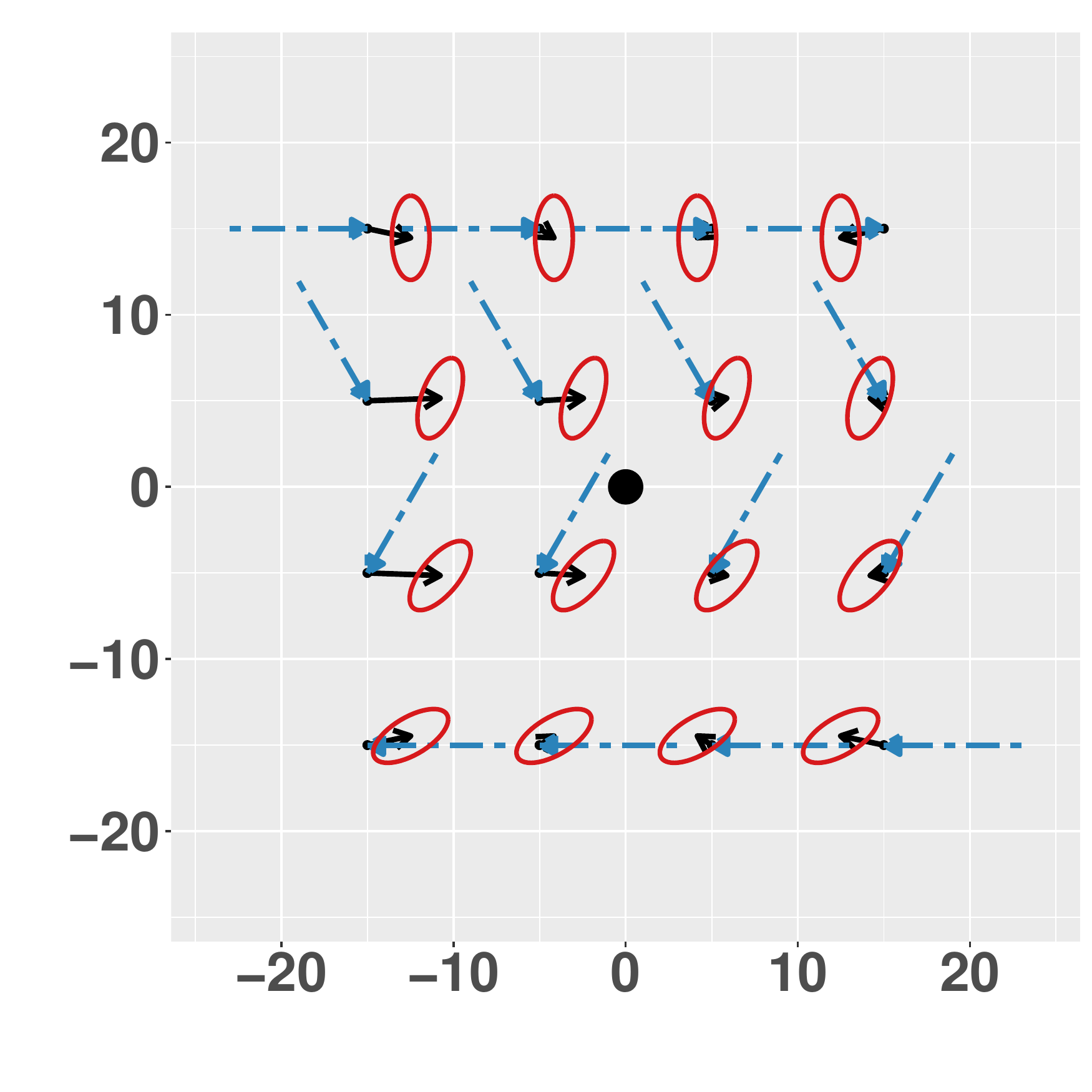}}}
    \caption{ Graphical representation of the conditional distribution of $\mathbf{s}_{{i+1}}$ for different possible values of $\mathbf{s}_{{i}}$ and  the previous directions.
    The dashed arrow represents the movement between  $\mathbf{s}_{{i-1}}$ and $\mathbf{s}_{{i}}$. The solid arrow is $\protect\vec{\mathbf{F}}_{{i}}$,   {while the ellipse is the area containing  95\% of the probability mass of  the  conditional  distribution of $\mathbf{s}_{{i+1}}$, computed using \eqref{eq:ell}.   In all the figures,   $\boldsymbol{\mu} = (0,0)'$, $\boldsymbol{\eta} = (0,6)'$,  $\tau=0.25$,
   $ \boldsymbol{\Sigma}= \left(
\begin{array}{cc}
0.2& 0\\
0& 1
\end{array} \right)$, and the central dot is the location $\boldsymbol{\mu}$, which  represents the attractor in the BRW and the BCRW.}
     }\label{fig:EsModel2}
\end{figure}


To define the CRW, which is one of the two components needed as the base  of our proposal, we model the displacement-coordinates $\mathbf{y}_{{i}}$ as
\begin{equation} \label{eq:yr}
    \mathbf{y}_{{i}} \sim N(\boldsymbol{\eta},\boldsymbol{\Sigma}),
\end{equation}
where $\boldsymbol{\eta} = (\eta_{1}, \eta_2)'\in \mathbb{R}^2$ is a vector of length 2 and $\boldsymbol{\Sigma}$ is a $2\times 2$ covariance matrix. From equation \eqref{eq:yr}, we can compute the following conditional distributions:
\begin{align}
    \mathbf{s}_{{i+1}} |\mathbf{s}_{{i}},\mathbf{s}_{{i-1}} &\sim N(\mathbf{s}_{{i}} +\mathbf{R}(\phi_{{i-1}})\boldsymbol{\eta},\mathbf{R}(\phi_{{i-1}})\boldsymbol{\Sigma}\mathbf{R}'(\phi_{{i-1}}) ) \label{eq:CondsD},\\
    \mathbf{v}_{{i}} |\mathbf{v}_{{i-1}}& \sim N(\mathbf{R}(\phi_{{i-1}})\boldsymbol{\eta},\mathbf{R}(\phi_{{i-1}})\boldsymbol{\Sigma}\mathbf{R}'(\phi_{{i-1}}) ) \label{eq:rds}.
\end{align}
 Since the bearing-angle $\phi_{{i}}$ is computed using  $\mathbf{s}_{{i}}$ and $\mathbf{s}_{{i-1}}$, see equation \eqref{eq:Phi}, the path is second-order Markovian. The distribution  \eqref{eq:CondsD} is in the form given by \eqref{eq:scond} with $\mathbf{M}_{{i}}= \mathbf{R}(\phi_{{i-1}})\boldsymbol{\eta}$
  and $\mathbf{V}_{{i}}= \mathbf{R}(\phi_{{i-1}})\boldsymbol{\Sigma}\mathbf{R}'(\phi_{{i-1}})$. {In this model, the parameters are $\boldsymbol{\eta}$, $\boldsymbol{\Sigma}$ and $\mathbf{s}_0$.}

The angle $\lambda_{{i}}$, which is the direction of  $\vec{\mathbf{F}}_{{i}}$,
 does not depend on the spatial location, but only on the bearing angle $\phi_{{i-1}}$ which rotates the  parameter $\boldsymbol{\eta}$. {In more detail, we have
 \begin{equation} \label{eq:crwd}
\lambda_i = \phi_{i-1}+ \text{atan}^*(\eta_2, \eta_1),
\end{equation}
where  $ \text{atan}^*(\eta_2, \eta_1)$ is the direction of $\boldsymbol{\eta}$.
Equation \eqref{eq:crwd} shows that $\lambda_i$, the direction of the expected movement,  is a rotation of the previous bearing-angle by an amount $\text{atan}^*(\eta_2, \eta_1)$, that  is a function of $\boldsymbol{\eta}$.  The value $\lambda_i$ depends  of the previous bearing-angle $\phi_{i-1}$,  and  $\phi_{i}$ and $\phi_{i-1}$ are, therefore,  correlated; this defines a CRW (see the beginning of Section \ref{sec:Model}). For example,
if $\boldsymbol{\eta} = (1,0)$ (i.e., $\text{atan}^*(\eta_2, \eta_1) = 0$), we obtain $\lambda_{{i}} =\phi_{{i-1}}$,  if $\boldsymbol{\eta} = (0,1)$  (i.e., $\text{atan}^*(\eta_2, \eta_1) = \pi/2$), we obtain
$\lambda_{{i}} =\phi_{{i-1}}+\pi/2$, and
$\lambda_{{i}} =\phi_{{i-1}}-\pi/2$
 if $\boldsymbol{\eta} = (0,-1)$ (i.e., $\text{atan}^*(\eta_2, \eta_1) = -\pi/2$).
The length of $\boldsymbol{\eta}$ is equal to $\xi_i$, since $||\mathbf{M}_{i}||_2 = ||\mathbf{R}(\phi_{{i-1}})\boldsymbol{\eta}||_2 = ||\boldsymbol{\eta}||_2$,
and it represents the distance between    $\mathbf{s}_{i}$ and $\mathbb{E}(\mathbf{s}_{i+1}|\mathbf{s}_{i})$. It should be noted that the covariance matrix of  $\mathbf{v}_{{i}}$ and
$\mathbf{s}_{{i+1}}$ rotates according to the bearing-angle $\phi_{{i-1}}$.
  }
The  step-length and  turning-angle, and variable $\mathbf{y}_{{i}}$, are i.i.d. and independent of the position in space and of the previous movement,   while the parameters of  the  distribution of the bearing-angle and $\mathbf{v}_{{i}}$
change according to $\phi_{{i-1}}$ (see equation \eqref{eq:rds}).

In Figure \ref{fig:EsModel2} (a),  we show the conditional distribution of $\mathbf{s}_{{i+1}}$ under different conditions, where  all the features we described can easily be seen. {Each vector $\vec{\mathbf{F}}_{{i}}$ (solid arrow) has the same length ($\xi_i= ||\boldsymbol{\eta}||_2= 6$), regardless of the position in space (the tail of $\vec{\mathbf{F}}_{{i}}$) and the previous movement (dashed arrow). The value of $\lambda_i$ depends on the previous direction, but the angle between the previous direction and $\vec{\mathbf{F}}_{{i}}$ is fixed and equal to
 $\text{atan}^*(\eta_2,\eta_1) = \pi/2$. The ellipse rotates with the bearing angle $\phi_{i-1}$.}

{It should be noted that  it is not possible to find a closed-form expression for the distribution of   $\mathbf{s}_{i+l}|\mathbf{s}_{i}, \mathbf{s}_{i-1}$  if $l>1$, for the proposed CRW. Hence, the CRW can be used if the time-interval is constant, since, even in the simple case in which there is no recorded observation for time $i+1$, we are unable to compute the conditional density $\mathbf{s}_{i+2}|\mathbf{s}_{i}, \mathbf{s}_{i-1}$  ($l=2$). Hence, in order to be able to compute the data likelihood in the MCMC algorithm, any missing data must be considered a further parameter and estimated during model fitting.
 }

\subsection{Biased random walk} \label{sec:brw}

 We use a two-dimensional AR(1) model for the  BRW component of our proposal
\begin{equation}
    \mathbf{s}_{{i+1}} |\mathbf{s}_{{i}} \sim N(\mathbf{s}_{{i}} +\tau \left(\boldsymbol{\mu}-  \mathbf{s}_{{i}} \right),\boldsymbol{\Sigma} ), \label{eq:CondsBRW}
\end{equation}
where
$\boldsymbol{\mu}$ is a two-dimensional vector and $\tau \in (0,1)$. The process is first-order Markovian and  can be expressed as equation \eqref{eq:scond} by setting $\mathbf{M}_{{i}}= \tau \left(\boldsymbol{\mu}-  \mathbf{s}_{{i}} \right)$ and $\mathbf{V}_{{i}}=\boldsymbol{\Sigma}$.
{The parameters, in this model,  are $\boldsymbol{\mu}$, $\tau$, and $\boldsymbol{\Sigma}$.}
The distributions of the displacement-coordinates are
\begin{align}
    \mathbf{v}_{{i}}| \mathbf{v}_{{i-1}},\dots ,\mathbf{v}_{{1}} & \sim N(\tau \left(\boldsymbol{\mu}-  \mathbf{s}_{{i}} \right),\boldsymbol{\Sigma} ) \label{eq:CondsBRW_V},\\
    \mathbf{y}_{{i}}|\mathbf{y}_{{i-1}},\dots ,\mathbf{y}_{{1}} & \sim N(\mathbf{R}'(\phi_{{i-1}})\tau \left(\boldsymbol{\mu}-  \mathbf{s}_{{i}} \right),\mathbf{R}'(\phi_{{i-1}})
    \boldsymbol{\Sigma}\mathbf{R}(\phi_{{i-1}}) ),\label{eq:CondsBRW_Y}
\end{align}
where $\mathbf{s}_{{i}}$  is a function of the previous displacement-coordinates through equations
\eqref{eq:pathV} and
\eqref{eq:disp1}.
{The direction of $\vec{\mathbf{F}}_{{i}}$  is equal to
\begin{equation} \label{eq:spatt}
\lambda_i = \text{atan}^*(\tau(\mu_{2}-s_{i,2}),\tau(\mu_{1}-s_{i,1})) = \text{atan}^*(\mu_{2}-s_{i,2},\mu_{1}-s_{i,1}),
\end{equation}
where it is clear that $\vec{\mathbf{F}}_{{i}}$ points to the parameter (or point in space) $\boldsymbol{\mu}$, and the former is therefore the spatial-attractor, i.e., we expect the movement to be in the direction of $\boldsymbol{\mu}$. Since the path described by equation \eqref{eq:CondsBRW_Y} is first-order Markovian, the distribution of $\phi_i$ is independent of the previous bearing angles  and only depends on the spatial coordinate $\mathbf{s}_{{i}}$ and  $\boldsymbol{\mu}$,  see  equations \eqref{eq:CondsBRW} and \eqref{eq:CondsBRW_V}. Hence, the  two-dimensional AR(1) model is a BRW (see the beginning of Section \ref{sec:Model}). The length of
$\vec{\mathbf{F}}_{{i}}$, $\xi_{{i}}$,
is equal to $\tau||\boldsymbol{\mu}-\mathbf{s}_{{i}} ||_2$,  and since $||\boldsymbol{\mu}-\mathbf{s}_{{i}} ||_2$ is the distance between the previous observation ($\mathbf{s}_{{i}}$) and the spatial attractor ($\boldsymbol{\mu}$), $\tau \in (0,1)$ indicates the fraction of that distance that is covered by $\vec{\mathbf{F}}_{{i}}$.
Parameter $\tau$ can then be interpreted as the strength of attraction, since  the larger  it is, the larger is  $\xi_{{i}}$.
In detail, we obtain  $\lim_{\tau \rightarrow 1}\mathbb{E}(\mathbf{s}_{i+1}|\mathbf{s}_{i}) =  \boldsymbol{\mu}$ and $\lim_{\tau \rightarrow 1} \xi_i = ||\boldsymbol{\mu}-\mathbf{s}_{{i}} ||_2$,
which indicates that if $\tau \rightarrow 1$ the mean of the conditional distribution of $\mathbf{s}_{i+1}$ is on the coordinate $\boldsymbol{\mu}$  and  the position at time $i+1$ is then centered on the attractor. On the other hand, if $\tau \rightarrow 0$, there is no attraction since
$\lim_{\tau \rightarrow 0} \mathbb{E}(\mathbf{s}_{i+1}|\mathbf{s}_{i}) = \mathbf{s}_{i} $ and $ \lim_{\tau \rightarrow 0}  \xi_i = 0$, and the model reduces to a random walk. Unlike the CRW,  the variability of the movement in the fixed reference frame is constant, as can be seen from  the covariance matrices of  $\mathbf{v}_{{i}}$ and $\mathbf{s}_{{i}}$.
}

{  Some examples of  the BRW can be seen in Figure \ref{fig:EsModel2} (b), where the described characteristics  are easy to identify. Each vector
$\vec{\mathbf{F}}_{{i}}$ (solid arrow) points to the spatial attractor $\boldsymbol{\mu} = (0,0)'$ (black dot),  its direction is independent of the previous  movement direction (the dashed arrow) and depends only on the relative position with respect to the attractor. Since $\tau = 0.25$, the length of any $\vec{\mathbf{F}}_{{i}}$  is equal to 25\% of the distance between $\mathbf{s}_i$ (the tail of $\vec{\mathbf{F}}_{{i}}$) and the spatial attractor (the dot). All the ellipses have the same shape, and are only translated in space, which means that the covariance matrix does not depend on the location $\mathbf{s}_i$ or on the previous bearing-angle $\phi_{i-1}$.}



\subsection{Step and turn with an attractive point} \label{sec:TheModel}

{In this section, we introduce the main contribution of this work, that  is a way to model a path that has the CRW and BRW of the previous two sections as special cases, and which can exhibit,  with different degrees, directional persistence and an attraction to a point in space.}
Our idea is to introduce a parameter $\rho \in [0,1]$, and to define  $\vec{\mathbf{F}}_{{i}}$ as a weighted mean  between  the vector $\vec{\mathbf{F}}_{{i}}$ in the CRW and the BRW, with weights equal to $\rho$ and $1-\rho$, respectively. We then obtain, in terms of equation  \eqref{eq:scond},
$\mathbf{M}_{{i}} = (1-\rho)\tau \left(\boldsymbol{\mu}-  \mathbf{s}_{{i}} \right)+\rho\mathbf{R}(\phi_{{i-1}})\boldsymbol{\eta}$, where it is clear that the values of $\rho$ that are close to 0  indicate a strong central place attraction and weak directional persistence, while values of $\rho$ that are close to 1 indicate a weak central place attraction and a strong directional persistence. The length and direction of $\vec{\mathbf{F}}_{{i}}$ change smoothly with $\rho$.

In order to specify the conditional distribution of  $\mathbf{s}_{{i+1}}$,
we  need its covariance matrix. As  shown in the previous sections, the covariance matrix for  the BRW is fixed (equation \eqref{eq:CondsBRW}), while  it depends on the previous bearing-angle for the CRW (equation \eqref{eq:CondsD}).
Our aim is for the covariance to change smoothly from $\boldsymbol{\Sigma}$ to $\mathbf{R}(\phi_{{i-1}})\boldsymbol{\Sigma}\mathbf{R}'(\phi_{{i-1}})$, on the basis of the value $\rho$, and since the rotation matrix reduces to an identity matrix when its argument is zero, and a  smaller argument than $\phi_{{i-1}}$ rotates the matrix by less than $\phi_{{i-1}}$,
we define the variance as
$\mathbf{V}_{{i}} = \mathbf{R}(\rho\phi_{{i-1}})\boldsymbol{\Sigma}\mathbf{R}'(\rho\phi_{{i-1}})$.
 The  conditional distribution of  the path   therefore becomes
 \begin{equation}
     \mathbf{s}_{{i+1}}|\mathbf{s}_{{i}},\mathbf{s}_{{i-1}} \sim  N(\mathbf{s}_{{i}} +(1-\rho )\tau\left(\boldsymbol{\mu}-  \mathbf{s}_{{i}} \right)+  \rho \mathbf{R}( \phi_{{i-1}})
     \boldsymbol{\eta} , \mathbf{R}(\rho \phi_{{i-1}}) \boldsymbol{\Sigma}\mathbf{R}'(\rho \phi_{{i-1}})),\label{eq:modelSTAP}
 \end{equation}
 which is second-order Markovian. {If the path follows equation \eqref{eq:modelSTAP}, we say that  it is  a STAP.  The parameters in  \eqref{eq:modelSTAP} are $\boldsymbol{\mu}$, $\boldsymbol{\eta}$, $\boldsymbol{\Sigma}$, $\tau$, $\rho$ and $\mathbf{s}_0$.
Examples  based on the CRW and BRW in Figures \ref{fig:EsModel2} (a) and (b)  are shown in Figures \ref{fig:EsModel2} (c) and (d), for two values of $\rho$.  We can see that  the STAP is closer to the BRW for $\rho=1/3$, i.e., the path is attracted to $\boldsymbol{\mu}$ and there is a modest directional persistence,  while the STAP resembles a CRW more   for $\rho=2/3$, i.e., there is a strong directional persistence but the movement is only slightly attracted to $\boldsymbol{\mu}$.}
Using equations \eqref{eq:scondV1} and \eqref{eq:scondY}, we can find the distribution of the displacement-coordinates
\begin{align}
&\mathbf{v}_{{i}} | \mathbf{v}_{{i-1}},\dots ,\mathbf{v}_{{1}}\sim     N((1-\rho )\tau\left(\boldsymbol{\mu}-  \mathbf{s}_{{i}} \right)+  \rho \mathbf{R}( \phi_{{i-1}})
\boldsymbol{\eta} , \mathbf{R}(\rho \phi_{{i-1}}) \boldsymbol{\Sigma}\mathbf{R}'(\rho \phi_{{i-1}})),\\
&\mathbf{y}_{{i}} | \mathbf{y}_{{i-1}},\dots ,\mathbf{y}_{{1}} \sim     N( (1-\rho )\mathbf{R}'( \phi_{{i-1}})\tau\left(\boldsymbol{\mu}-  \mathbf{s}_{{i}} \right)+  \rho\boldsymbol{\eta} , \mathbf{R}(\phi_{{i-1}}(\rho-1)) \boldsymbol{\Sigma}\mathbf{R}'(\phi_{{i-1}} (\rho-1))),
\end{align}
where, for the conditional distribution of  $\mathbf{y}_{{i}}$ we use the following properties of the rotation matrix: $\mathbf{R}^{-1}(w)= \mathbf{R}'(w)= \mathbf{R}(-w)$ and $\mathbf{R}(w_1)\mathbf{R}(w_2)= \mathbf{R}(w_1+w_2)$. From the above equation, it is clear that  the displacement-coordinates are  not identically distributed and the movement-metrics are location-dependent.

{It should be noted that the STAP can only be defined for observations that are  equally-spaced in time. This is due to its CRW component which, as discussed in Section \ref{sec:crw}, presents challenges for the definition of the conditional density for non constant time-intervals.}

\subsection{Hidden Markov model} \label{sec:stap-hmm}

{If we assume that the animal exhibits different behaviors in the observed time-window, the STAP model, as proposed in  \eqref{eq:modelSTAP}, may not be   flexible enough to be used to model the data, since different movement characteristics  should be modeled by different  parameters, while  the parameters  in equation \eqref{eq:modelSTAP} do not change over time.}
For this reason, trajectory tracking  data are  generally assumed to originate from a mixture-type model in which the latent cluster-membership  variable $z_{{i}} \in {Z} \subseteq \mathbb{N}$  is  interpreted as a behavior indicator, so that  $z_{{i}}=j$ means that, at time ${i}$, the animal is following $j$-th behavior.
We can easily link the behavior to the STAP parameters by assuming $ \mathbf{s}_{{i+1}} |\mathbf{s}_{{i}},\mathbf{s}_{{i-1}}  \sim N\left(\mathbf{s}_{{i}} +\mathbf{M}_{{i}, {z}_{{i}}}, \mathbf{V}_{{i}, {z}_{{i}}} \right)$
with
\begin{align}
 \mathbf{M}_{{i},{z_{{i}}}} &= (1-\rho_{z_{{i}}} )\tau_{z_{{i}}}\left(\boldsymbol{\mu}_{z_{{i}}}-  \mathbf{s}_{{i}} \right)+  \rho_{z_{{i}}} \mathbf{R}(     \phi_{{i-1}} )\boldsymbol{\eta}_{z_{{i}}}, \label{eq:par1}\\
\mathbf{V}_{{i},z_{{i}}} &= \mathbf{R}(\rho_{z_{{i}}}\phi_{{i-1}} )\boldsymbol{\Sigma}_{z_{{i}}}  \mathbf{R}'(\rho_{z_{{i}}}\phi_{{i-1}} ).\label{eq:par2}
\end{align}
In both  \eqref{eq:par1} and \eqref{eq:par2}, the parameters are indexed by the latent variable $z_{{i}}$ and they  therefore  change according to the animal's behavior at time ${i}$.
The set $(\rho_{j},\tau_j,\boldsymbol{\mu}_{j},\boldsymbol{\eta}_{j},\boldsymbol{\Sigma}_j)$ contains the STAP parameters that describe the $j$-th behavior and, for example, if $\rho_1=1$,  $\rho_2=0$  and $\rho_3 = 0.5$, the first behavior is a BRW, the second a CRW and the third is a BCRW.

 Different  approaches have been proposed to model  $z_{{i}}$, and the most commonly used is the HMM where the temporal evolution of $z_{{i}}$ is generated  by a first-order  Markov process, so that $\mathbb{P}(z_{{i}}=j|z_{{i-1}},\dots, z_{{1}}) = \mathbb{P}(z_{{i}}=j|z_{{i-1}}) = \pi_{z_{{i-1}},j}$ for $i\in \{1,\dots , T-1\}$, where the probability  depends on the previous behavior and  $z_{0}$ is
 here equal to 1. Generally,   $z_{{i}}$ is assumed to have values in the set $\{1,2,\dots, K^*\}$, where $K^*$ indicates the maximum number of behaviors that we expect to see between time ${1}$ and ${T-1}$, and the optimal $K^*$
 is chosen by considering the  opinion of experts or information criteria.
 Instead of using a model with a finite $K^*$, we propose modeling the data with the sHDP-HMM of \cite{fox2011}, using the STAP as the emission-distribution.  In the sHDP-HMM,  we have $Z \equiv \mathcal{N}$, i.e., there are infinite possible behaviors but, with a finite number of time-points, only a finite number  of them, $K$, can be observed (these are called ``non-empty states'' \citep{Schnatte2019}), and   the random variable $K$   is used to estimate the number of behaviors. The HMM component of the model is non-parametric,  since the number of possible behaviors  is assumed to be countable and unbounded,   and K is  a random variable that can be estimated.
 The proposed HMM model, which we call  STAP-HMM, is
\begin{align}
  \mathbf{s}_{{i+1}} |\mathbf{s}_{{i}},\mathbf{s}_{{i-1}}, \mathbf{M}_{{i}, {z}_{{i}}}, \mathbf{V}_{{i}, {z}_{{i}}} & \sim N\left(\mathbf{s}_{{i}} +\mathbf{M}_{{i}, {z}_{{i}}}, \mathbf{V}_{{i}, {z}_{{i}}} \right)\label{eqlike1} ,\\
  z_{{i}}|z_{{i-1}}, \boldsymbol{\pi}_{z_{{i-1}}} &\sim \text{Multinomial}(1, \boldsymbol{\pi}_{z_{{i-1}}}),\label{eq:z}\\
  \boldsymbol{\pi}_j|\alpha,\kappa, \boldsymbol{\beta}& \sim \text{Dir}\left(\alpha+\kappa, \frac{\alpha \boldsymbol{\beta}+\kappa  \delta_j}{\alpha+\kappa} \right), \\
  \boldsymbol{\beta}|\gamma &\sim \text{GEM}\left(\gamma \right),\\
  \rho_{j},\tau_j,\boldsymbol{\mu}_{j},\boldsymbol{\eta}_{j},\boldsymbol{\Sigma}_j &\sim  H,
\end{align}
where $\boldsymbol{\pi}_j$ and $\boldsymbol{\beta}$ are infinite-dimensional probability vectors,  $\text{GEM}(\cdot)$ indicates the  Griffiths-Engen-McCloskey distribution \citep{Ishwaran2002},  $\delta_j$ is the Dirac-delta, and $H$ is a distribution over the likelihood parameters which acts as a prior.
 We assume  a uniform distribution over the observed domain $\mathcal{D}$ for parameter $\mathbf{s}_{{0}}$.

For  distribution $H$, we  assume independence between the parameters  with a normal distribution   for $\boldsymbol{\mu}_j$ and $\boldsymbol{\eta}_j$, an inverse-Wishart for $\boldsymbol{\Sigma}_j$, and  a uniform distribution over $(0,1)$ for $\phi_j$.
{We want to be able to tell whether a behavior is a CRW ($\rho_j=1$), a BRW ($\rho_j=0$) or a BCRW ($\rho_j\in (0,1)$), which means that we must allow $\rho_j$  to assume,  a posteriori, not only values in $(0,1)$, but also 0 or 1. This can be done by assuming a mixed-type distribution over $\rho_j$, which then becomes a mixed-type random variable composed of a discrete distribution with probabilities $w_0>0$ and $w_1>0$ on 0 and 1, respectively, and  a distribution over $(0,1)$, which, in our case, is the uniform distribution, with an associated weight $w_{(0,1)}>0$; in order to obtain a proper distribution, it is necessary to have $w_0+w_1+w_{(0,1)}=1$.  The cumulative distribution function of the distribution over   $\rho_j$
is
\begin{equation}
\mathbb{P}(\rho_j \leq d) =
\begin{cases} \label{eq:rhodis}
0  & \text{ if } d <0,\\
w_0  & \text{ if } d =0,\\
w_0+ w_{(0,1)} d  & \text{ if } 0<d <1,\\
w_0+ w_{(0,1)}+w_1=1  & \text{ if } d \geq 1.
\end{cases}
\end{equation}
We consider the use of a mixed-type distribution over $\rho_j$ as being an important contribution of this work, since it allows us  to  easily  detect the type of behavior. } We also assume  $(\alpha, \kappa, \gamma)$ to be random variables and,  following \cite{fox2011}, closed-form expressions for the updating of these parameters  can be achieved with the following priors: $\alpha+\kappa \sim G(a_{1}, b_{1})$, $\kappa/(\alpha+\kappa) \sim B\left(a_{2}, b_{2}\right)$, and
$\gamma \sim G(a_{3}, b_{3})$, where $B(\cdot, \cdot)$ stands for the Beta distribution and $G(\cdot, \cdot)$ for the Gamma. Missing observations   are also  considered  as   parameters that have to be estimated. {Since we are  directly modeling the path,  the posterior distribution of the missing data is coherent with the observed locations, and we do not need any further  assumptions, as instead are needed in  \cite{Jonsen2005} and  \cite{McClintock2012}, where the missing locations are assumed to be on the line connecting the previous  and subsequent observed locations.}

The  suggested distributions allow the  implementation of MCMC updates  based  on Gibbs steps, with the only exceptions being $\rho_j$ and the missing data, as  shown in Appendix \ref{sec:imp}.

\subsection{Differences in  and similarities between the STAP-HMM and the  \cite{McClintock2012} proposal} \label{sec:diff}
The  \cite{McClintock2012}  model is a general framework for  BCRWs that generalizes most of the other approaches. It shares  some similarities with our proposal and  can be considered its  main competitor,  since it is an HMM with an emission distribution that has directional persistence and attractive points as it most prominent features.  Instead of coordinates,  it models the step-length and  bearing-angle assuming independence between the two components; the same model can be expressed using the turning-angle  (see equation \eqref{eq:theta}). {Even though independence between the step-length and  bearing-angle is not strictly necessary,} and their framework can still be used without it, there are very few distributions over the cylinder that can accommodate  dependence between circular and linear variables, see, for example,  \cite{ABE201791}, \cite{MASTRANTONIO2018} and references therein.  {The main idea of \cite{McClintock2012}} is to define the circular mean $\lambda_{{i},j}^{\text{mc}}$ of the angular variable, in the $j$-th behavior,  as a weighted average between  the previous direction and the direction of the spatial attractor: $\lambda_{{i},j}^{\text{mc}} = \rho_j^{\text{mc}} \phi_{{i-1}}+(1-\rho_j^{\text{mc}})\zeta_{{i}}$,
where $\zeta_{{i}}$ is the angle of the vector connecting $\mathbf{s}_{{i}}$ and the attractor, and  $\rho_j^{\text{mc}}$ is the weight of the $j$-th behavior.  Indeed, as in our proposal, with $\rho_j^{\text{mc}}=1$ the model reduces to a CRW and $\rho_j^{\text{mc}}=0$ defines a BRW, while a behavior with
$\rho_j^{\text{mc}} \in (0,1)$ has both characteristics.
Other parameters of the movement-metrics   are modeled to take into account that other movement characteristics, such as   circular variance and mean velocity, can be affected by the distance from the attractor.
A more detailed description of one of the possible  formulations of the model can be found in Appendix \ref{sec:mc}.

Although the \cite{McClintock2012} approach is similar to our proposal and it introduces the same features,
there are  some differences. The  \cite{McClintock2012} approach  is defined using  unimodal and  symmetric circular distributions, and such distributions are necessary  to be able to interpret the results. {To understand why, let us suppose that the  movement is ruled by a  bimodal circular distribution,  which  is   a mixture of two wrapped-Cauchy with means $\lambda_{{i},j}^{\text{mc}}+\pi/3$ and $\lambda_{{i},j}^{\text{mc}}-\pi/3$, small circular variances,  and equal weights. In this case, the circular mean is $\lambda_{{i},j}^{\text{mc}}$ and, if  $\rho_j^{\text{mc}}=0$, the model reduces to a BRW, and
$\lambda_{{i},j}^{\text{mc}} = \zeta_{{i}}$ is therefore the direction of the attractor.  Since there are two modes and the circular variance of the two components is small, even though    the model is a BRW, the density of the circular distribution over the direction  $\zeta_{{i}}$  is almost zero, which means that the animal never goes toward the attractor, but is more likely to be in  a direction close to   $\zeta_{{i}}+\pi/3$ or $\zeta_{{i}}-\pi/3$. We then lose the possibility of modeling and interpreting  the attractor in a meaningful  way. } A similar issue can arise if the circular mean is modeled with a non symmetrical distribution.
 On the other hand, the circular distribution of our proposal is highly flexible, as shown in Figure  \ref{fig:ProjN}, and we do not have the same problem as \cite{McClintock2012},  since we are modeling the coordinates, and  the mean of the  conditional normal density of $\mathbf{s}_{{i}+1}$ depends on the location of the attractor. In our proposal, the characteristics of the distribution of the movement-metrics (e.g., means and variances) change according to the distance from the attractor and  the previous bearing-angle (see Section \ref{sec:TheModel}), but there is no need to model them directly.

{We also want  to show that a bimodal circular distribution can be   helpful to describe an animal's path, especially in a CRW. Even if, for a given time-interval, for example 10 minutes,  the  turning-angle of a CRW is distributed as  a unimodal and symmetric distribution, e.g., a wrapped-Cauchy, if we record the data with a different time-interval, e.g., 20 minutes, and we use the recorded data to compute the turning-angle, this may be distributed as  a multimodal or asymmetric distribution, as  is possible to see  in the simulated data in Appendix \ref{sec:Bimod}. To further clarify this point, we here show  a simple example.  Let us suppose that the true  path  of an animal for  a time-interval of 10 minutes is described by a wrapped-Cauchy with a circular mean $\pi$ (half a circle),  the step-length is Weibull  distributed with a mean of 1, and both distributions have a  small variance. If $\mathbf{s}_1 = (0,0)'$, and $\phi_0=0$,  we can expect $\mathbf{s}_2$ to be close to the spatial point $(-1,0)'$,    $\mathbf{s}_3$ to be close to $(0,0)'$,  and,  since the variances of the distributions are small, we expect  $s_{3,2} \approx 0$.  This means that $\mathbb{P}(s_{3,1}<0) \approx 0.5$,
i.e., half of the time the step-length of time $i=1$ is larger than the step-length of time $i=2$, and $\mathbb{P}(s_{3,1}\geq 0) \approx 0.5$, i.e., half of the time the opposite is true.
 If we record the data   with a time-interval of 20 minutes, the first turning-angle is  computed as
 $$
\theta_{1} = \text{atan}^*(s_{3,2}-s_{1,2},s_{3,1}-s_{1,1})-\phi_0 = \text{atan}^*(s_{3,2}-s_{1,2},s_{3,1}-s_{1,1}),
 $$
see equation \eqref{eq:theta_atan}, because the second recorded coordinate is $\mathbf{s}_3$. Then, since $s_{3,2} \approx 0$,  $\mathbb{P}(s_{3,1}<0) \approx 0.5$ and $\mathbb{P}(s_{3,1}\geq 0) \approx 0.5$,
we obtain  $\mathbb{P}(\theta_{1} \in \partial 0) \approx 0.5$ and $\mathbb{P}(\theta_{1} \in \partial \pi) \approx 0.5$, where  $\partial 0$ and  $\partial \pi$ denote small sections  around  0 and $\pi $, respectively. This means that the circular distribution is unimodal for a time-interval of  10 minutes, while, if we record the data with a  time-interval of 20 minutes, it is bimodal with two modes at $\approx 0$ and $ \approx \pi$.  When data are available, we cannot know  if the turning-angle distribution recorded at the given time-interval is unimodal and symmetric and we should  therefore use a distribution that can  also take into account asymmetry and multimodality, which is not possible with the model of \cite{McClintock2012}, but is, instead, a by-product of our proposal, see Figure  \ref{fig:ProjN}.
 }

{ The \cite{McClintock2012}  framework  in the original Bayesian model and the new frequentist implementation    in  the \emph{momentuHMM} \textit{R}-package \citep{momentuHMM2018},  both require the number of behaviors to be fixed in order  to estimate the model,  but this is not necessary in the STAP-HMM. Moreover, the  prior on $\rho_j$ allows us to detect the behavior-type, while the  \cite{McClintock2012} approach  requires  an
RJMCMC to obtain the same result, or the use of informational criteria, such as the AIC in the \emph{momentuHMM} \textit{R}-package.}  The RJMCMC  is an algorithm which, apart from other problems,  presents challenges in its implementation and  in the designing of a valid trans-dimensional proposal \citep{Hastie2012}, especially since it is not possible to find a closed-form full conditional for most likelihood parameters,  and Metropolis steps need to be used. On the other hand, our MCMC  is mostly based on Gibbs steps, and it does not need a trans-dimensional MCMC.
{Moreover, working directly with  coordinates instead of  movement-metrics  allows us to easily estimate any missing  locations  as part of the model fitting, and without imposing any constraints, apart from the missing observations to only have a positive density   in $\mathcal{D}$,  while \cite{McClintock2012} assumes    linearity between non-missing points. }

\section{Real data} \label{sec:realdata}

\begin{figure}[t]
    \centering
    {\includegraphics[trim = 0 10 0 40,scale=0.22]{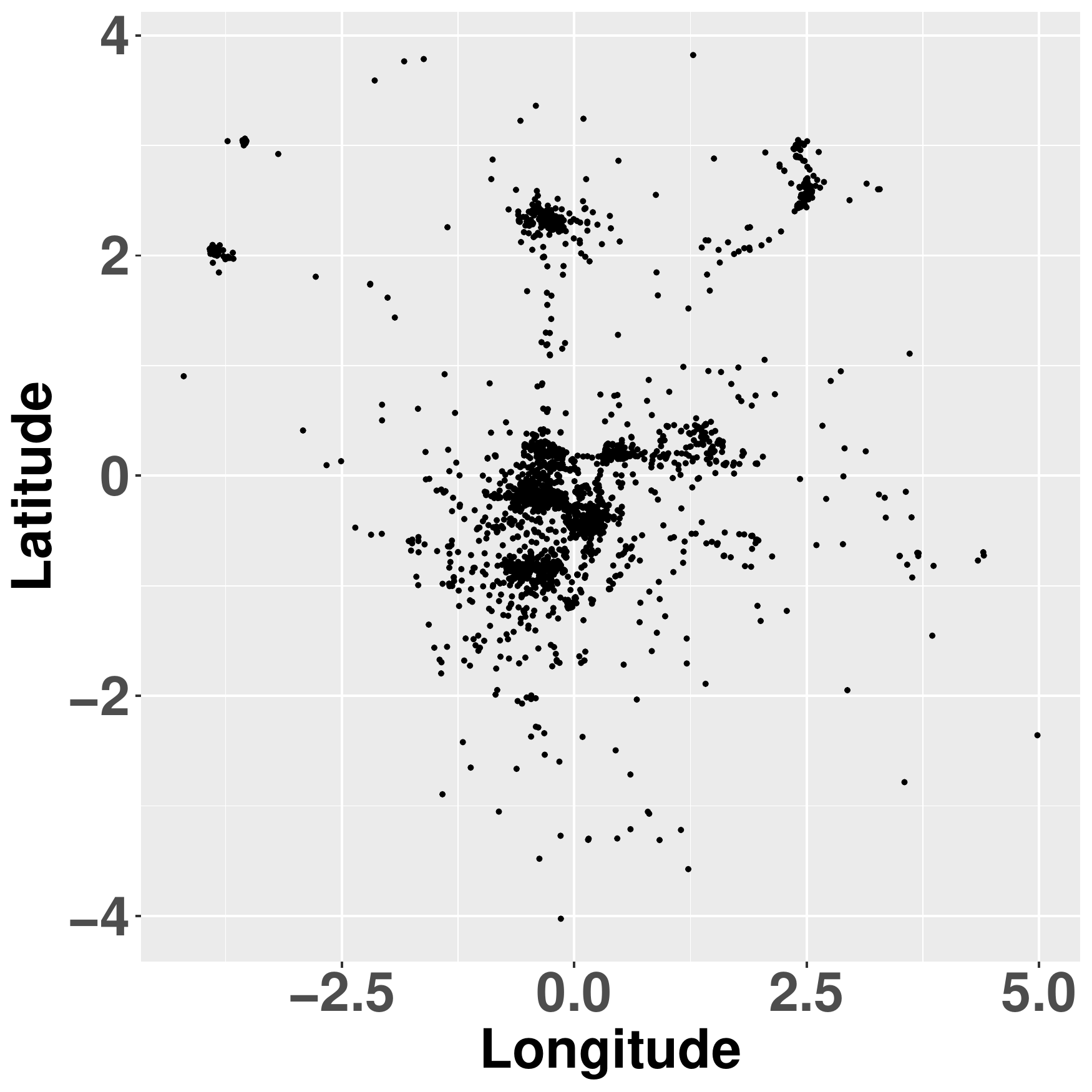}}
    {\subfloat[]{\includegraphics[trim = 0 10 0 40,scale=0.22]{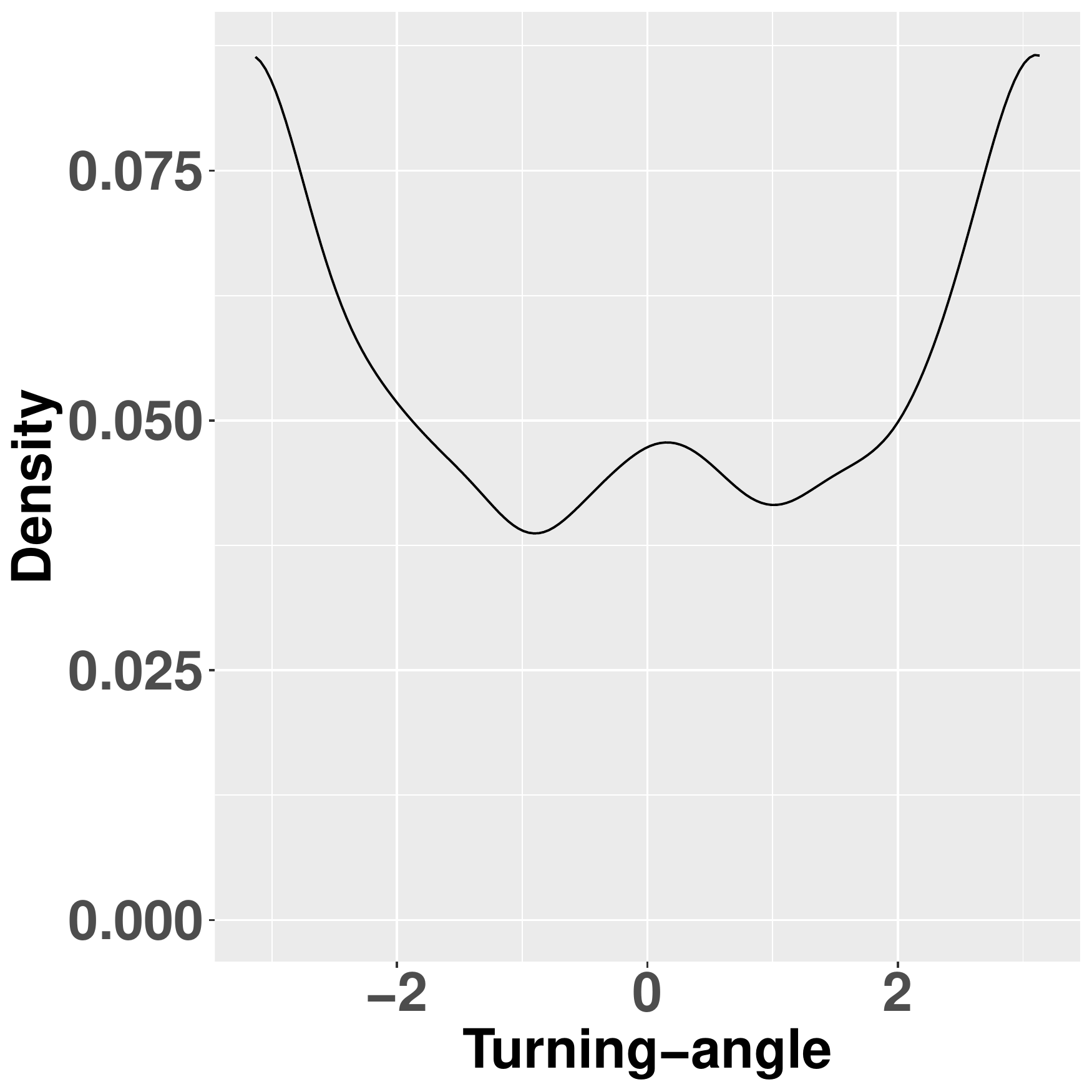}}}
    {\subfloat[]{\includegraphics[trim = 0 10 0 40,scale=0.22]{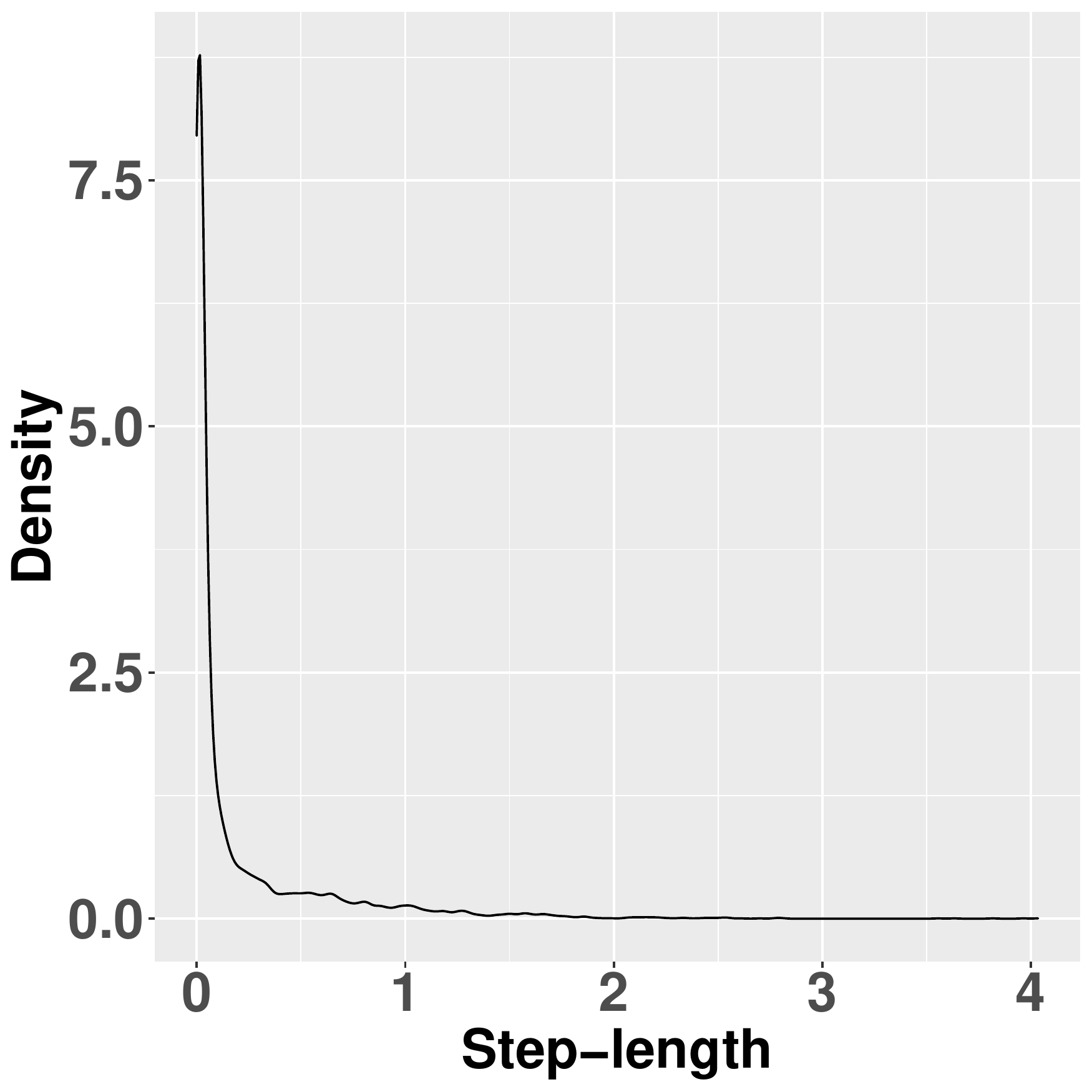}}}
    \caption{Observed coordinates (a) and kernel density estimates of movement-metric distributions ((b) and (c)).}\label{fig:ObsData}
\end{figure}



Maremma Sheepdogs are  large-breed dogs, originating from Italy, that have been used for centuries all over Europe and Asia to protect livestock from predators and thieves. These dog  are trained, from an early age,  to  live with livestock, thereby creating  a strong bond, and   adult dogs therefore view  the livestock  as their social companions and protect them from threats  for the rest of their lives. Recent studies have proved that Sheepdogs are effective in protecting livestock from a wide range of potential predators \citep{Gehring,bommel2}.
The dogs
can be fence-trained, to  remain in the proximity of the paddock where the livestock are confined, but they are generally allowed to cross  fences, move  freely  and  roam over large areas.
The use of Maremma Sheepdogs (or  general livestock guardian dogs) is relatively new in Australia and outside the country of origin, and  interest in their use is increasing \citep{van2010guardian}. Owing  to the extension of the  properties, which can be  several thousand squared hectares, it is hard, or even   impossible, for an owner to  supervise  dogs \citep{Bommel3}, which are visited rarely, and sometimes only once a week.

The dataset\footnote{available on   \url{https://www.datarepository.movebank.org/handle/10255/move.395}} contains GPS locations of Maremma Sheepdogs and sheep on three properties \citep{SheepdogRep}.
 From the available data, we selected a subset of temporally-contiguous observations of one female dog, called ``Bindi'', who belongs to the ``Rivesdale'' property, situated in North-East Victoria. Bindi's locations were recorded every 30 minutes, starting from
 2010-03-13  at 18:30, to
 2010-07-23 at  17:33.
 The data, which consist of 6335 time-points with 196 missing observations, are plotted in Figure \ref{fig:ObsData} (a); it should be noted that, to facilitate the specification of the priors,  the coordinates are centered and divided by a pooled variance.
  The kernel density  estimates of the observed movement-metrics are depicted in  Figures  \ref{fig:ObsData} (b) and (c), where we can see bimodality in both the log step-length and turning-angle, thus suggesting  a CRW, while  attractive points are suggested from Figure \ref{fig:ObsData} (a), e.g., around coordinates (0,0), which could lead to a  BRW.

Owing to the still recent use of these dogs in Australia, and given the  extension of the property, the owner is often   unaware of the dog's movements and  how she  behaves \citep{van2010guardian}.
In this work, we are  interested in finding how many behaviors the dog has and  to  describe  them.  We are  interested in evaluating  the behavior  when she is looking after the livestock, at the core of the home-range, and  at what time of the day she is more likely to leave the sheep unattended.  The description of the movement outside the central core area is also interesting, because it can be related to the dog protecting livestock from predators and be considered evidence of the effectiveness of  such dogs.

{To the best  of our knowledge, this is the first attempt that has been made  to identify the number and characteristics of the  behavior of a  Sheepdog,  using a modeling approach. Other  works have only described the behavior of a Maremma   qualitatively, or have used
 descriptive statistics \citep{van2010guardian} and tests \citep{Bommel3}.}

\subsection{Model comparison}
\begin{table}[!t]
  \scriptsize
  \centering
\begin{tabular}{c|cccccc}
  \hline
  & STAP-HMM & $\text{MC-HMM}_{4,5}$ & $\text{MC-HMM}_{7,2}$ & $\text{MC-HMM}_{5,4}$ & $\text{MC-HMM}_{3,6}$ & $\text{MC-HMM}_{6,2}$ \\ \hline
 $\text{DIC}_5$&  \textbf{-17290.37}& -16634.912  & -16557.948 &    -16317.222 & -16012.851 & -15727.025\\
 ICL&  \textbf{13245.672}& 9082.2480 & 8818.4409 & 8673.5212 & 8773.3074  & 8421.2723\\
 \hline \hline
\end{tabular} \caption{The table shows the $\text{DIC}_5$ and the ICL  for the  STAP-HMM and the 5 best $\text{MC-HMM}_{n_{b}^{\text{mc}},n_{c}^{\text{mc}}}$. The model selected by each index is indicated in bold.} \label{tab:dicDD}
\end{table}


Our aim here is to show that our proposal outperforms the model of \cite{McClintock2012} when estimated on  the motivating dataset;
the  performance is evaluated using the $\text{DIC}_5$ of  \cite{celeux2006deviance} and the ICL of \cite{Biernacki:2000}, which can  easily be computed  if no measurement error is assumed.

We use the code  provided in the supplementary material of \cite{McClintock2012} to estimate their model, modifying it   to remove the measurement error.
{The code estimates  an HMM with a given number of  BCRW and    CRW (which \cite{McClintock2012} called ``exploratory state'') and we   indicate  as $\text{MC-HMM}_{n_{b}^{\text{mc}},n_{c}^{\text{mc}}}$
the model  of \cite{McClintock2012}  which is based on $n_{b}^{\text{mc}}$ BCRW   and $n_{c}^{\text{mc}}$ CRW behaviors.} An RJMCMC is exploited to choose between different parametrizations that allow  a BCRW to be reduced to a BRW, thus making it  possibility to determine the behavior-type.
The emission distribution is the product of a Weibull distribution on the step-length and  of a wrapped-Cauchy for the bearing-angle;  these distributions are commonly used, see, for example, \cite{morales2004} and \cite{Michelot2016}. The directional persistence and attraction to the central location is introduced by modeling the wrapped-Cauchy circular mean, and  the distance from the center-of-attraction is also used to model the circular concentration and the two  Weibull parameters.
We use the priors suggested in the original paper, which are shown in  Appendix \ref{sec:mc} together with more details on the model formalization.

For our proposal, we use weakly informative priors for the likelihood parameters, namely   $N(\mathbf{0}_2,1000\mathbf{I}_2)$ for $\boldsymbol{\mu}_j$ and $\boldsymbol{\eta}_j$, $\boldsymbol{\Sigma}_j \sim IW(3,\mathbf{I}_2)$, a uniform over $(0,1)$ for $\phi_j$. {The  three weights $w_0$, $w_1$,
and $w_{(0,1)}$, which define the prior of $\rho_j$,
(see equation \eqref{eq:rhodis})
 are all equal to 1/3.}
{We also assume the  priors
$\alpha+\kappa \sim G(0.1, 1)$,
$\kappa/(\alpha+\kappa) \sim B\left(10,1\right)$,
$\gamma \sim G(0.1, 1)$,
which are defined to avoid the  production of  redundant behaviors, i.e., behaviors with similar vectors of the  likelihood parameters  (see \cite{fox2011}), since they induce  a distribution over $K$ which   is almost fully concentrated on 1.  The distribution over $K$  has been evaluated by simulating datasets  from the STAP-HMM with the same number of temporal points as the data we are modeling. } This  means that we are  a priori assuming   that  only one behavior is observed, i.e., the model is not a mixture, and  if the data support the hypothesis of different behaviors, the posterior of $K$ will move  away  from the prior. The spatial domain where the path  is recorded  is defined as the square $[-5,5]^2 \equiv  \mathcal{D}$.
The STAP-HMM    is estimated using an MCMC algorithm, implemented in  Julia 1.3 \citep{bezanson2017julia}. All the models have 5000 posterior samples, based on 125000 iterations, burnin 75000, thin 10.
The codes that can be  used to replicate the results of our proposal are available at \url{https://github.com/GianlucaMastrantonio/STAP_HMM_model}.

We decided to test $\text{MC-HMM}_{n_{b}^{\text{mc}},n_{c}^{\text{mc}}}$ with a number of latent behaviors between 2 and 9 and  all possible combinations of BCRW and CRW, e.g., with 4 behaviors  we tested the following:
$\text{MC-HMM}_{0,4}$, $\text{MC-HMM}_{1,3}$, $\text{MC-HMM}_{2,2}$,  $\text{MC-HMM}_{3,1}$, and $\text{MC-HMM}_{4,0}$. We selected 9 as the maximum number of behaviors since it is reasonable to assume that 10 or more behaviors are not likely to be observed. In fact, such a large number is
somewhat uncommon,  and the models in literature are usually tested on a maximum of  2 to 5 behaviors \citep[see, for example,][]{Langrock2012,Pohle2017,Michelot2019}. Moreover, we experienced convergence issues and extremely long computation times for models with more  than 9 behaviors.

Both $\text{DIC}_5$ and $ICL$ are likelihood-based indices and   the likelihood  of the two approaches must be  evaluated on the same data for a fair comparison. For this reason, instead of  locations $\mathbf{s}_{{i+1}}$
we evaluated the likelihood of the STAP-HMM  using $(r_{{i+1}}, \psi_{{i+1}})$, whose density can   easily be  determined from the displacement-coordinates distribution using standard results \citep[see, for example,][]{MASTRANTONIO2018}.
 We show the computed $\text{DIC}_5$ and $ICL$  for the STAP-HMM and the best five $\text{MC-HMM}_{n^{\text{b}},n^{\text{c}}}$ in Table  \ref{tab:dicDD}, which are the same for both indices. ICL and $\text{DIC}_5$ select our proposal as the best model  and  4 out of 5  $\text{MC-HMM}_{n^{\text{b}},n^{\text{c}}}$ have 9 behaviors, while the other has 8.
In order to compute the estimated number of behaviors with the STAP-HMM, we can use the posterior samples.
Let $K^b$ be the number of unique values of $\{{z}_{{i}}\}_{i=1}^{T-1}$ in the b-th MCMC iteration,  $K^b$  is therefore  a posterior sample of  $K$ and the set $\{K^b\}_{1}^{5000}$ can be used to estimate the distribution of the number of observed behaviors ($K$);  we found that  $\mathbb{P}(K=5|\mathbf{s}) \approx 0.996$ and $\mathbb{P}(K=6|\mathbf{s}) \approx 0.004$.  {These results and the ones in  Appendix \ref{sec:priors}, where we show that the number of behaviors with the highest posterior probability does not change for a wide range of STAP-HMM hyperparameters priors,
indicate that only 5 behaviors are needed to describe the data.}

The STAP-HMM is not only preferable from the information criteria point of view, but it is  also  more parsimonious in terms of number of   estimated behaviors. \cite{Pohle2017} showed, in a large simulation study,   that if the emission distribution is not flexible enough, information criteria can select a number of behaviors that is much greater than the real one. {Since, as discussed in Section \ref{sec:diff},
the circular distributions used in the general framework of \cite{McClintock2012}  are not flexible, i.e.,  they are unimodal and symmetric, and independence is assumed between the step-length and bearing-angle (or turning-angle),} we believe that this is   why a large number of behaviors are selected. On the other hand, the distributions induced on the movement-metrics by our proposal, especially in terms of turning- and bearing-angle, are quite flexible.


%

\subsection{The results}

\begin{table}[!t]
\scriptsize
  \centering
\begin{tabular}{c|ccccc}
  \hline
 & j=1& j=2 & j=3  & j=4 & j=5\\
 \hline \hline
 $\hat{\mu}_{j,1}$   &  -1.26  &  -0.4  &  8.317  &  -0.35  &  0.179  \\
 (CI)  & (-63.744 60.201) & (-63.476 60.828) & (-2.394 46.037) & (-0.413 -0.286) & (0.169 0.189)  \\
 $\hat{\mu}_{j,2}$   &  0.579  &  0.064  &  4.209  &  -0.328  &  -0.404  \\
 (CI)  & (-60.586 61.246) & (-62.861 62.019) & (-6.944 32.779) & (-0.4 -0.261) & (-0.413 -0.395)  \\
 $\hat{\eta}_{j,1}$   &  -0.005  &  -0.045  &  23.83  &  -0.068  &  -0.037  \\
 (CI)  & (-0.006 -0.004) & (-0.06 -0.03) & (3.257 64.705) & (-60.042 61.38) & (-62.237 61.557)  \\
 $\hat{\eta}_{j,2}$   &  0  &  -0.001  &  -5.239  &  -0.261  &  0.275  \\
 (CI)  & (-0.001 0.001) & (-0.012 0.01) & (-21.373 2.178) & (-60.86 62.259) & (-61.478 62.021)  \\
 $\hat{\tau}_{j}$   &  0.499  &  0.492  &  0.027  &  0.628  &  0.985  \\
 (CI)  & (0.024 0.97) & (0.026 0.973) & (0.001 0.079) & (0.574 0.685) & (0.973 0.997)  \\
 $\hat{\rho}_{j}$   &  1  &  1  &  0.018  &  0  &  0  \\
 (CI)  & [1 1] & (1 1] & (0.003 0.071) & [0 0) & [0 0]  \\
 $\hat{\boldsymbol{\Sigma}}_{j,1,1}$   &  0.001  &  0.029  &  0.791  &  0.128  &  0.006  \\
 (CI)  & (0.001 0.001) & (0.022 0.039) & (0.67 0.953) & (0.104 0.158) & (0.005 0.008)  \\
 $\hat{\boldsymbol{\Sigma}}_{j,1,2}$   &  0  &  0  &  -0.078  &  0.033  &  0.001  \\
 (CI)  & (0 0) & (-0.002 0.002) & (-0.139 -0.024) & (0.018 0.049) & (0 0.001)  \\
 $\hat{\boldsymbol{\Sigma}}_{j,2,2}$   &  0.001  &  0.018  &  0.373  &  0.168  &  0.005  \\
 (CI)  & (0.001 0.001) & (0.014 0.023) & (0.315 0.444) & (0.14 0.201) & (0.004 0.006)  \\
 $\hat{{\pi}}_{1,j}$   &  0.848  &  0.073  &  0.023  &  0.044  &  0.011  \\
 (CI)  & (0.835 0.861) & (0.063 0.084) & (0.016 0.031) & (0.035 0.055) & (0.005 0.019)  \\
 $\hat{{\pi}}_{2,j}$   &  0.286  &  0.388  &  0.145  &  0.094  &  0.086  \\
 (CI)  & (0.242 0.336) & (0.335 0.44) & (0.109 0.185) & (0.057 0.135) & (0.057 0.116)  \\
 $\hat{{\pi}}_{3,j}$   &  0.066  &  0.202  &  0.513  &  0.162  &  0.056  \\
 (CI)  & (0.037 0.102) & (0.143 0.268) & (0.441 0.581) & (0.114 0.218) & (0.027 0.092)  \\
 $\hat{{\pi}}_{4,j}$   &  0.169  &  0.206  &  0.001  &  0.419  &  0.205  \\
 (CI)  & (0.125 0.218) & (0.145 0.272) & (0 0.018) & (0.343 0.493) & (0.161 0.252)  \\
 $\hat{{\pi}}_{5,j}$   &  0.873  &  0  &  0.008  &  0.003  &  0.116  \\
 (CI)  & (0.768 0.976) & (0 0.003) & (0 0.035) & (0 0.027) & (0.018 0.217)  \\
 $\hat{{\beta}}_{j}$   &  0.248  &  0.169  &  0.14  &  0.172  &  0.219  \\
 (CI)  & (0.079 0.484) & (0.039 0.373) & (0.023 0.341) & (0.039 0.375) & (0.065 0.439)  \\
 \hline \hline
 & $\alpha$ & $\kappa$  & $\gamma$   \\
 \hline
 $\hat{}$   &  0.284  &  2.701  &  1.177  \\
 (CI)  & (0.007 1.073) & (1.408 4.447) & (0.296 2.754)  \\
 \hline \hline
\end{tabular} \caption{Posterior means and 95\% CIs of the STAP-HMM parameters for  K=5.} \label{tab:resSTAP}
\end{table}

\begin{figure}[t]
    \centering
    {\subfloat[First behavior]{\includegraphics[trim = 0 10 0 20,scale=0.25]{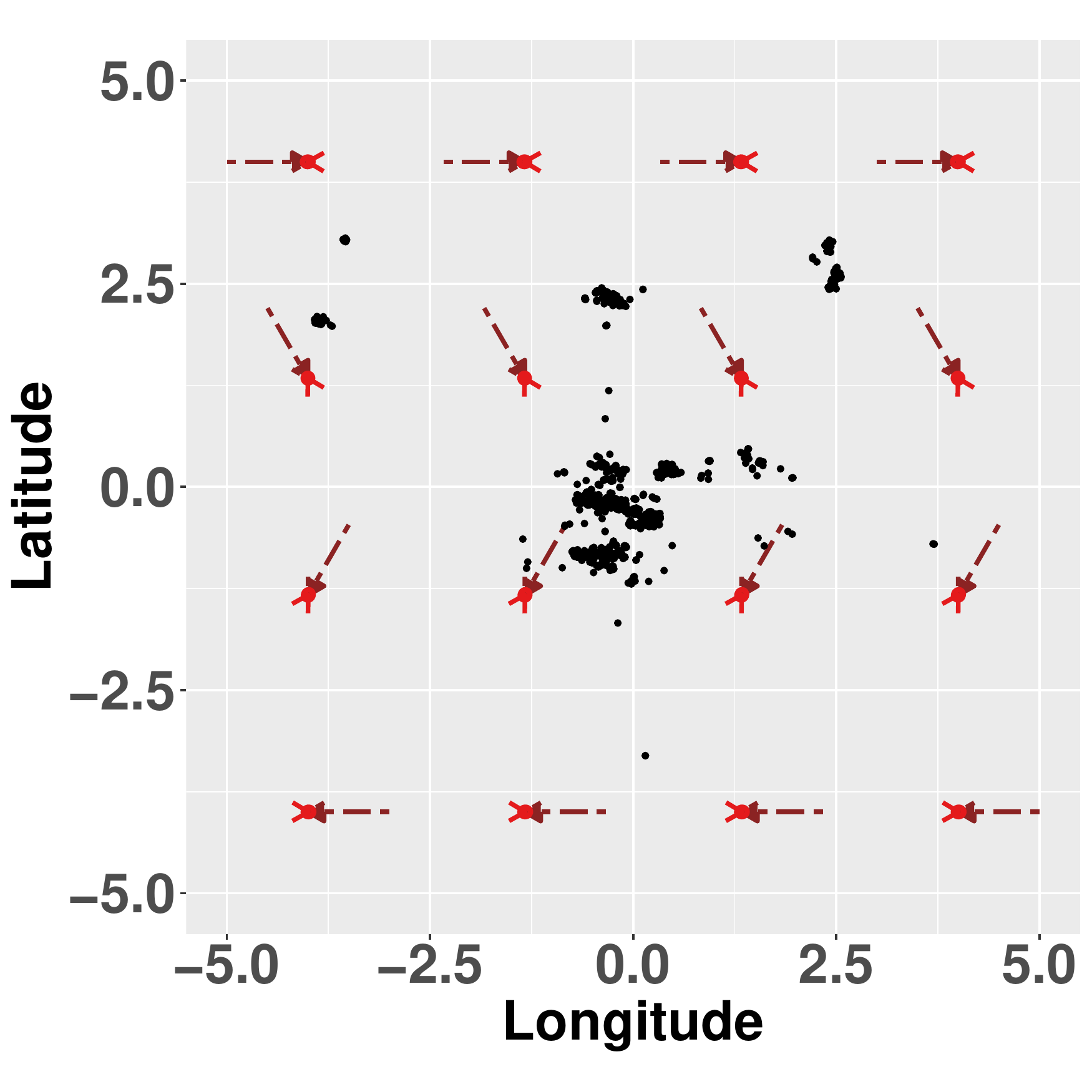}}}
    {\subfloat[Second behavior]{\includegraphics[trim = 0 10 0 20,scale=0.25]{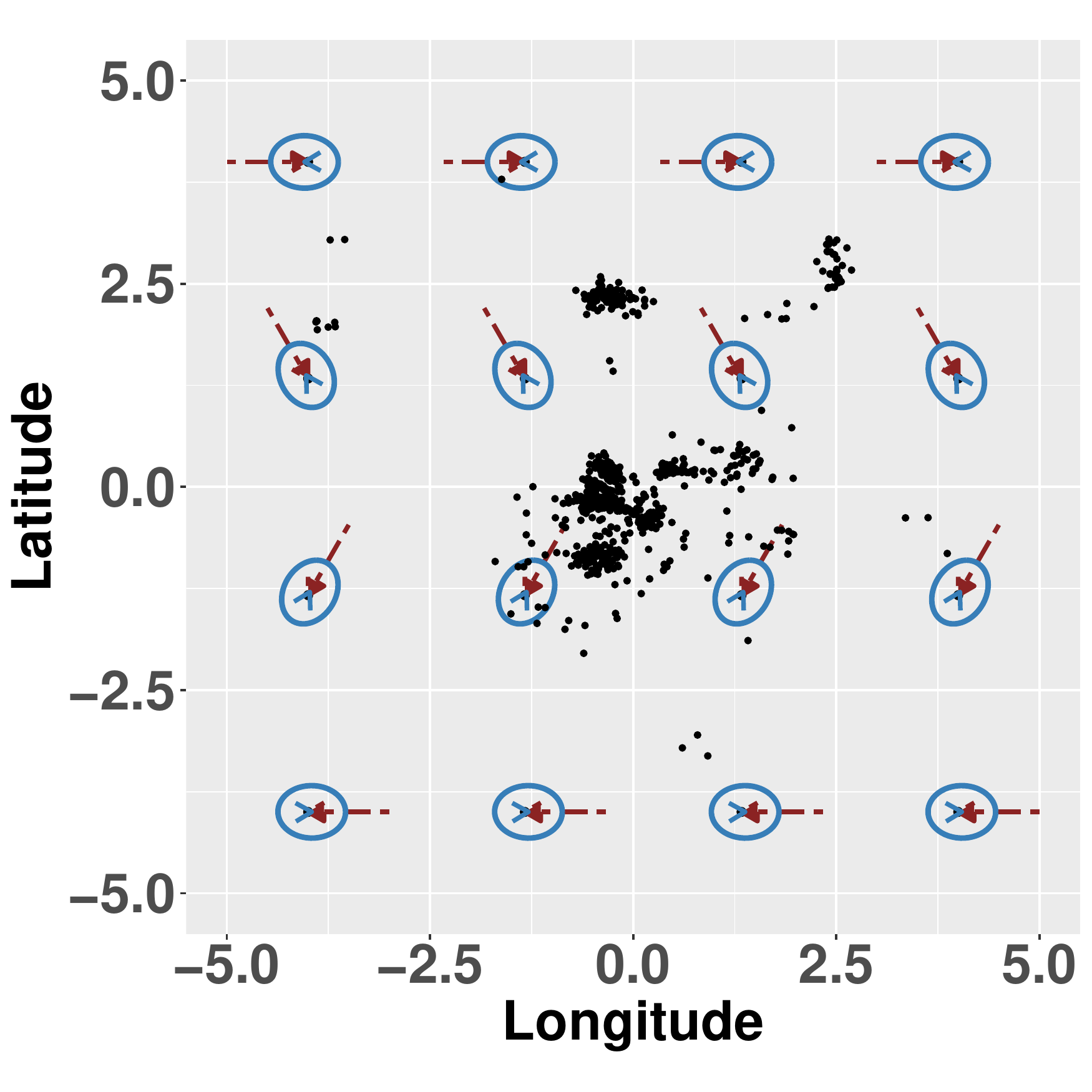}}}
    {\subfloat[Third behavior]{\includegraphics[trim = 0 10 0 20,scale=0.25]{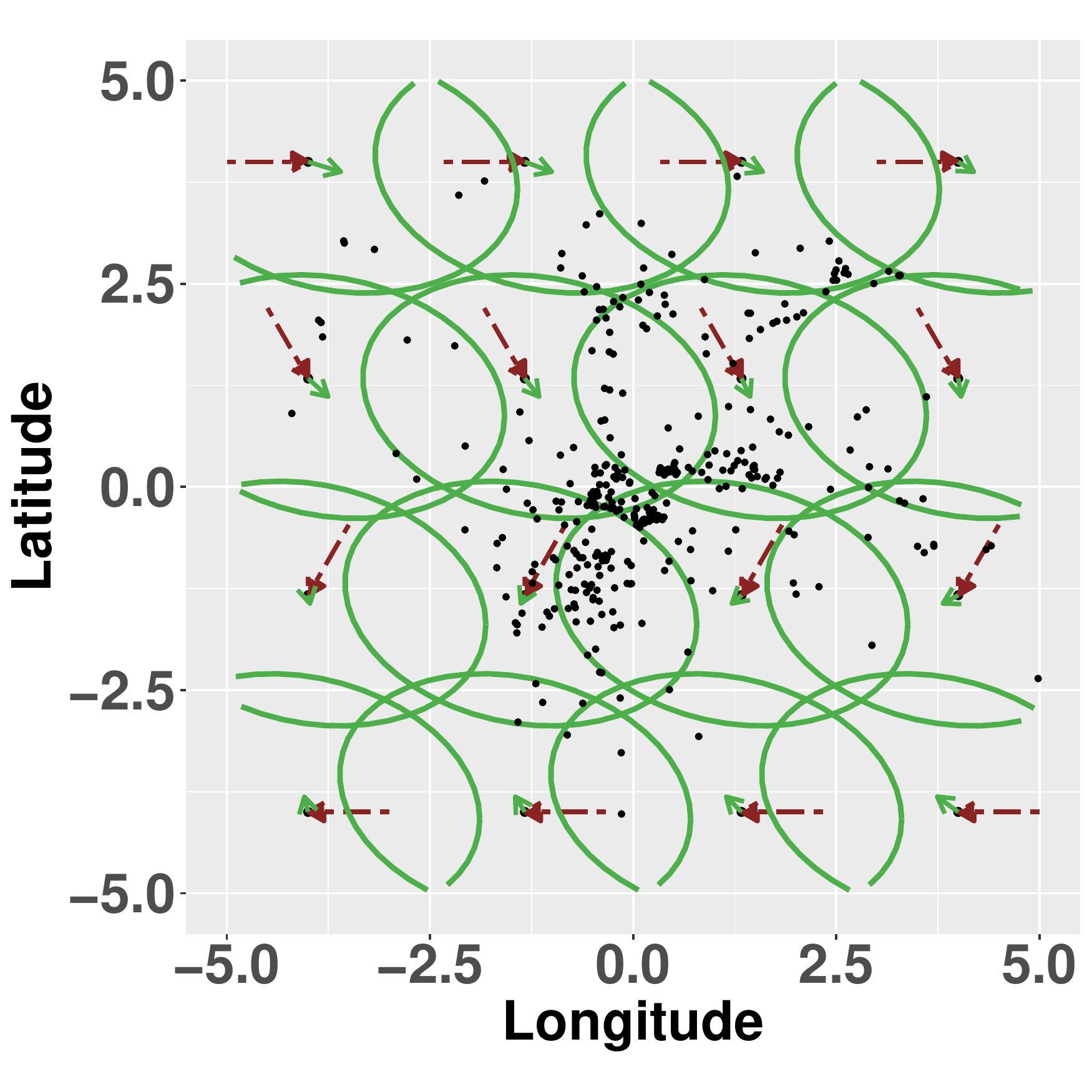}}}\\
    {\subfloat[Fourth behavior]{\includegraphics[trim = 0 10 0 20,scale=0.25]{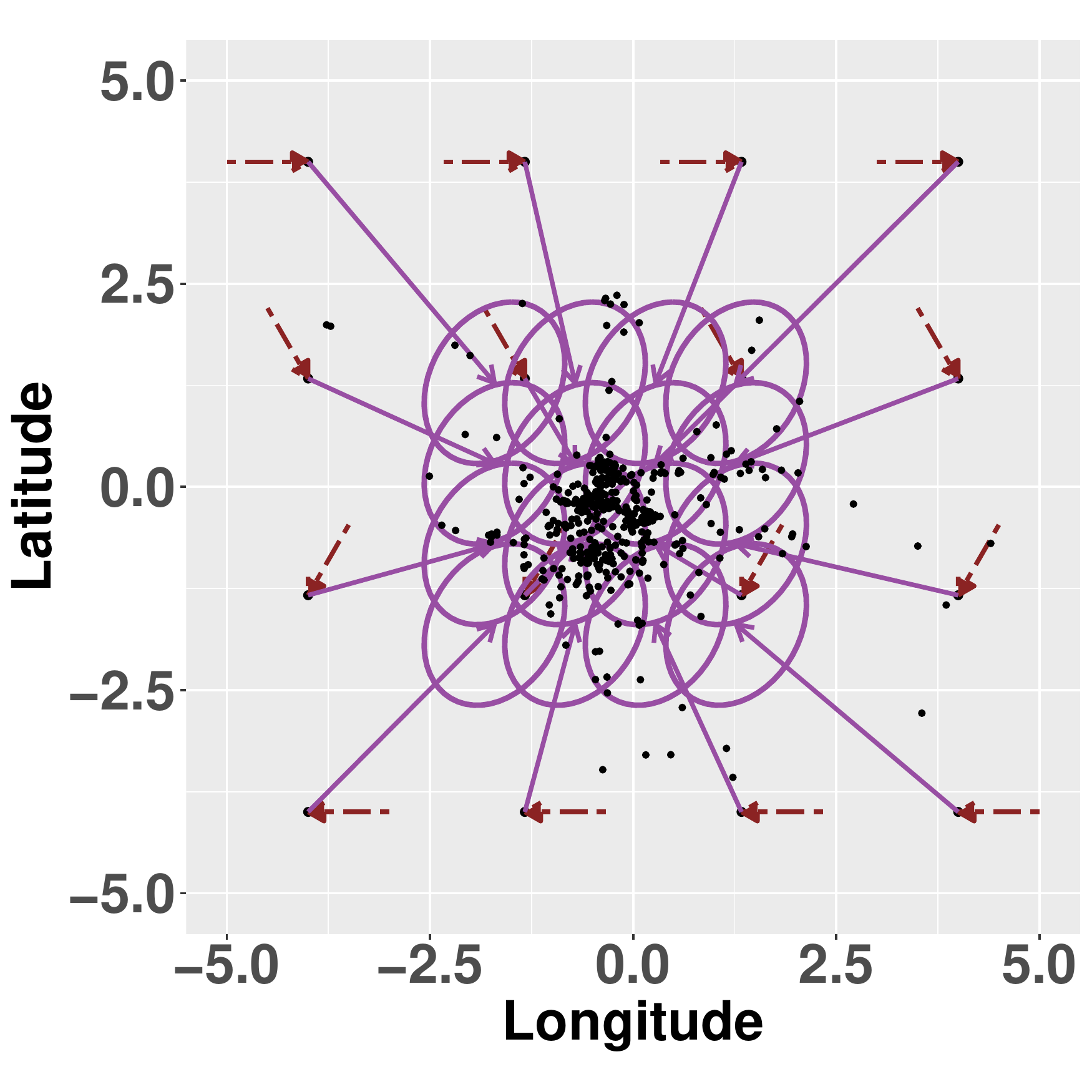}}}
    {\subfloat[Fifth behavior]{\includegraphics[trim = 0 10 0 20,scale=0.25]{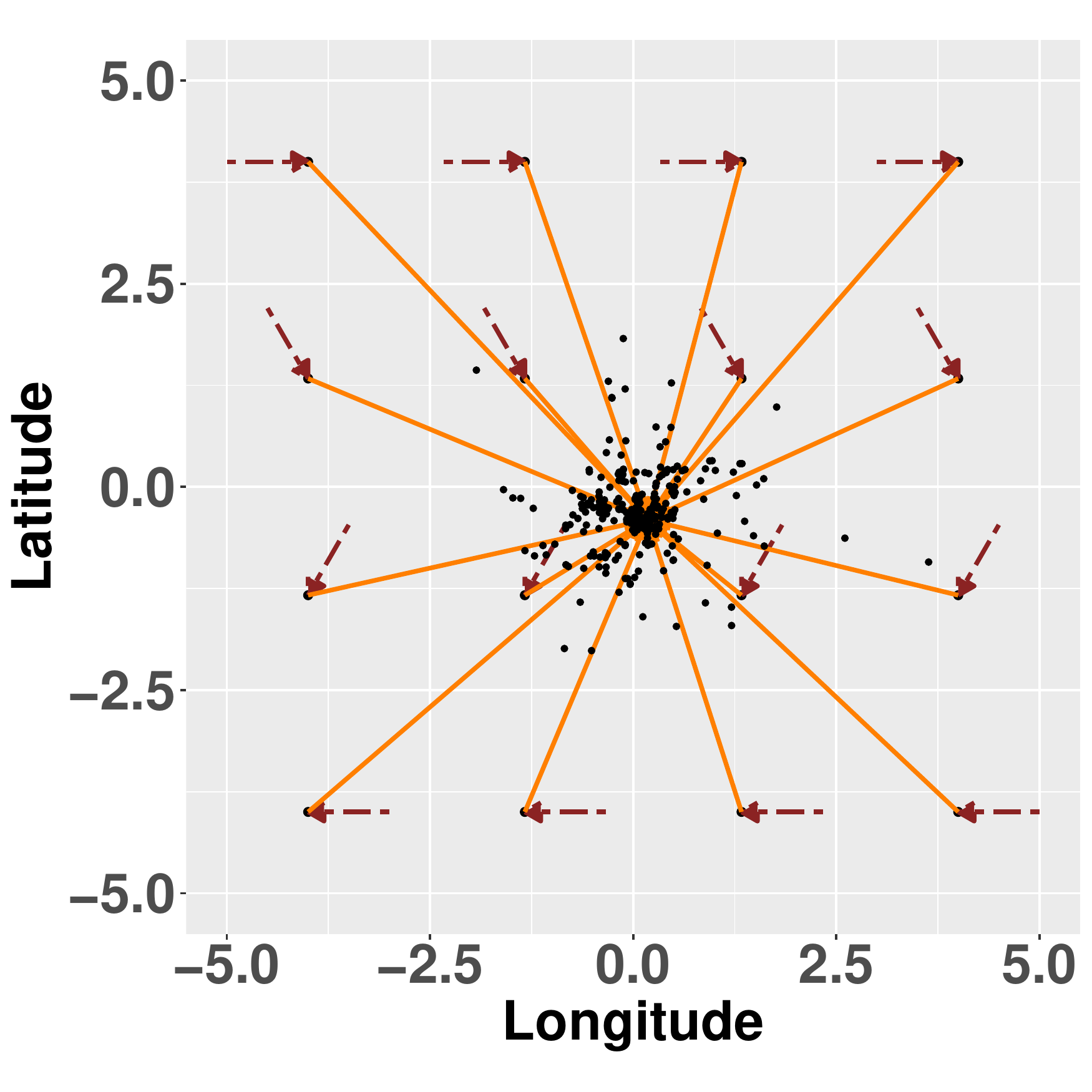}}}
    \caption{ Graphical representation of the predictive conditional distribution of $\mathbf{s}_{{i+1}}$ for different possible values of $\mathbf{s}_{{i}}$ and  previous directions.
    The dashed arrow represents the movement between  $\mathbf{s}_{{i-1}}$ and $\mathbf{s}_{{i}}$.
     The solid arrow is $\protect\vec{\mathbf{F}}_{{i}}$,  while the ellipse represents an area containing  95\% of the probability mass of  the  predictive conditional  distribution of $\mathbf{s}_{{i+1}}$, computed using \eqref{eq:ell}. The dots for the $j$-th figure are the coordinates for which $\hat{z}_{{i}}=j$. }\label{fig:PathModelSTAP}
\end{figure}
\begin{figure}[t]
    \centering
    {\subfloat[First behavior]{\includegraphics[trim = 0 10 0 20,scale=0.3]{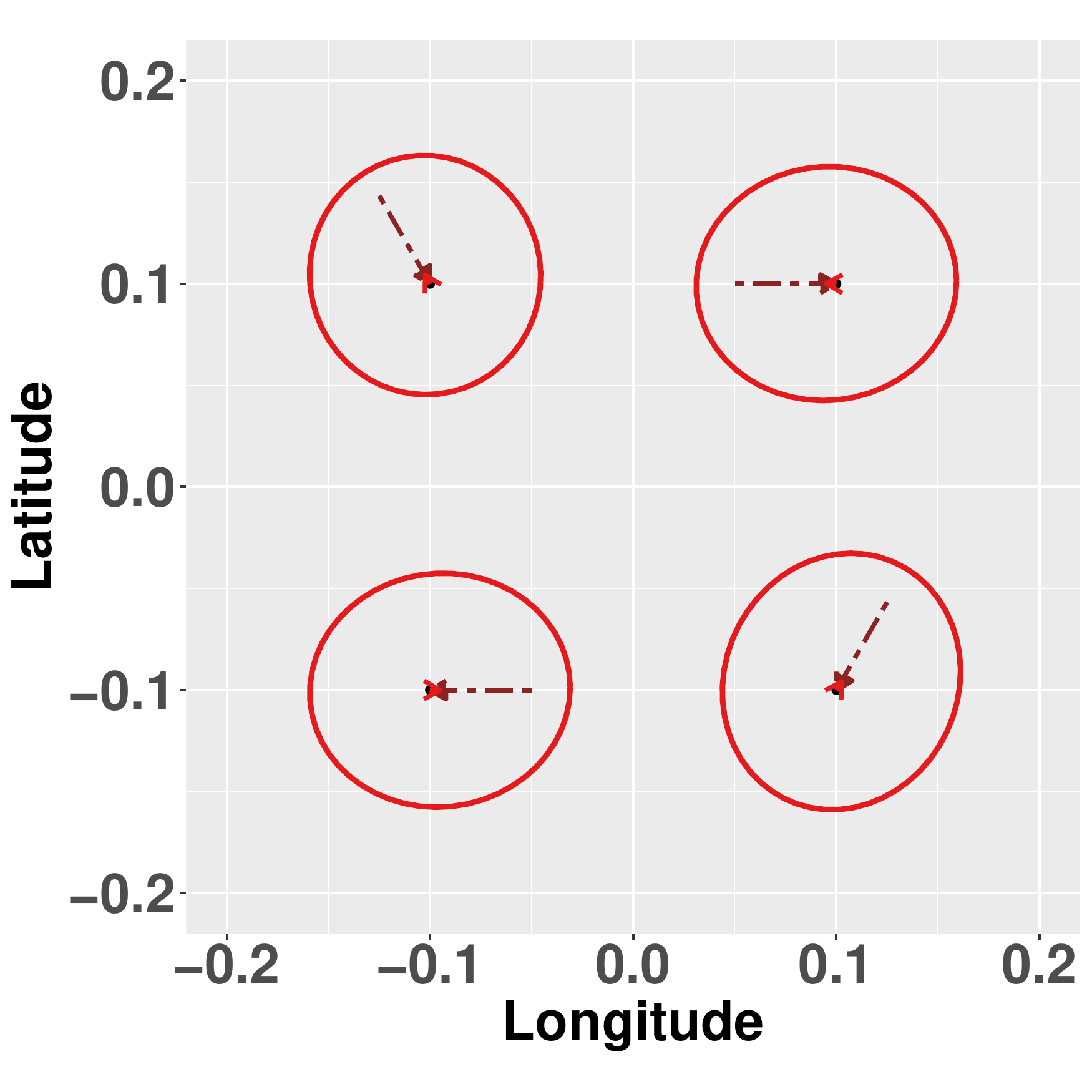}}}
    {\subfloat[Second behavior]{\includegraphics[trim = 0 10 0 20,scale=0.3]{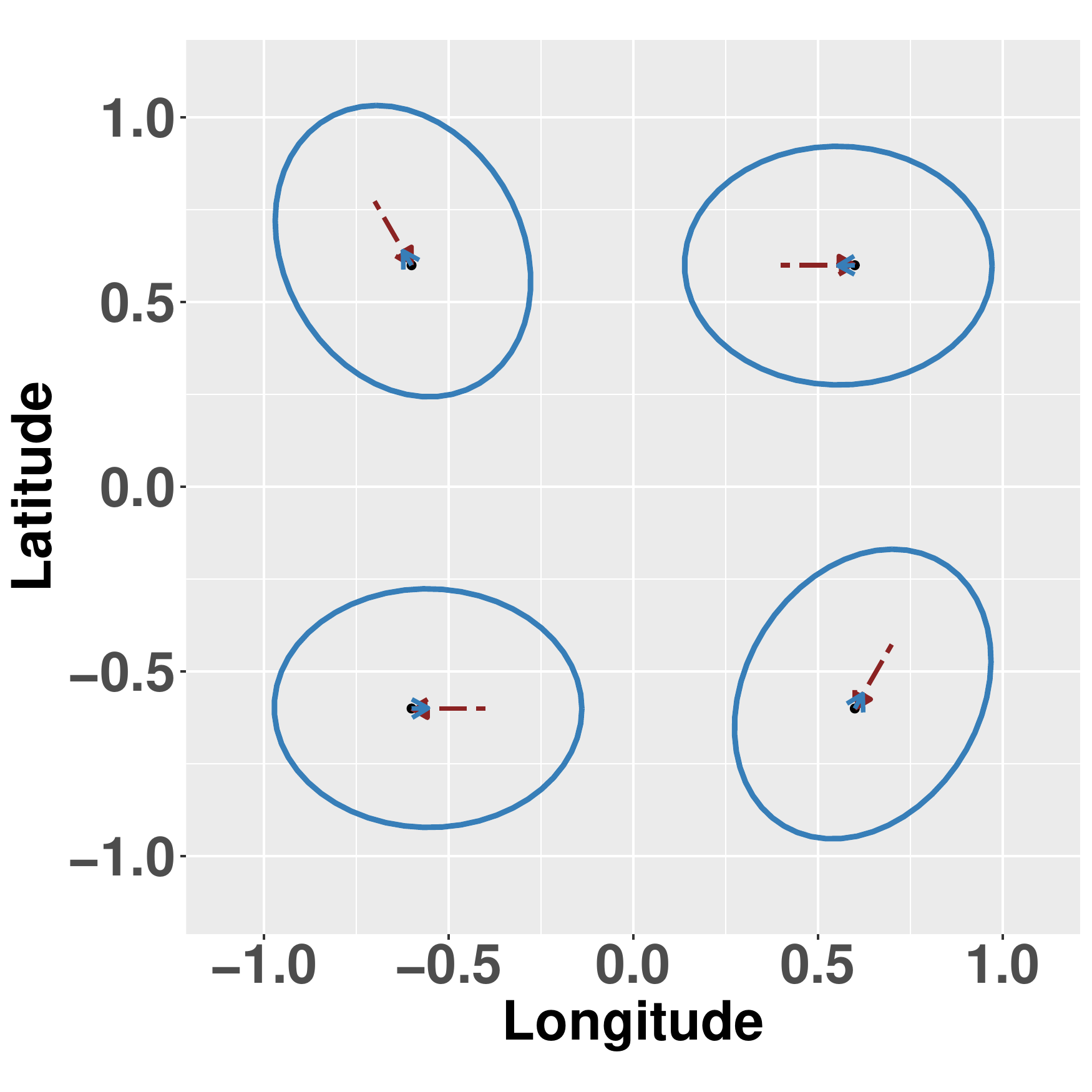}}}
    \caption{
    Graphical representation of the predictive conditional distribution of $\mathbf{s}_{{i+1}}$ for different possible values of $\mathbf{s}_{{i}}$ and  previous directions  for the first two behaviors at  greater spatial detail than in Figure \ref{fig:PathModelSTAP}.
    The dashed arrow represents the movement between  $\mathbf{s}_{{i-1}}$ and $\mathbf{s}_{{i}}$.
     The solid arrow is $\protect\vec{\mathbf{F}}_{{i}}$,  while the ellipse represents an area containing  95\% of the probability mass of  the  predictive conditional  distribution of $\mathbf{s}_{{i+1}}$, computed using \eqref{eq:ell}.   }\label{fig:PathModelSTAP_v2}
\end{figure}

%
%
%
%
%
%

\begin{figure}[t]
    \centering
    {\subfloat[First behavior]{\includegraphics[scale=0.2]{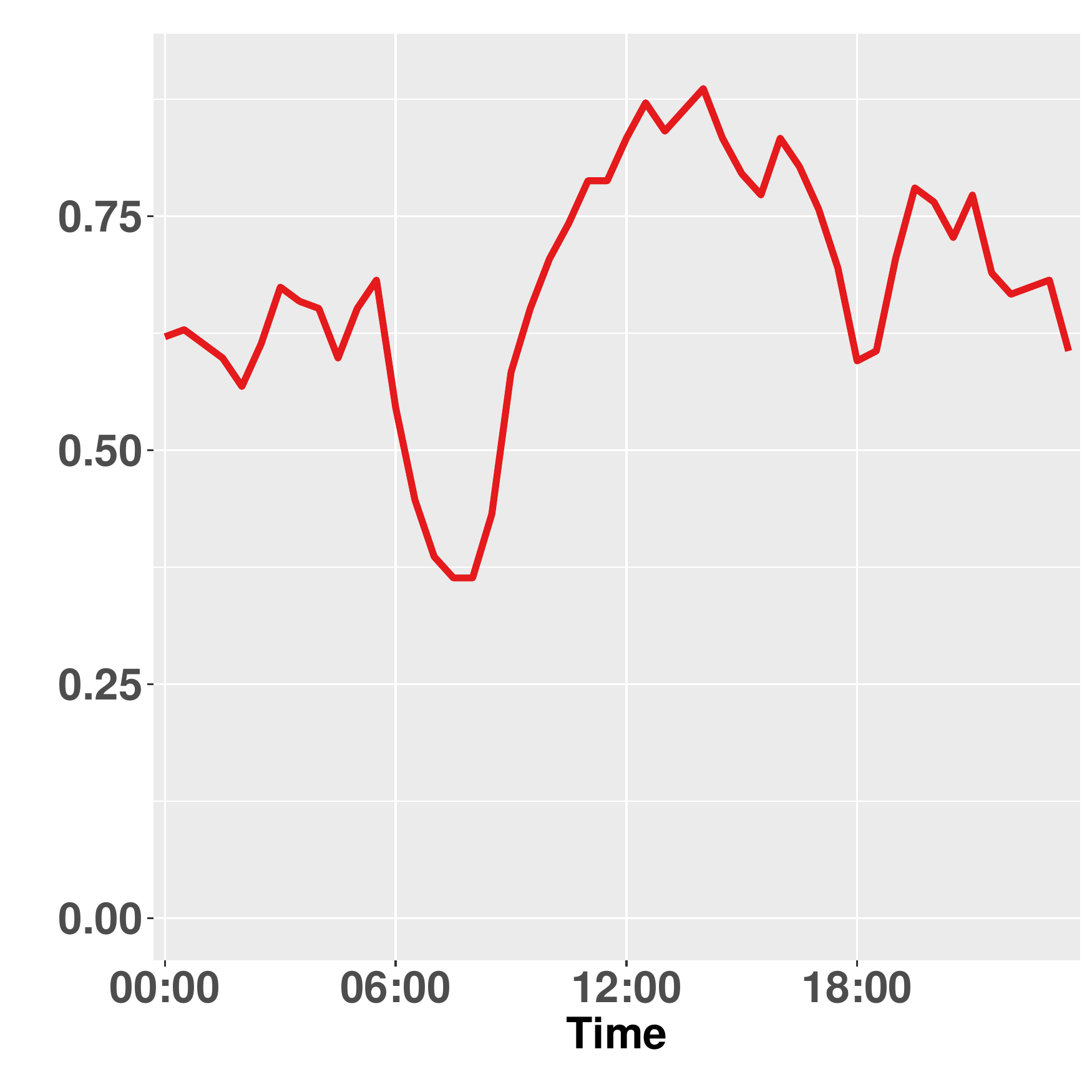}}}
    {\subfloat[Second behavior]{\includegraphics[scale=0.2]{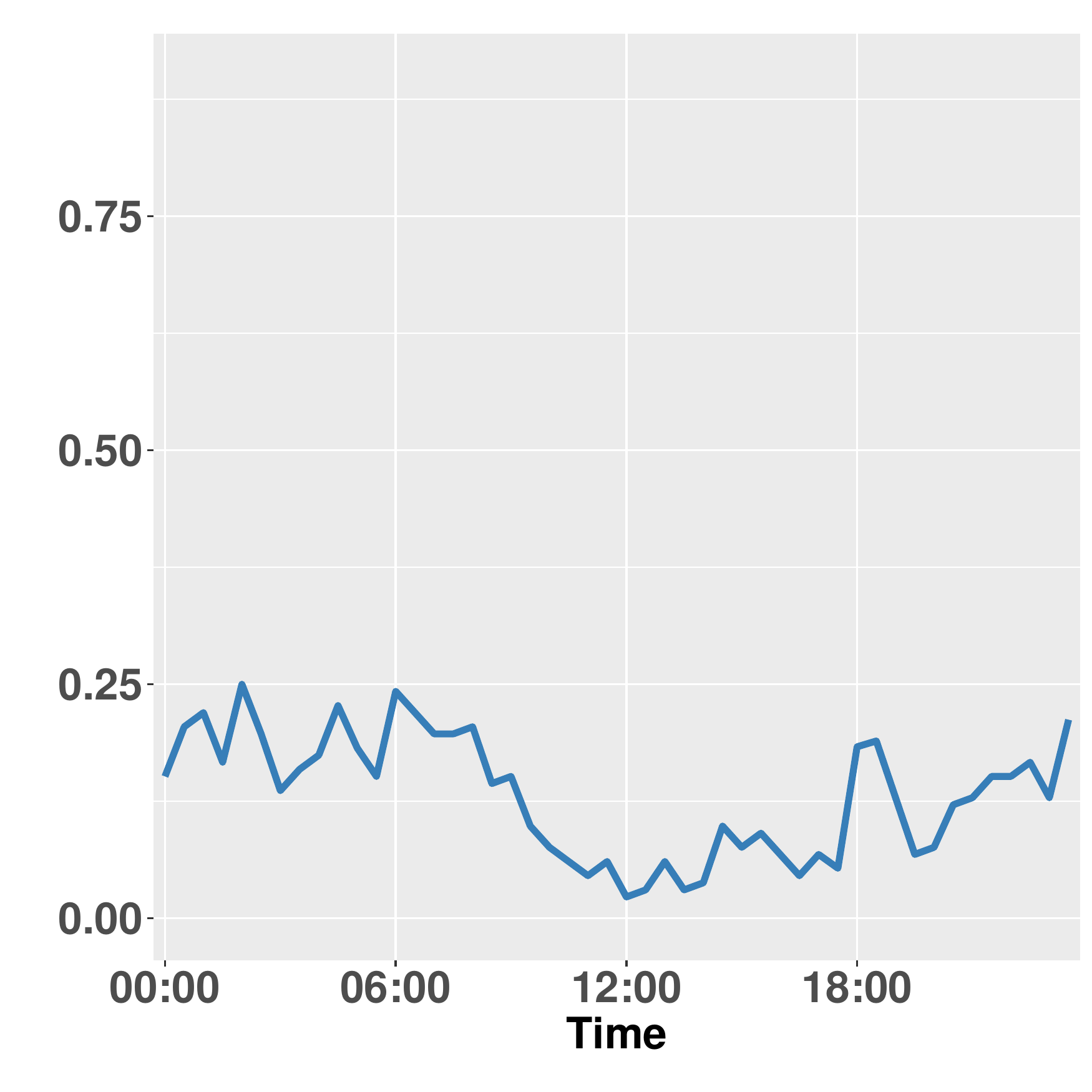}}}
    {\subfloat[Third behavior]{\includegraphics[scale=0.2]{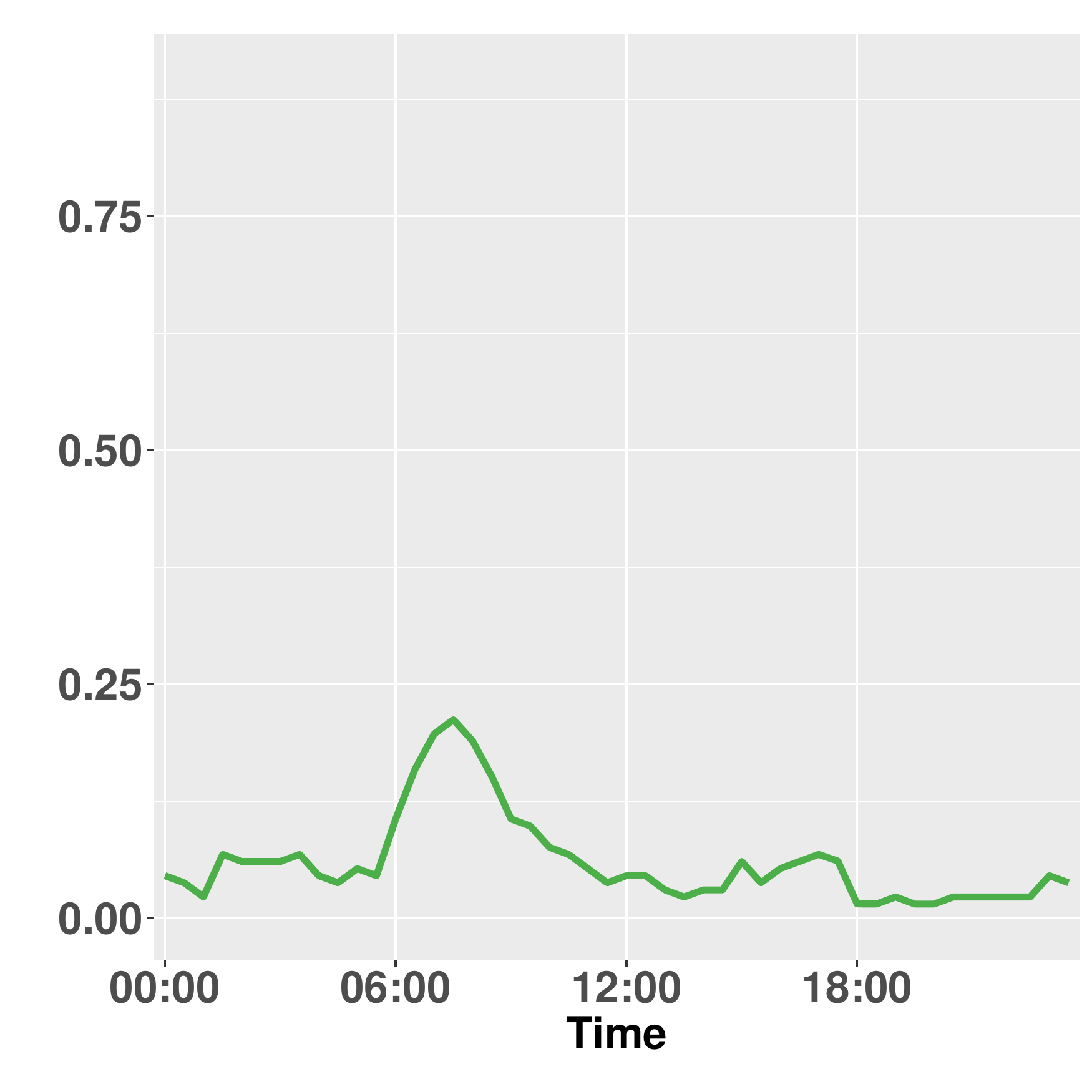}}}\\
    {\subfloat[Fourth behavior]{\includegraphics[scale=0.2]{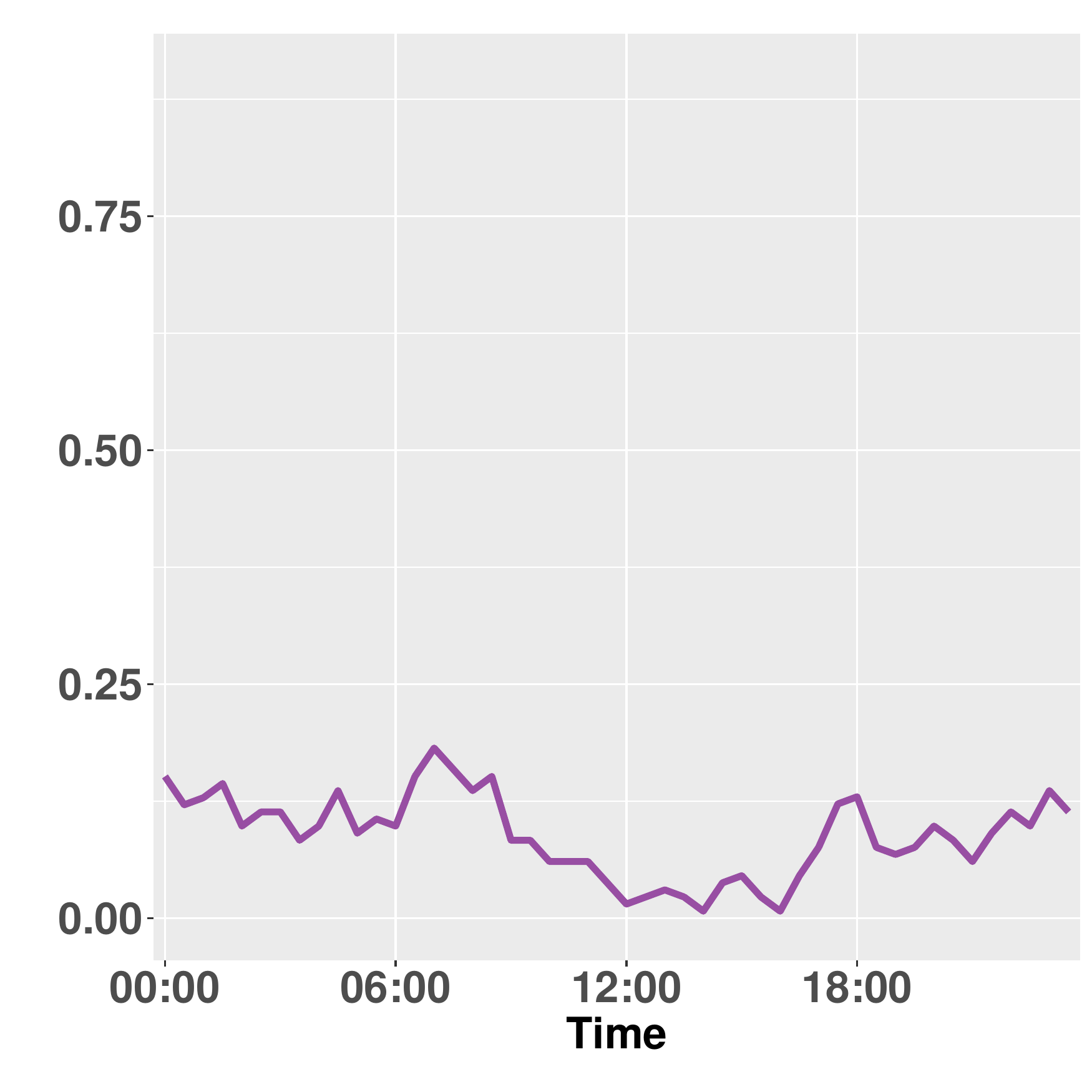}}}
    {\subfloat[Fifth behavior]{\includegraphics[scale=0.2]{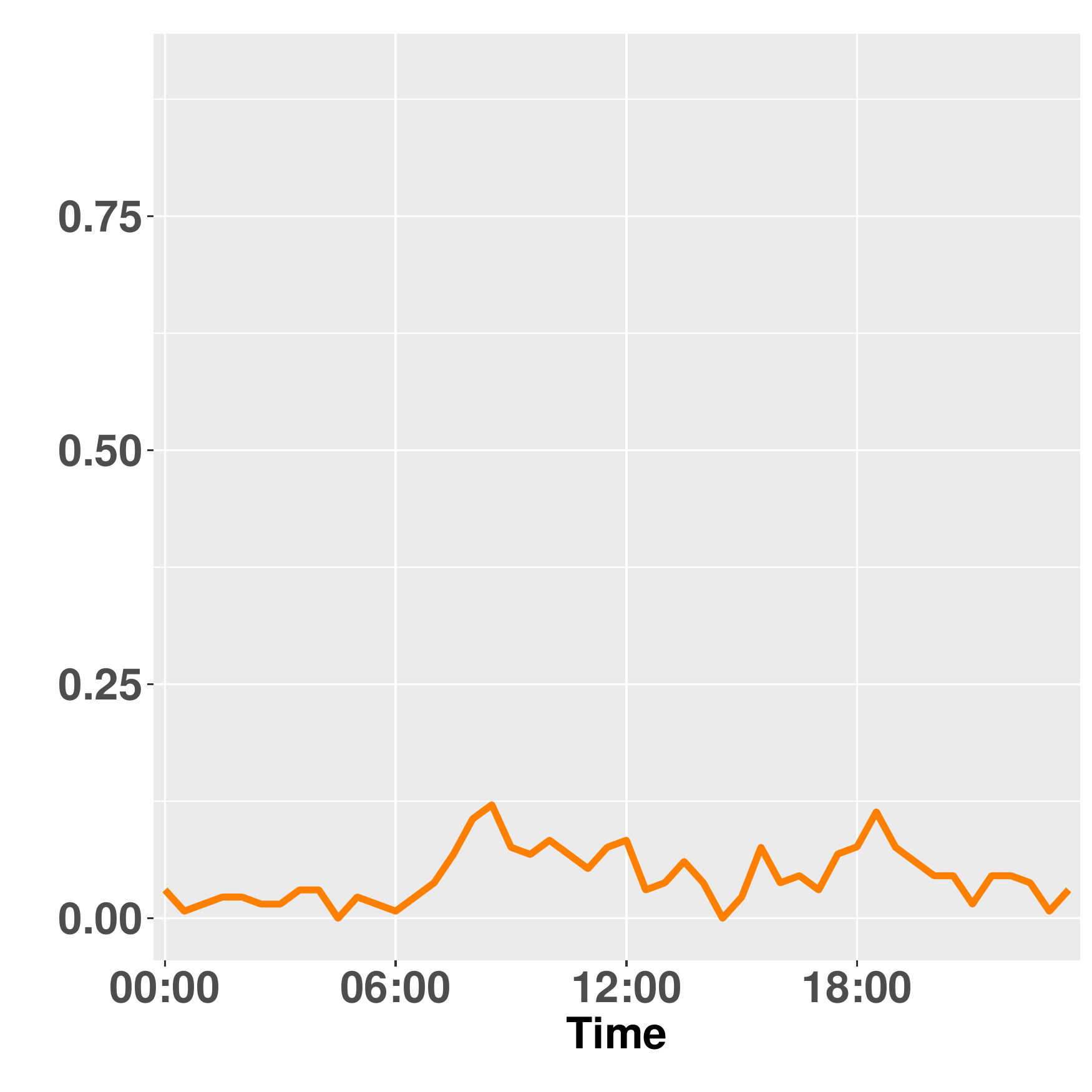}}}
    \caption{  The values of the  lines at each time (x-axis) represent the proportion of the times  that   the STAP-HMM MAP behaviors are observed (y-axis).}\label{fig:btimeSTAP}
\end{figure}

\begin{figure}[t]
    \centering
    {\subfloat[]{\includegraphics[trim = 0 10 0 40, scale=0.25]{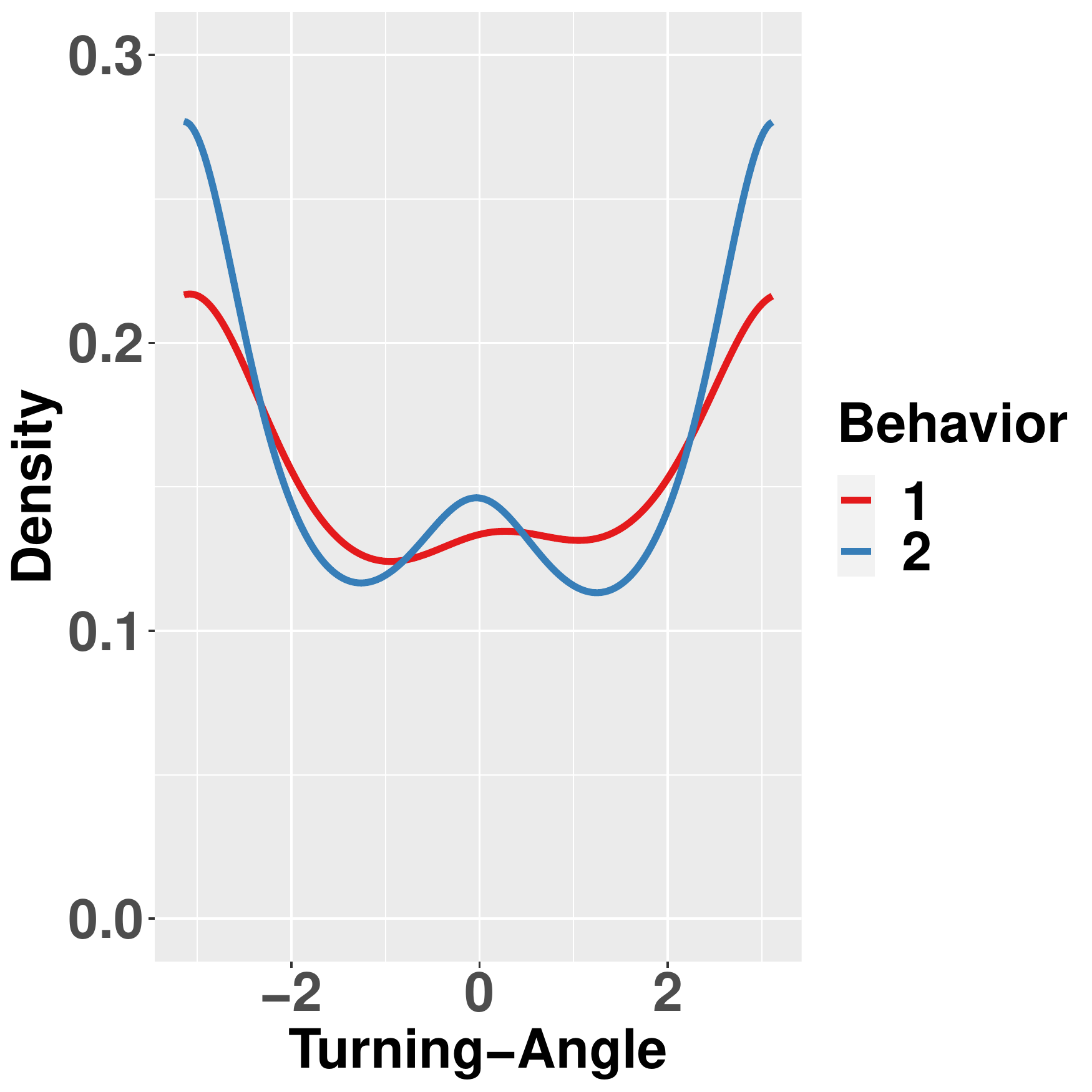}}}
    {\subfloat[]{\includegraphics[trim = 0 10 0 40, scale=0.25]{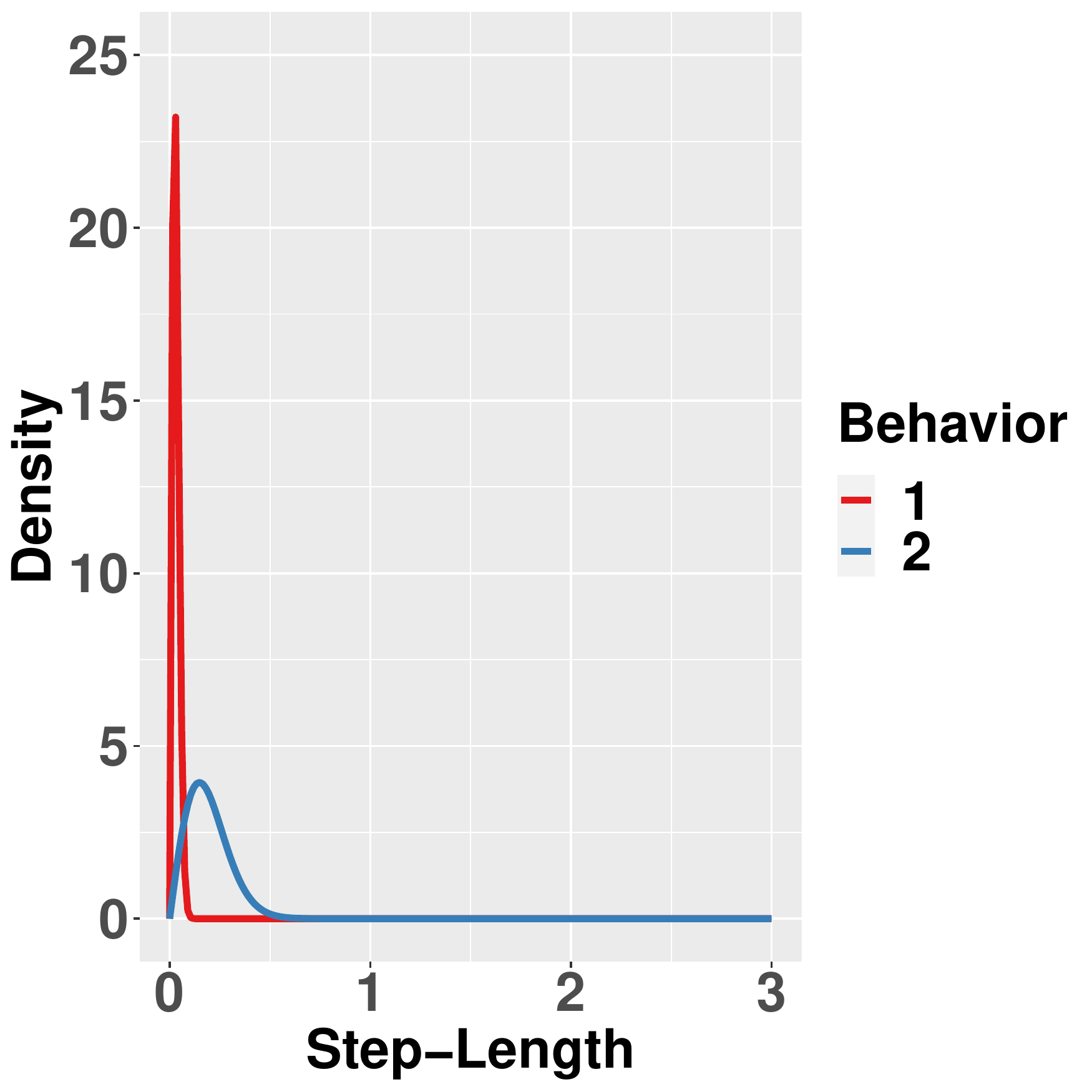}}}\\
    \caption{Turning-angle (a) and log step-length (b) predictive distributions for the first two  behaviors of the  STAP-HMM. The distributions are computed using only the CRW component. }\label{fig:steptunrSTAP}
\end{figure}


Since the   posterior mode of $K$ is almost fully concentrated on 5,  we only describe the results obtained with the set of samples that has $K^b = 5$ and, in order  to simplify the discussion, we indicate  the $j$-th latent behavior with $\text{LB}_{j}$.
We show  the posterior means (indicated using the hat notation $\hat{\cdot}$) and the 95\% credible intervals (CIs) for all the model parameters in Table    \ref{tab:resSTAP}.
 We compute the maximum-at-posteriori estimate of $ {z}_{{i}}$ for each time-point, which  we indicate with $\hat{ {z}}_{{i}}$ and we call it the MAP behavior at time ${i}$.
We can then associate a MAP behavior  to each spatial location    and  use it to see wheter there are any differences in the  spaces utilized by the animal; this information is indicated with  dots in Figure  \ref{fig:PathModelSTAP}.
The vectors $\vec{\mathbf{F}}_{{i}}$  and the predictive conditional distributions of $\mathbf{s}_{{i+1}}$ are also shown in the same figure.
It should be noted that, since the  ellipses in Figure \ref{fig:PathModelSTAP} are too small for the first two behaviors, we plot the same objects on a smaller spatial scale in Figure \ref{fig:PathModelSTAP_v2}.
The MAP behavior is also used in Figure \ref{fig:btimeSTAP} where, for a given time of the day (half an hour apart), we show  the proportion of occurrence of  the values of the MAP behaviors.


  We can see from $\hat{\rho}_j$  in  Table    \ref{tab:resSTAP}, that the first two behaviors  are  CRW, and the last two are BRW,  with  a strength of attraction that is moderate in the fourth ($\hat{\tau}_4=0.628$) and very strong in the fifth ($\hat{\tau}_5=0.985$). The third behavior is close to a  random walk since $\hat{\tau}_3 = 0.027$,  with a slight direction persistence ($\hat{\rho}_3=0.018$). The same information can be deduced from Figures \ref{fig:PathModelSTAP}
 and \ref{fig:PathModelSTAP_v2}, which also give   more insight into the movement characteristics.
 The predictive distribution of the  turning-angle and log step-length are depicted  in Figure  \ref{fig:steptunrSTAP}, albeit only  for  the behaviors that have a strong CRW component ($\text{LB}_{1}$ and $\text{LB}_{2}$) since,  as discussed in Sections \ref{sec:brw} and \ref{sec:TheModel},  the distribution of the movement metrics in BRWs and BCRWs  are location-dependent.

 The strong attraction present in
 $\text{LB}_{5}$ can be seen in Figure \ref{fig:PathModelSTAP}, since  the heads of the vectors $\vec{\mathbf{F}}_{{i}}$ are all on the same coordinates, regardless of the spatial location.  It is also possible to see the moderate attraction  of $\text{LB}_{4}$ in the same figure.  We can see from the CIs of $\boldsymbol{\mu}_4$ and $\boldsymbol{\mu}_5$ that the two attractors are well defined in space and different from each other.
 On the other hand,  the  length  of $\vec{\mathbf{F}}_{{i}}$  is approximatively zero in the first two behaviors, see Figure \ref{fig:PathModelSTAP_v2}, but the variance of the conditional distribution of $\mathbf{s}_{{i+1}}$ rotates with the bearing-angle (see the direction of the  major axis of the ellipses) which is a characteristic of  directional persistence. Figure   \ref{fig:steptunrSTAP} shows that $\text{LB}_{1}$
 has  a slower speed than  $\text{LB}_{2}$ and both have  a bimodal circular distribution with the  major mode at around $-\pi$ and a smaller one at  0. These modes  indicate that the dog tends to move in the opposite direction to the previous movement (mode at $-\pi$) or in the same direction (mode at $0$).
 The   vector   $\vec{\mathbf{F}}_{{i}}$ in $\text{LB}_{3}$ changes slightly according to the spatial location, but the attractor is not well defined in space since the CI of $\boldsymbol{\mu}_3$ is large and contains the entire  space where the animal has been observed. It should be noted that the major axis of the ellipses rotate  slightly in Figure \ref{fig:PathModelSTAP} (c), thus suggesting  a very small  directional persistence.

 In  Appendix  \ref{sec:realOtherRes}, we show simulated coordinates from the STAP-HMM, obtained using  the  posterior means in  Table    \ref{tab:resSTAP} as parameters, where we can see that  the coordinates  have the data characteristic, for example the central bulk of observations, as well as the extension of the ``explored space''.
 The results that are obtained when we only consider   CRW  (CRW-HMM) or  BRW behaviors (BRW-HMM) are also discussed in Appendix \ref{sec:realOtherRes}.

\paragraph*{Behavior description}
It is clear from the posterior mean of $\pi_{j,k}$ and  from   Figure \ref{fig:btimeSTAP}  that the animal spends most of her time in $\text{LB}_{1}$.
By comparing the     spatial coordinates of  $\text{LB}_{1}$, shown in Figure \ref{fig:PathModelSTAP} (a), with the property boundaries   in  Figure \ref{fig:PlosOne}, we can see that   the animal mostly stays inside the property boundaries.  This consideration, together with the low speed and the changes in direction  (Figure \ref{fig:steptunrSTAP}), allows us interpret it as
boundary-patrolling or scent-marking behavior, which has already been  observed for other  livestock guardian dogs   \citep{MCgrew}. The probability of observing $\text{LB}_{1}$  drops in the early morning and late afternoon, and  it is smaller  during the night than  for the central hours  of the day.

Regarding $\text{LB}_{2}$,
the higher speed and the smaller variance in the turning-angle distribution with respect to $\text{LB}_{1}$, the spatial coordinates   that are very similar to those of $\text{LB}_{1}$,
and the fact that it is more likely during the night and the afternoon,  are all clues that, with this behavior,  the animal could be defending the territory from predators, which  is  coherent with the  known behavior of potential predators that are present in the area where the data were collected \citep{Brook2012,Walton2017}.

There is also an increase in the probability of $\text{LB}_{3}$ in  the first hours of the day,  which is otherwise very small. This is the behavior with the largest variability in the observed coordinates  (Figure \ref{fig:PathModelSTAP} (c)) and an almost total absence of structure in the movement, from both the CRW and BRW points-of-view. With this behavior, the animal is exploring, going outside the property boundaries and also visiting nearby sheep flocks  \citep[see][]{Sheepdog}.
If the dog leaves $\text{LB}_{3}$, the probability of switching   to   $\text{LB}_{1}$ and $\text{LB}_{5}$  is small ($\hat{\pi}_{3,1}/(1-\hat{\pi}_{3,3})=0.135$ and $\hat{\pi}_{3,4}/(1-\hat{\pi}_{3,3})=0.115$),
while the probability of going to $\text{LB}_{4}$  ($\hat{\pi}_{3,4}/(1-\hat{\pi}_{3,3})=0.333$) and $\text{LB}_{2}$ ($\hat{\pi}_{3,2}/(1-\hat{\pi}_{3,3})=0.415$) is higher; the high probability of $\text{LB}_{2}$ is likely due to the possibility of spotting a  potential predator  during the exploration.

$\text{LB}_{4}$  is one of the behaviors with a well defined attractor, and its spatial coordinates are located in the central area   where the sheep are (for completeness, the observed coordinates of the sheep are depicted in the Appendix,  Figure \ref{fig:Sheep}). The spatial attractor has a posterior mean $\hat{\boldsymbol{\mu}}_4= (-0.35,-0.328)'$, which is located in the center of the  livestock paddock  \citep[for details, see][]{Sheepdog}.
When the dog is close to its attractor, the 95\% probability area, represented by the ellipses in Figure \ref{fig:PathModelSTAP} (d), is large enough to encompass most of the sheep locations,  which means that the dog is moving between the sheep.  This is more probable during the night, and two  probability spikes can be observed in the  early morning and late afternoon when there is a drop in the $\text{LB}_{1}$ probability.
It should be noted that all the previously described  behaviors have some coordinates that overlap those of the sheep, but not all of them do, which means that the dog is moving through the space according to a CRW ($\text{LB}_{1}$ and $\text{LB}_{2}$) or an RW ($\text{LB}_{3}$) and  the path sometimes overlaps  the sheep locations, probably as a way of continuously guarding the livestock while engaging in other activities.
On the other hand, when the dog shows this behavior ($\text{LB}_{4}$), she is attracted to the sheep paddock, and  is more likely to show this behavior during the night, when the sheep are more vulnerable to attacks.
This can easily  be  interpreted as the dog attending the livestock, and the high probability of moving to this behavior from $\text{LB}_{3}$ ($\hat{\pi}_{3,4}/(1-\hat{\pi}_{3,3})=0.333$), i.e., the exploring one, can be interpreted  as a way of checking on the  livestock  after
 leaving the sheep unsupervised.

In the final behavior, there is a well located attractor  and the variability of the movement is very low (the size of the ellipses in Figure \ref{fig:PathModelSTAP} (e)), which means that, for any point in space, the animal moves precisely  on the coordinate of the attractor.
 The probability of continuing  this behavior is very low ($\hat{\pi}_{5,5} = 0.116$)   and the only switching probability with a high value is $\hat{\pi}_{5,1}= 0.873$.
{In order to interpret this behavior, we can look at Figure \ref{fig:PlosOne}, which shows that there is a self-feeder,  located approximatively at the spatial location $\hat{\boldsymbol{\mu}}_5$,  where the dog can obtain dry food ad-lib.}
 As a result of these considerations,
we believe that this can be interpreted as a feeding behavior and, after feeding, the dog switches, with a high probability,  to boundary-patrolling.

\section{Final remarks} \label{sec:final}

{In this work, we have proposed an HMM that is based on a new emission distribution, which we have called STAP, that can be used to model animal movements with attractive points and directional persistence. The STAP belongs to the BCRW family.}
We have described the similarities to and differences from the model proposed by \cite{McClintock2012}, which is a generalization of previous approaches  that are aimed at combining
  attraction and directional persistence. Our proposal is easy to implement, is very flexibile, especially in terms of the  angular variable distribution, and is easily  interpretable. We also introduce a distribution over the likelihood parameters that has allowed us to detect the type of behavior of the dog.

The proposal was estimated on the motivating example, where the spatial locations of a Maremma Sheepdog were recorded.
{To the best of our knowledge, this work is the first to have used a mixture-model to identify the characteristics of the behavior  of a Maremma, and  the results we have obtained are clear and easy to interpret.
The model is able to find 5 different behaviors, two  CRW ($\text{LB}_1$ and $\text{LB}_2$), two BRW ($\text{LB}_4$ and $\text{LB}_5$) and an RW, with  slightly directional persistence ($\text{LB}_3$), which is therefore a BCRW. These behaviors are easily characterizable,  and they appear to be very different.}
 We found that the dog is on boundary-patrolling for most of the time ($\text{LB}_1$), and when she changes behavior, because she is defending the property ($\text{LB}_2$), exploring the space ($\text{LB}_3$),  staying within the sheep paddock ($\text{LB}_4$) or going to the self-feeder  ($\text{LB}_5$),  she switches back to  boundary-patrolling after a few time-points.
 {The information obtained with our model can be used to better understand how to employ these dogs and it   suggests that they may be effective in protecting livestock, since the recorded  dog  constantly guarded the property and  livestock and  never left  for an extensive period of time.}

{From the obtained results, we can infer   that a mixture-type model with a  BCRW emission distribution is needed to model such data.  As shown in Appendix \ref{sec:priors},
regardless of the priors chosen for the HMM hyper-parameters, 5 was always the   number of  behaviors  with the highest posterior probability, which points out that  a mixture-type model is the right approach. The STAP has the CRW and the  BRW as  special cases, and they can be estimated if the parameter $\rho$ is equal to 1 or 0, respectively. Since between the five behaviors that were found  we have CRWs, BRWs and also a BCRW,  a density that is able to model all of the three is necessary.}

{The two CRW behaviors have bimodal posterior predictive distributions, see Figure  \ref{fig:steptunrSTAP}.
Since the STAP with the right set of parameters  can have a unimodal and symmetric
circular density,   bimodality was needed to better fit these behaviors.  The informational criteria in Table \ref{tab:dicDD}  show that the STAP-HMM is preferable to the model of  \cite{McClintock2012}, and it requires a smaller number of behaviors to describe the data.  Hence, there is  evidence  that the STAP is superior, in terms of  description ability, to the emission distributions  proposed in  \cite{McClintock2012}.}
{
We have   shown, in  Appendix \ref{sec:Bimod}, that it is not possible  to assume a unimodal and symmetric turning-angle distribution for  all   time-intervals, which, we believe, is one of the reasons why the model of \cite{McClintock2012}  needs  a large number of behaviors to describe the data.
On the other hand, we have shown that the number of estimated behaviors of our proposal does not change  even when changing the priors over the STAP-HMM hyperparameters,
see  Appendix \ref{sec:priors}, which  suggests that the animal exhibited 5 different types of behavior in the observed time-window. }



In the future, we will extend our model to incorporate several animals, as in \cite{langrock2014b},
we will use more flexible temporal dynamics for the latent behavior switching, as in \cite{Harris201329} or  \cite{mastrantonio2019}, and we will also introduce covariates into the emission distribution.

\section*{Acknowledgements}
The author would like  to thank the Editor-in-Chief, the Associate Editor and the two anonymous reviewers for their comments that have greatly improved the manuscript.
This work has partially been developed under the MIUR grant Dipartimenti di Eccellenza 2018 - 2022 (E11G18000350001), conferred to the Dipartimento di Scienze Matematiche - DISMA, Politecnico di Torino.

\begin{appendices}
\counterwithin{figure}{section}
\counterwithin{table}{section}

\section{Implementation details} \label{sec:imp}
In this section, we describe how to sample   the STAP parameters and missing observations in the MCMC algorithm, by showing the full conditional distribution, when it can be derived in closed-form, or the proposal distribution used in the Metropolis step, when the full conditional is not available in closed-form. For the  sHDP-HMM parameters, the reader  can refer to the original paper of \cite{fox2011}, where different sample schemes are explained;  we used
the  \emph{degree L weak limit approximation}  \citep{Ishwaran2002}  with $L=200$ (algorithm 4).

Let  $\mathcal{I}_j $ be the  set  of indices $i$, so that $z_{{i}}=j$,  and let  $n_j$ be the number of elements in $\mathcal{I}_j$; $\mathcal{I}_j$ can be empty. Given the conditional specification  \eqref{eqlike1}, if $n_j>0$,   the   likelihood contribution to the  full conditional of a STAP parameter, for the $j-th$  behavior, is proportional to
\begin{equation} \label{eq:likefull}
\prod_{i \in \mathcal{I}_j} \left| \boldsymbol{\Sigma}_j \right|^{-\frac{1}{2}}\exp\left(-\frac{ \left(\mathbf{s}_{{i+1}}-\mathbf{s}_{{i}}- \mathbf{M}_{{i},j}   \right)' \mathbf{V}_{{i},j}^{-1}  \left(\mathbf{s}_{{i+1}}-\mathbf{s}_{{i}}- \mathbf{M}_{{i},j}   \right)  }{2} \right),
\end{equation}
where we used the relation $\left| \boldsymbol{\Sigma}_j \right| =\left| \mathbf{R}(\rho_j \phi_{{i-1}}) \boldsymbol{\Sigma}_j\mathbf{R}'(\rho_j \phi_{{i-1}}) \right|$.
It should be noted that if $n_j=0$, the full conditional of a parameter
is equal to its prior.

\paragraph*{Parameter $\boldsymbol{\mu}_j$}
Since we assume  $\boldsymbol{\mu}_j\sim N\left(\mathbf{B}_{\boldsymbol{\mu}},\mathbf{W}_{\boldsymbol{\mu}}\right)$, and $\boldsymbol{\mu}_j$ enters  linearly into the mean of the conditional distribution of $\mathbf{s}_{{i+1}}$, standard results  \citep{Gelman95bayesiandata} show  that the full conditional is $\boldsymbol{\mu}_j \sim N\left(\mathbf{B}_{\boldsymbol{\mu},j},\mathbf{W}_{\boldsymbol{\mu},j}\right)$
with
\begin{align}
\mathbf{W}_{\boldsymbol{\mu},j}& =  \left( (1-\rho_j)^2\tau_j^2  \sum_{i \in \mathcal{I}_j}\mathbf{V}_{{i},j}^{-1}   +\mathbf{W}_{\boldsymbol{\mu}}^{-1}  \right)^{-1},\\
\mathbf{B}_{\boldsymbol{\mu},j}& =  \mathbf{W}_{\boldsymbol{\mu},j} \left(  (1-\rho_j)\tau_j  \sum_{i \in \mathcal{I}_j}\mathbf{V}_{{i},j}^{-1}\left(\mathbf{s}_{{i+1}}- \mathbf{s}_{{i}} +(1-\rho_j )\tau_j  \mathbf{s}_{{i}} - \rho_j \mathbf{R}
\left(\phi_{{i-1}}\right) \boldsymbol{\eta}_j    \right)  +\mathbf{W}_{\boldsymbol{\mu}}^{-1}\mathbf{B}_{\boldsymbol{\mu}}  \right).
\end{align}

\paragraph*{Parameter $\boldsymbol{\eta}_j$}

If   $\boldsymbol{\eta}_j\sim N\left(\mathbf{B}_{\boldsymbol{\eta}},\mathbf{W}_{\boldsymbol{\eta}}\right)$, then,  for the same reason as above, the full conditional of  $\boldsymbol{\eta}_j$ is $ N\left(\mathbf{B}_{\boldsymbol{\eta},j},\mathbf{W}_{\boldsymbol{\eta},j}\right)$
with
\begin{align}
\mathbf{W}_{\boldsymbol{\eta},j}& =  \left( \rho_j^2  \sum_{i \in \mathcal{I}_j}\mathbf{R}'(  \phi_{{i-1}}) \mathbf{V}_{{i},j}^{-1}\mathbf{R}( \phi_{{i-1}})   +\mathbf{W}_{\boldsymbol{\eta}}^{-1}  \right)^{-1},\\
\mathbf{B}_{\boldsymbol{\eta},j}& =  \mathbf{W}_{\boldsymbol{\eta},j} \left(  \rho_j  \sum_{i \in \mathcal{I}_j}\mathbf{R}'( \phi_{{i-1}}) \mathbf{V}_{{i},j}^{-1}\left(\mathbf{s}_{{i+1}}- \mathbf{s}_{{i}} -(1-\rho_j )\tau_j (\boldsymbol{\mu}_j-\mathbf{s}_{{i}})     \right)  +\mathbf{W}_{\boldsymbol{\eta}}^{-1}\mathbf{B}_{\boldsymbol{\eta}}  \right).
\end{align}

\paragraph*{Parameter $\tau_j$}

Similarly to  $\boldsymbol{\mu}_j$ and  $\boldsymbol{\eta}_j$,  $\tau_j$ also enters  linearly into the mean of the conditional distribution of $\mathbf{s}_{{i+1}}$. Since its domain is (0,1) and the prior is  uniform,  the full conditional is therefore a truncated normal   $N\left(\mathbf{B}_{\tau,j},\mathbf{W}_{\tau,j}\right)I(0,1)$ with
\begin{align}
\mathbf{W}_{\tau,j}& =  \left( (1-\rho_j)^2  \sum_{i \in \mathcal{I}_j} (\boldsymbol{\mu}_j-\mathbf{s}_{{i}})'  \mathbf{V}_{{i},j}^{-1}(\boldsymbol{\mu}_j-\mathbf{s}_{{i}})    \right)^{-1},\\
\mathbf{B}_{\tau,j}& =  \mathbf{W}_{\tau,j} \left(  (1-\rho_j)  \sum_{i \in \mathcal{I}_j}(\boldsymbol{\mu}_j-\mathbf{s}_{{i}})' \mathbf{V}_{{i},j}^{-1}\left(\mathbf{s}_{{i+1}}- \mathbf{s}_{{i}} -    \rho_j \mathbf{R}
\left(\phi_{{i-1}}\right) \boldsymbol{\eta}_j     \right)   \right).
\end{align}

\paragraph*{Parameter $\boldsymbol{\Sigma}_j$}

We can express equation \eqref{eq:likefull} as
\begin{equation}    \label{fakeeq:full sigma}
\prod_{i \in \mathcal{I}_j} \left| \boldsymbol{\Sigma}_j \right|^{-\frac{1}{2}}\exp\left(- \frac{\left(\mathbf{R}'(\rho_j \phi_{{i-1}})   \left(\mathbf{s}_{{i+1}}- \mathbf{M}_{{i},j}   \right)\right)'\boldsymbol{\Sigma}_j^{-1}
\left(\mathbf{R}'(\rho_j \phi_{{i-1}})  \left(\mathbf{s}_{{i+1}}- \mathbf{M}_{{i},j}   \right)\right)  }{2} \right)
\end{equation}
which shows that $\boldsymbol{\Sigma}_j$ is the covariance of  normal densities. Given that the prior is $IW(a_{\boldsymbol{\Sigma}},\mathbf{C}_{\boldsymbol{\Sigma}})$,  the full conditional is $IW(a_{\boldsymbol{\Sigma},j},\mathbf{C}_{\boldsymbol{\Sigma},j})$  \citep{Gelman95bayesiandata} with
\begin{align}
a_{\boldsymbol{\Sigma},j}& = n_j+a_{\boldsymbol{\Sigma}},\\
\mathbf{C}_{\boldsymbol{\Sigma},j} & =\mathbf{C}_{\boldsymbol{\Sigma}}+ \sum_{i \in \mathcal{I}_j} \left(\mathbf{R}'(\rho_j \phi_{{i-1}})  \left(\mathbf{s}_{{i+1}}- \mathbf{M}_{{i},j}   \right)\right)\left(\mathbf{R}'(\rho_j \phi_{{i-1}})  \left(\mathbf{s}_{{i+1}}- \mathbf{M}_{{i},j}   \right)\right)'.
\end{align}

\paragraph*{Parameter $\rho_j$}

Parameter $\rho_j$ is part of the  rotation matrix argument, and as  it is therefore not possible to find a closed-form full conditional,  it  must be  sampled using a Metropolis step.
 Our suggestion is to propose a new value sampling from a mixed-type distribution composed of   two bulks of probability  in 0 and 1,  and a uniform distribution in $(max(0, \rho_j^*-c),min(1, \rho_j^*+c) )$ with $c>0$,  where $\rho_j^*$ is the last accepted $\rho_j$  in the MCMC algorithm.  The bulks of probability ensure that $\rho_j$ can assume  values 0 and 1 a posteriori, while the parameters of the uniform  are defined in order to be able to propose a new value that is close to the previously accepted one.

\paragraph*{Missing observations and $\mathbf{s}_{{0}}$}

A closed-form full conditional cannot be found for  the missing observations, and a Metropolis should therefore be used. We suggest  proposing a new value of the missing $\mathbf{s}_{{i+1}}$  sampling from its conditional distribution  $N(\mathbf{s}_{{i}}+\mathbf{M}_{{i},z_{{i}}},\mathbf{V}_{{i},z_{{i}}})$, since it will simplify the Metropolis ratio.
The initial value  $\mathbf{s}_{{0}}$ is sampled using a Metropolis step with a  proposal that is a normal distribution with  a mean equal to the previously accepted value of $\mathbf{s}_{{0}}$.

\section{Simulated examples} \label{sec:sim}

\begin{figure}[t]
    \centering
    {\subfloat[First simulated data]{\includegraphics[scale=0.25]{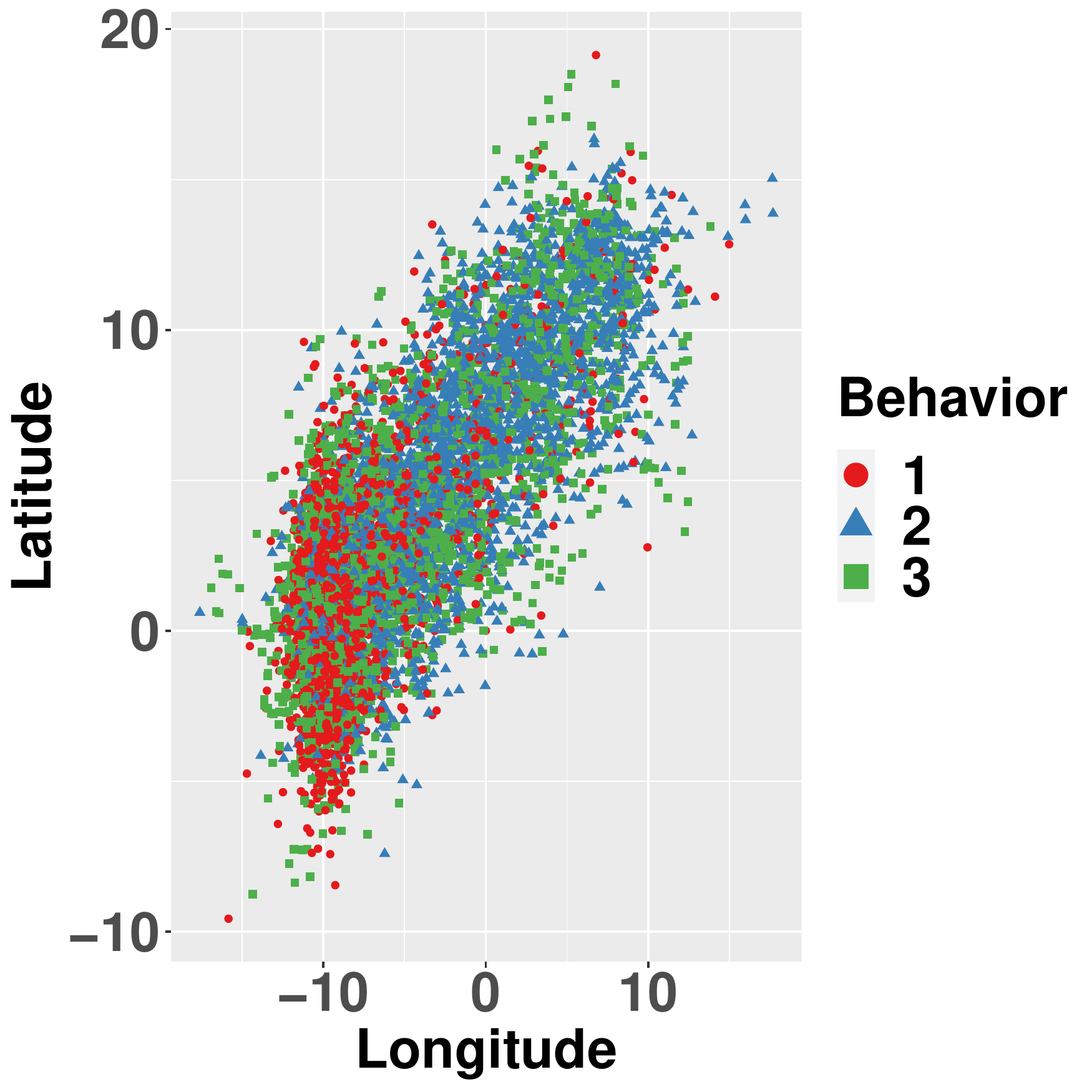}}}
    {\subfloat[Second simulated data]{\includegraphics[scale=0.25]{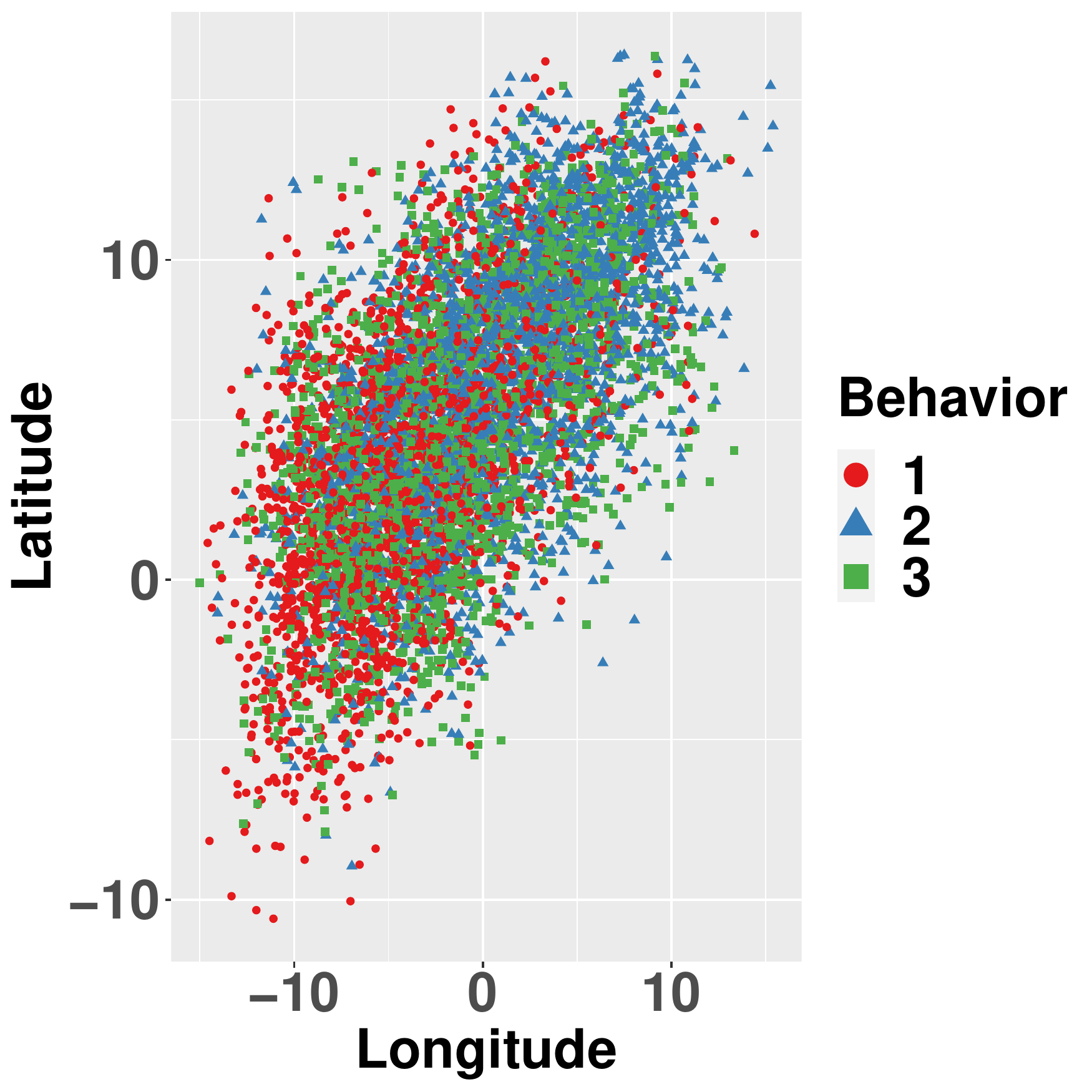}}}
    {\subfloat[Third simulated data]{\includegraphics[scale=0.25]{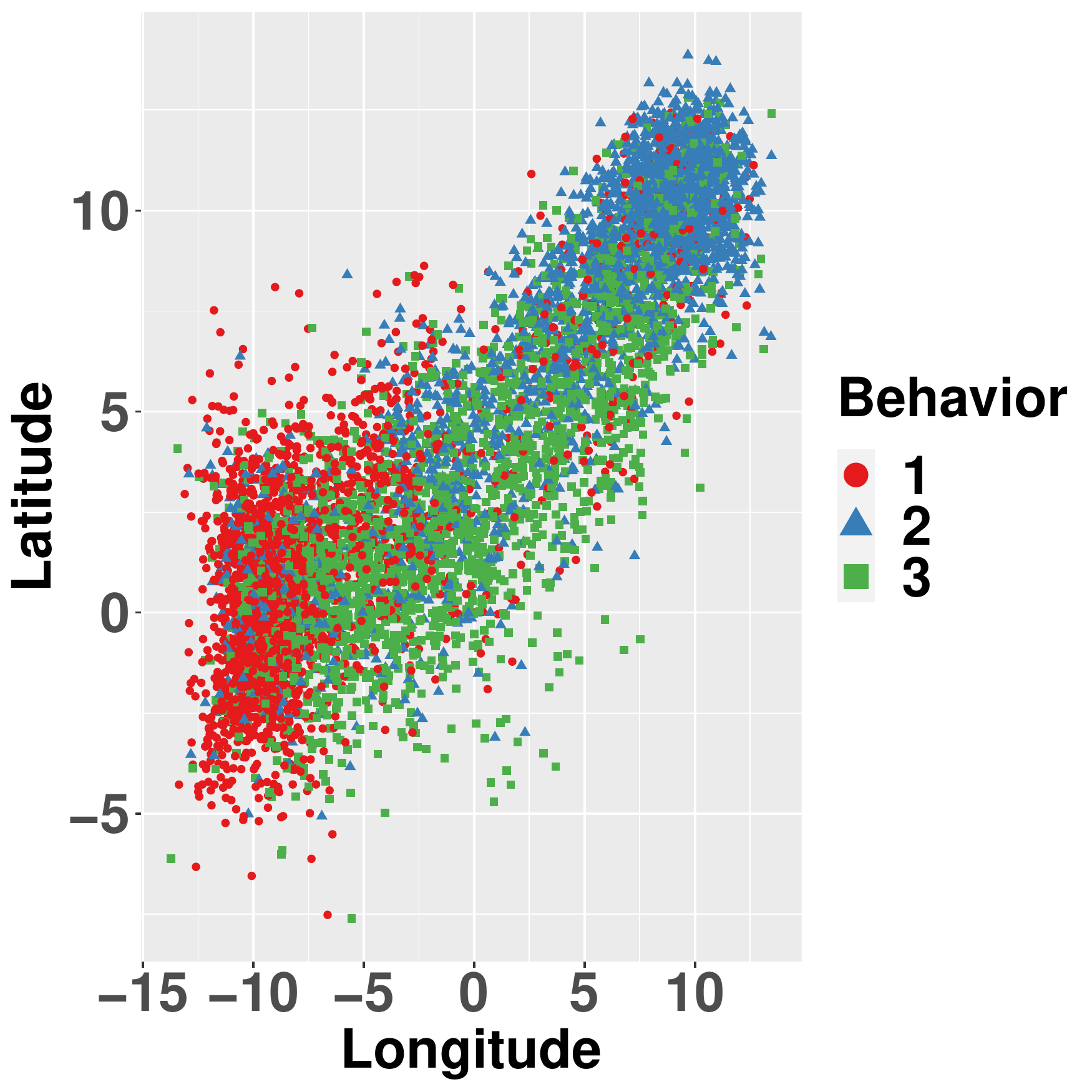}}}
    \caption{Locations of the simulated datasets. }\label{fakefig:Sim}
\end{figure}

We here want to  show that the parameters of the STAP-HMM can easily be  estimated.
For this reason,  we  simulate synthetic data from an  HMM with a fixed number of behaviors and STAP emission distribution, and after model fitting  we compare the 95\% CI of each parameter with the true value; if the latter is inside the CI, we consider the parameter ``well estimated''.
We simulate three datasets with 7000 observations, and  3 behaviors ($K^*=3$), with $\pi_{j,j}=0.8$,  $\pi_{j,j'}=0.1$ if $j \neq j'$,
and different sets of STAP parameters. In the first dataset, we assume
\begin{equation}
\boldsymbol{\mu}_1  =  \left(
\begin{array}{c}
-10\\
0
\end{array}
\right), \,\,
\boldsymbol{\mu}_2=    \left(
\begin{array}{c}
10\\
10
\end{array}
\right), \,\,
\boldsymbol{\mu}_3  =  \left(
\begin{array}{c}
0\\
0
\end{array}
\right), \,\,
\boldsymbol{\eta}_1  =  \left(
\begin{array}{c}
0\\
0
\end{array}
\right), \,\,
\boldsymbol{\eta}_2=    \left(
\begin{array}{c}
1\\
1
\end{array}
\right), \,\,
\boldsymbol{\eta}_3  =  \left(
\begin{array}{c}
-2\\
0
\end{array}
\right), \,\,
\end{equation}
\begin{equation}
\boldsymbol{\Sigma}_1  =  \left(
\begin{array}{cc}
0.5& 0.0\\
0.0 &5.0
\end{array}
\right), \,\,
\boldsymbol{\Sigma}_2=    \left(
\begin{array}{cc}
1.0& -0.25\\
-0.25& 0.5
\end{array}
\right), \,\,
\boldsymbol{\Sigma}_3  =  \left(
\begin{array}{cc}
1.0 &0.25\\
0.25 &1.0
\end{array}
\right),
\end{equation}
and $\tau_1=0.5$, $\tau_2 = 0.2$, $\tau_3 = 0$, $\rho_1=0$, $\rho_2 = 0.5$, $\rho_3 = 1$.
With this dataset we want to show that our model is able to detect whether a behavior is a BRW (behavior 1), a CRW (behavior 3), or a STAP (behavior 2). In the second dataset, we assume  $\tau_1=\tau_2 = \tau_3 = 0.2$, $\rho_1=0.2$, $\rho_2 = 0.5$, $\rho_3 = 0.8$, while all the other parameters remain the same. With this dataset, we want to show that we are able to learn the strength of attraction, even when its value is moderate ($\tau=$0.2) and the model is closer to a CRW than a BRW (behavior 3, $\rho=0.8$). In the last dataset, we have   $\tau_1=\tau_2 = \tau_3 = 0.8$, while all the other parameters are the same as those of the second dataset. With this example, we are interested in showing that, even though the strength of attraction is strong, we can still  detect directional persistence. We use the same priors and iterations as Section \ref{sec:realdata} for the model implementation. We show the simulated datasets in Figure \ref{fakefig:Sim}, where we can see that the spatial locations of the 3 behaviors are clearly overlapping.

The posterior estimates of the parameters are shown in Tables \ref{tab:simres1}, \ref{tab:simres2} and \ref{tab:simres3}, and  the parameters that are not well estimated are highlighted in bold. As can be seen, as expected most of the parameters are inside the intervals,   and the ones that are not well  estimated have intervals that are close to  the real values. The posterior probabilities of $K=3$ are 0.985, 0.969, and 0.98,       respectively,  for the three datasets. The accuracy of  the prediction of  $z_{{i}}$, based on the MAP estimator, and the true behavior at time ${i}$ is 0.942 for the first dataset, 0.916 for the second one, and 0.978 for the third one. Given these results, we can assume that the STAP-HMM parameters can be estimated with the proposed algorithm, which  is also able to detect the right number of latent behaviors.

\begin{table}[!t]
\scriptsize
  \centering
\begin{tabular}{c|ccc}
  \hline
 & j=1& j=2 & j=3  \\
 \hline \hline
 $\hat{\mu}_{j,1}$   &  -10.007  &  10.247  &  0.139  \\
 (CI)  & (-10.08 -9.937) & (9.602 10.947) & (-60.403 63.725)  \\
 $\hat{\mu}_{j,2}$   &  0.08  &  9.764  &  0.645  \\
 (CI)  & (-0.116 0.273) & (9.246 10.278) & (-60.098 61.962)  \\
 $\hat{\eta}_{j,1}$   &  -0.062  & {1.157}  &  -2.026  \\
 (CI)  & (-62.867 62.613) & (0.932 1.301) & (-2.075 -1.978)  \\
 $\hat{\eta}_{j,2}$   &  0.003  &  {1.126}  &  0.022  \\
 (CI)  & (-61.274 60.611) & (0.076 1.261) & (-0.022 0.068)  \\
 $\hat{\tau}_{j}$   &  0.502  &  {0.181}  &  0.499  \\
 (CI)  & (0.495 0.509) & (0.167 0.196) & (0.024 0.975)  \\
 $\hat{\rho}_{j}$   &  0  &  {0.459}  &  1  \\
 (CI)  & [0 0) & (0.422 0.493) & (1 1]  \\
 $\hat{\boldsymbol{\Sigma}}_{j,1,1}$   &  0.482  &  0.974  &  0.983  \\
 (CI)  & (0.451 0.516) & (0.916 1.037) & (0.919 1.054)  \\
 $\hat{\boldsymbol{\Sigma}}_{j,1,2}$   &  0.02  &  -0.232  &  0.267  \\
 (CI)  & (-0.046 0.088) & (-0.264 -0.201) & (0.22 0.315)  \\
 $\hat{\boldsymbol{\Sigma}}_{j,2,2}$   &  4.912  &  0.487  &  0.991  \\
 (CI)  & (4.625 5.202) & (0.458 0.518) & (0.931 1.056)  \\
 $\hat{{\pi}}_{1,j}$   &  0.796  &  0.101  &  0.103  \\
 (CI)  & (0.777 0.815) & (0.087 0.116) & (0.088 0.12)  \\
 $\hat{{\pi}}_{2,j}$   &  0.085  &  0.821  &  0.094  \\
 (CI)  & (0.073 0.097) & (0.804 0.837) & (0.081 0.107)  \\
 $\hat{{\pi}}_{3,j}$   &  0.115  &  0.087  &  0.799  \\
 (CI)  & (0.1 0.131) & (0.074 0.101) & (0.78 0.818)  \\
 $\hat{\beta}_{j}$   &  0.312  &  0.312  &  0.308  \\
 (CI)  & (0.052 0.681) & (0.049 0.676) & (0.049 0.688)  \\
 \hline \hline
 & $\alpha$ & $\kappa$  & $\gamma$   \\
 \hline
 $\hat{}$   &  0.342  &  3.46  &  0.728  \\
 (CI)  & (0.009 1.347) & (1.184 6.992) & (0.075 2.303)  \\
 \hline \hline
\end{tabular} \caption{Posterior means and 95\% CIs of the  parameters of the first simulated dataset for K=3.
}\label{tab:simres1}
\end{table}
%
%
%

\begin{table}[!t]
\scriptsize
  \centering
\begin{tabular}{c|ccc}
  \hline
 & j=1& j=2 & j=3  \\
 \hline \hline
 $\hat{\mu}_{j,1}$   &  -10.131  &  10.01  &  -0.39  \\
 (CI)  & (-10.495 -9.787) & (9.447 10.578) & (-1.571 0.816)  \\
 $\hat{\mu}_{j,2}$   &  0.556  &  10  &  0.743  \\
 (CI)  & (-0.052 1.152) & (9.517 10.483) & (-0.787 2.062)  \\
 $\hat{\eta}_{j,1}$   &  -0.037  &  1.088  &  -2.029  \\
 (CI)  & (-0.324 0.248) & (0.974 1.205) & (-2.186 -1.895)  \\
 $\hat{\eta}_{j,2}$   &  0.235  &  1.033  &  -0.004  \\
 (CI)  & (-0.037 0.508) & (0.927 1.144) & (-0.062 0.055)  \\
 $\hat{\tau}_{j}$   &  0.198  &  0.191  &  0.196  \\
 (CI)  & (0.19 0.207) & (0.177 0.207) & (0.142 0.264)  \\
 $\hat{\rho}_{j}$   &  0.201  &  0.48  &  0.795  \\
 (CI)  & (0.193 0.208) & (0.449 0.511) & (0.741 0.844)  \\
 $\hat{\boldsymbol{\Sigma}}_{j,1,1}$   &  0.478  &  0.972  &  0.959  \\
 (CI)  & (0.445 0.514) & (0.909 1.039) & (0.886 1.036)  \\
 $\hat{\boldsymbol{\Sigma}}_{j,1,2}$   &  0.036  &  -0.241  &  0.265  \\
 (CI)  & (-0.034 0.107) & (-0.275 -0.21) & (0.215 0.314)  \\
 $\hat{\boldsymbol{\Sigma}}_{j,2,2}$   &  4.993  &  0.492  &  1.005  \\
 (CI)  & (4.692 5.301) & (0.462 0.525) & (0.937 1.077)  \\
 $\hat{{\pi}}_{1,j}$   &  \textbf{0.811} &  0.096  &  0.107  \\
 (CI)  & (0.802 0.818) & (0.081 0.112) & (0.088 0.127)  \\
 $\hat{{\pi}}_{2,j}$   &  {0.084}  &  {0.821}  &  0.095  \\
 (CI)  & (0.071 0.099) & (0.783 0.838) & (0.08 0.111)  \\
 $\hat{{\pi}}_{3,j}$   &  0.113  &  0.093  &  0.794  \\
 (CI)  & (0.096 0.131) & (0.079 0.11) & (0.771 0.815)  \\
 $\hat{\beta}_{j}$   &  0.316  &  0.307  &  0.311  \\
 (CI)  & (0.052 0.675) & (0.051 0.672) & (0.049 0.675)  \\
 \hline \hline
 & $\alpha$ & $\kappa$  & $\gamma$   \\
 \hline
 $\hat{}$   &  0.351  &  3.382  &  0.73  \\
 (CI)  & (0.008 1.419) & (1.143 6.927) & (0.088 2.278)  \\
 \hline \hline
\end{tabular} \caption{Posterior means and 95\% CIs of the  parameters of the second simulated dataset for K=3. The parameter whose 95\% CI does not contain the true value is shown in bold.}\label{tab:simres2}
\end{table}
%
%
%

\begin{table}[!t]
\scriptsize
  \centering
\begin{tabular}{c|ccc}
  \hline
 & j=1& j=2 & j=3  \\
 \hline \hline
 $\hat{\mu}_{j,1}$   &  -9.997  &  10.028  &  -0.019  \\
 (CI)  & (-10.057 -9.938) & (9.923 10.137) & (-0.274 0.236)  \\
 $\hat{\mu}_{j,2}$   &  0.054  &  10.003  &  0.22  \\
 (CI)  & (-0.084 0.188) & (9.895 10.113) & (-0.054 0.493)  \\
 $\hat{\eta}_{j,1}$   &  0.01  &  0.994  &  -1.977  \\
 (CI)  & (-0.233 0.254) & (0.908 1.082) & (-2.07 -1.908)  \\
 $\hat{\eta}_{j,2}$   &  0.015  &  0.962  &  0.005  \\
 (CI)  & (-0.244 0.265) & (0.873 1.054) & (-0.049 0.056)  \\
 $\hat{\tau}_{j}$   &  0.806  &  0.816  &  0.915  \\
 (CI)  & (0.796 0.816) & (0.777 0.862) & (0.77 0.996)  \\
 $\hat{\rho}_{j}$   &  0.205  &  0.511  &  0.824  \\
 (CI)  & (0.197 0.211) & (0.488 0.537) & (0.791 0.84)  \\
 $\hat{\boldsymbol{\Sigma}}_{j,1,1}$   &  0.483  &  0.982  &  0.975  \\
 (CI)  & (0.454 0.512) & (0.928 1.04) & (0.917 1.037)  \\
 $\hat{\boldsymbol{\Sigma}}_{j,1,2}$   &  0.006  &  -0.244  &  0.264  \\
 (CI)  & (-0.06 0.074) & (-0.275 -0.215) & (0.222 0.308)  \\
 $\hat{\boldsymbol{\Sigma}}_{j,2,2}$   &  4.962  &  0.495  &  0.976  \\
 (CI)  & (4.679 5.259) & (0.468 0.524) & (0.917 1.039)  \\
 $\hat{{\pi}}_{1,j}$   &  0.799  &  0.101  &  0.099  \\
 (CI)  & (0.781 0.817) & (0.089 0.114) & (0.086 0.114)  \\
 $\hat{{\pi}}_{2,j}$   &  0.09  &  \textbf{0.817}  &  0.093  \\
 (CI)  & (0.079 0.102) & (0.801 0.832) & (0.081 0.105)  \\
 $\hat{{\pi}}_{3,j}$   &  0.107  &  0.092  &  0.801  \\
 (CI)  & (0.094 0.121) & (0.08 0.105) & (0.783 0.818)  \\
 $\hat{\beta}_{j}$   &  0.317  &  0.304  &  0.31  \\
 (CI)  & (0.053 0.673) & (0.052 0.675) & (0.047 0.668)  \\
 \hline \hline
 & $\alpha$ & $\kappa$  & $\gamma$   \\
 \hline
 $\hat{}$   &  0.327  &  3.377  &  0.724  \\
 (CI)  & (0.007 1.279) & (1.16 6.848) & (0.076 2.315)  \\
 \hline \hline
\end{tabular} \caption{Posterior means and 95\% CIs of the  parameters of the third simulated dataset for K=3. The parameter whose 95\% CI does not contain the true value is shown in bold.}\label{tab:simres3}
\end{table}

%
%
%
%

\section{Details on the model of \cite{McClintock2012}} \label{sec:mc}

%

The model  is an HMM with a fixed number of behaviors and an emission distribution which  is the product of a wrapped Cauchy over the bearing-angle and a Weibull for the step-length.
The temporal evolution of the random variable  $z_{{i}}^{\text{mc}} \in \{1,2,\dots , K^{\text{mc}}\}$ that represents the behavior at time ${i}$, follows a first-order Markov process with
$
z_{{i}}^{\text{mc}} |z_{{i-1}}^{\text{mc}},\boldsymbol{\pi}_{z_{{i-1}}^{\text{mc}} }^{\text{mc}}
  \sim  \text{Multinomial}(1, \boldsymbol{\pi}_{z_{{i-1}}^{\text{mc}} }^{\text{mc}}).
$
If  $z_{{i}}^{\text{mc}}=j$, the distributions over the movement-metric are
\begin{align}
\phi_{{i}} |\phi_{{i-1}}, z_{{i}}^{\text{mc}}&\sim \text{wCauchy}(\lambda_{{i},j}^{\text{mc}},\epsilon_{{i},j}^{\text{mc}})
\label{eq:bearLO},\\
r_{{i}}|z_{{i}}^{\text{mc}} & \sim \text{Weibull}(a_{{i},j}^{\text{mc}},b_{{i},j}^{\text{mc}}) .
\end{align}
If the $j$-th behavior at time  ${i}$ is a  BCRW,  we have
\begin{align}
\lambda_{{i},j}^{\text{mc}} &= \rho_j^{\text{mc}} \phi_{{i-1}}+(1-\rho_j^{\text{mc}})\zeta_{{i}}, \\
\epsilon_{{i},j}^{\text{mc}} &= \text{logit}^{-1}(m_{j}^{\text{mc}}+ g_j^{\text{mc}} \psi_{{i}}+q_{j}^{\text{mc}} \psi_{{i}}^2),\\
a_{{i},j}^{\text{mc}} &= a_{1,j}^{\text{mc}} ( 1- I_{[0, d_j^{\text{mc}}  )}(\psi_{{i}}) )+ a_{2,j}^{\text{mc}}I_{[0, d_j^{\text{mc}} )}(\psi_{{i}}) ,\\
b_{{i},j}^{\text{mc}} &= b_{1,j}^{\text{mc}} ( 1- I_{[0, d_j^{\text{mc}}  )}(\psi_{{i}}) )+ b_{2,j}^{\text{mc}}I_{[0, d_j^{\text{mc}} )}(\psi_{{i}}),
\end{align}
while, if it is a CRW, we obtain
$\lambda_{{i},j}^{\text{mc}} = \phi_{{i-1}}$,
$\epsilon_{{i},j}^{\text{mc}} =\nu_{j}^{\text{mc}} $,
$a_{{i},j}^{\text{mc}} = a_{j}^{\text{mc}}  $,
$b_{{i},j}^{\text{mc}} = b_{j}^{\text{mc}}$,
 where  $\psi_{{i}}$ is the distance between the observation $\mathbf{s}_{{i}}$ and the attractor.
The prior distributions  are:  $m_{j}^{\text{mc}},  g_{j}^{\text{mc}}, q_{j}^{\text{mc}} \sim N(0, \tau^{\text{mc}})$, with  $\tau^{\text{mc}} \sim G(2,3)$, $\rho_{j}^{\text{mc}}, \nu_{j}^{\text{mc}} \sim U(0,1)$,
$\phi_0 \sim U(0, 2\pi)$,
$a_j^{\text{mc}},a_{1,j}^{\text{mc}},a_{2,j}^{\text{mc}}\sim U(0,14000)$,
$b_j^{\text{mc}},b_{1,j}^{\text{mc}},b_{2,j}^{\text{mc}}\sim U(0,30)$,
$d_j^{\text{mc}}\sim U(0,14000)$, and
$\boldsymbol{\pi}_{j}^{\text{mc}} \sim \text{Dir}(1,\dots , 1)$.

%
%

\section{{Turning-angle distribution for different time-intervals}}\label{sec:Bimod}

\begin{figure}[t]
    \centering
    {\subfloat[set1 - Wrapped-Cauchy]{\includegraphics[scale=0.20]{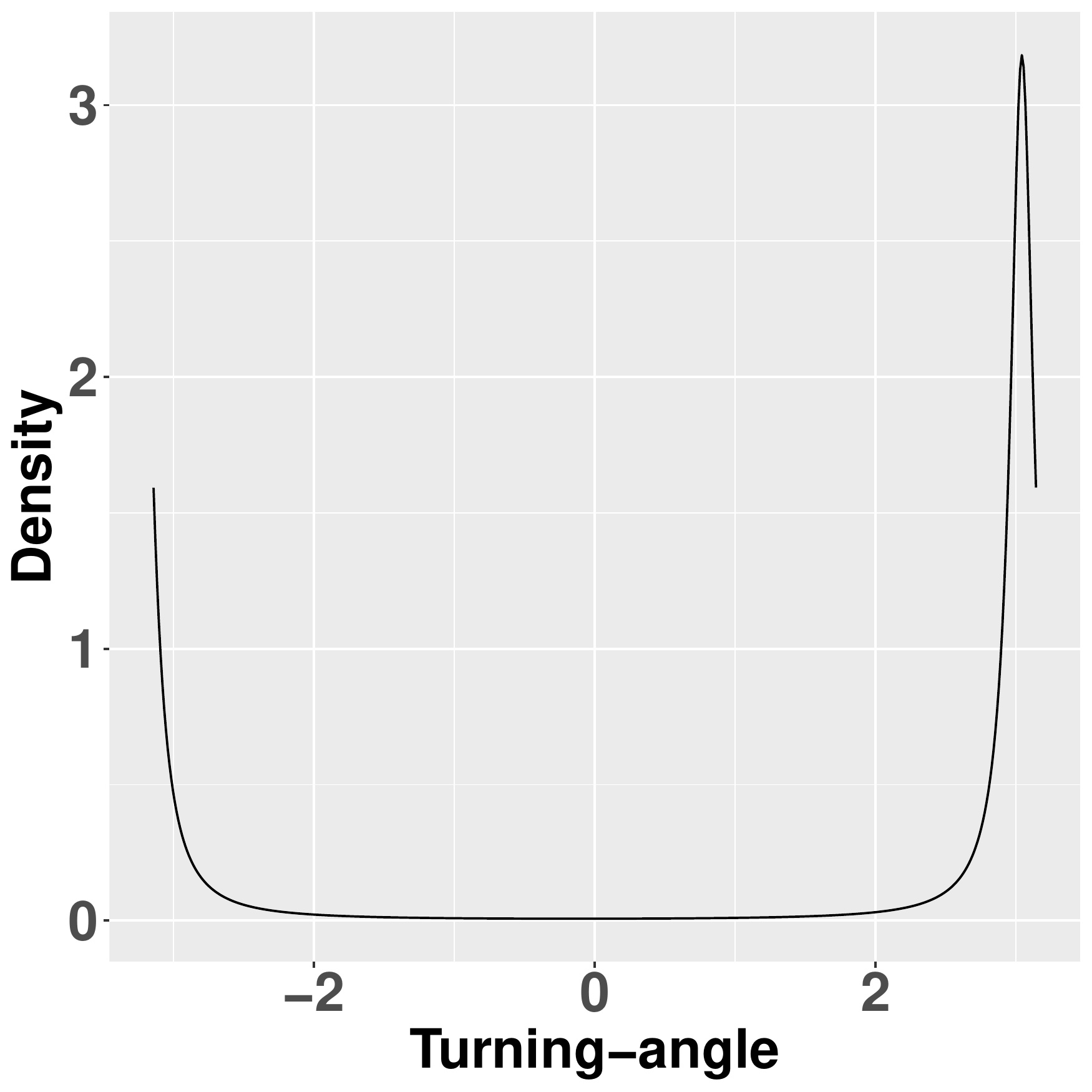}}}
    {\subfloat[set2 - Wrapped-Cauchy]{\includegraphics[scale=0.20]{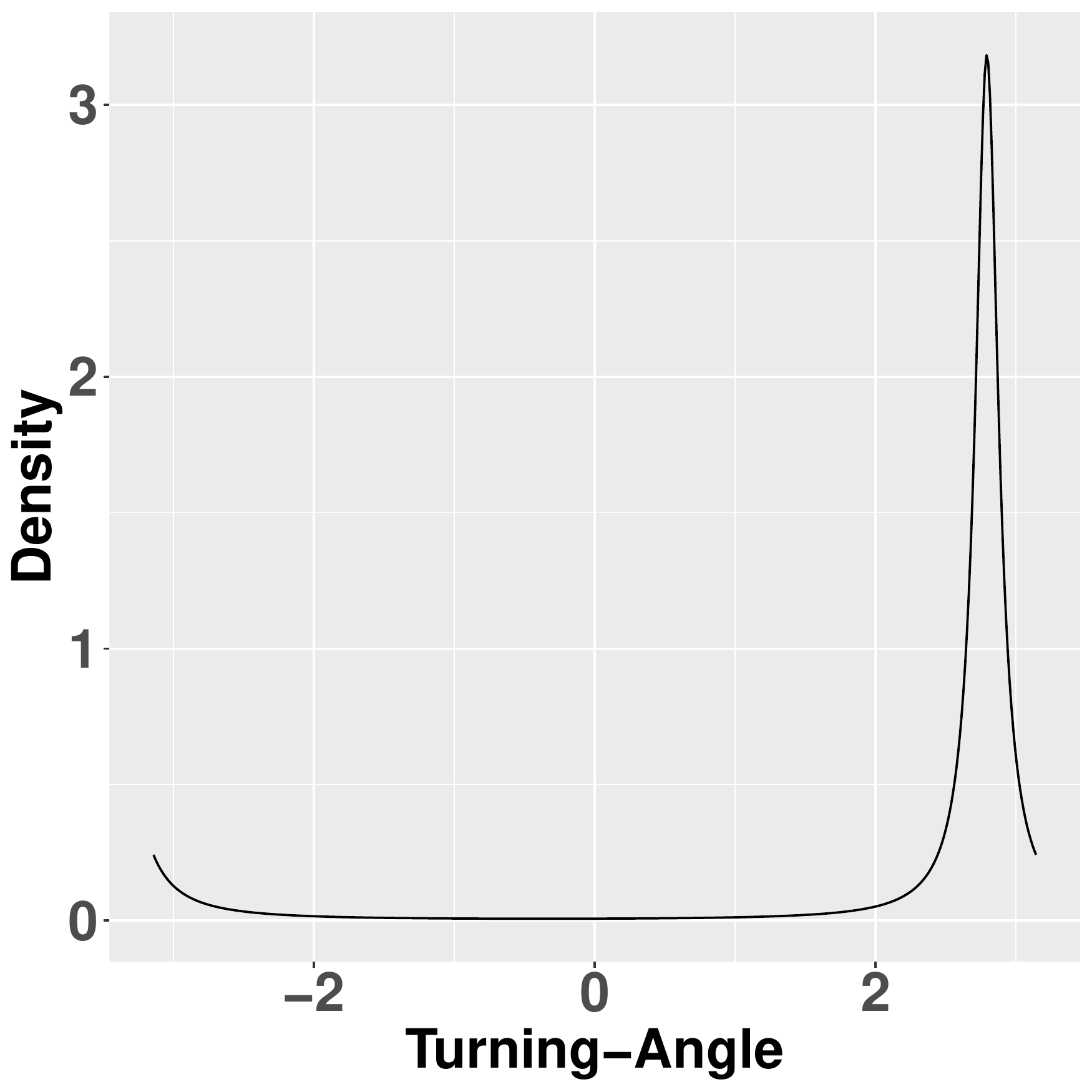}}}
    {\subfloat[set3 - Wrapped-Cauchy]{\includegraphics[scale=0.20]{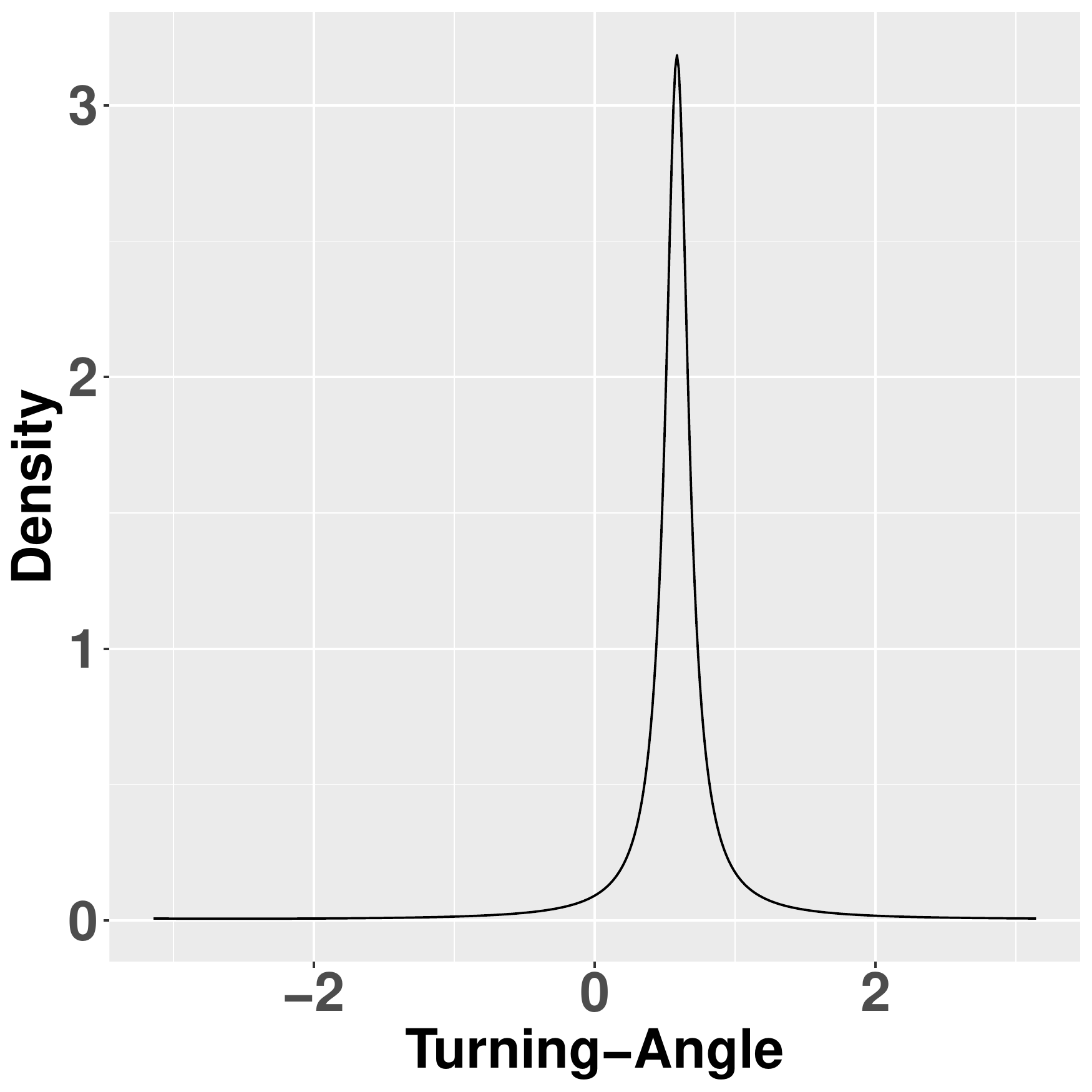}}}\\
    {\subfloat[set1 - Kernel density estimate]{\includegraphics[scale=0.20]{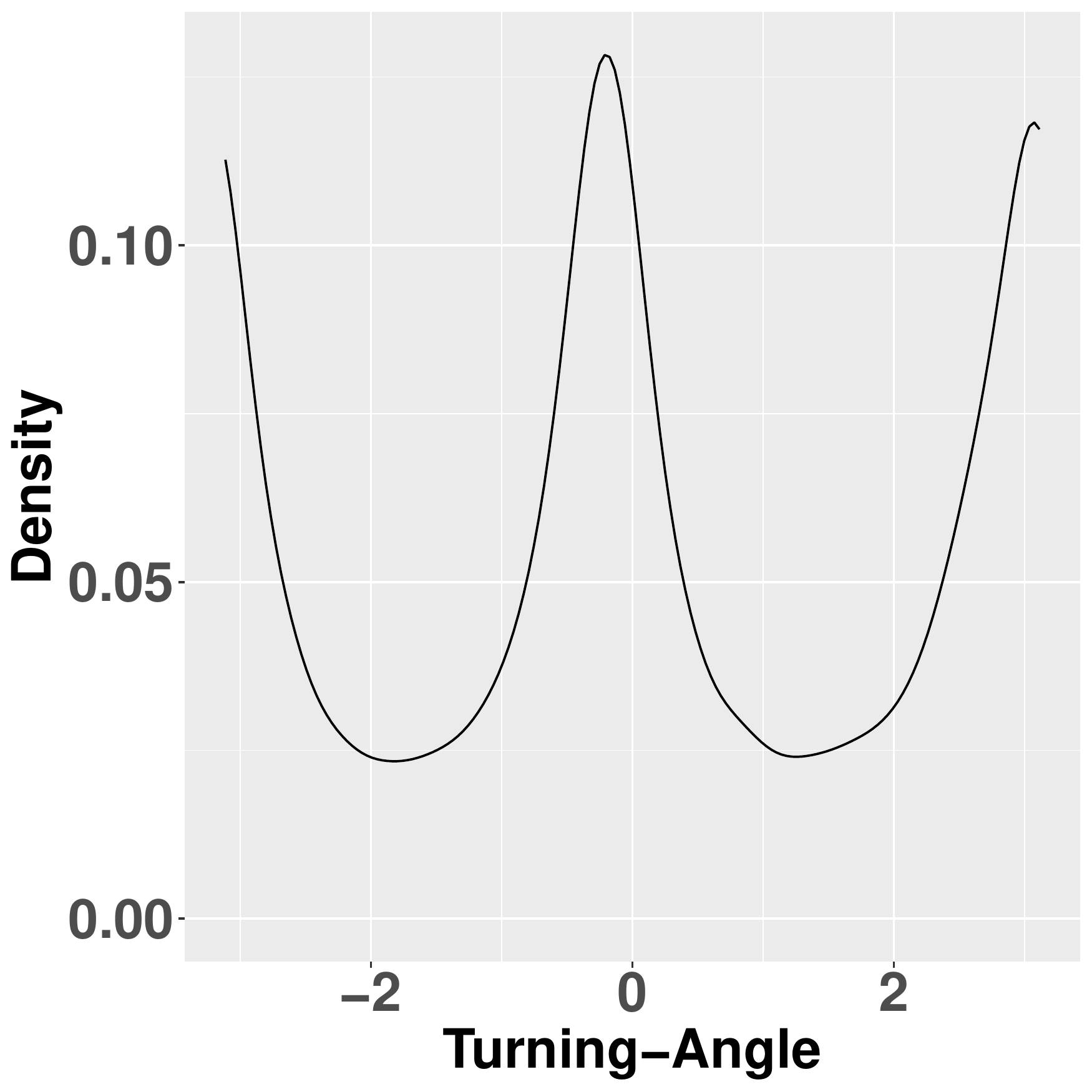}}}
    {\subfloat[set2 - Kernel density estimate]{\includegraphics[scale=0.20]{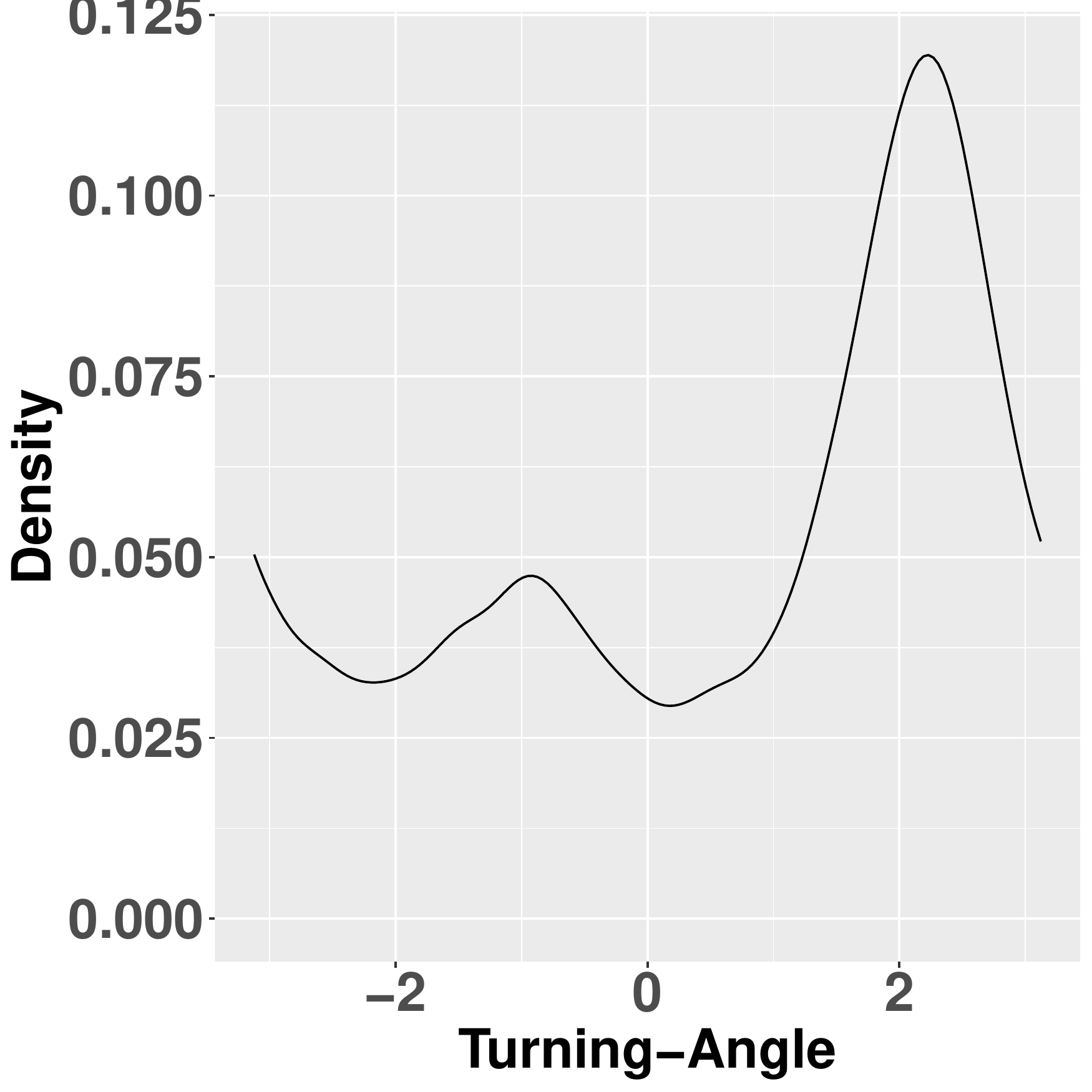}}}
    {\subfloat[set3 - Kernel density estimate]{\includegraphics[scale=0.20]{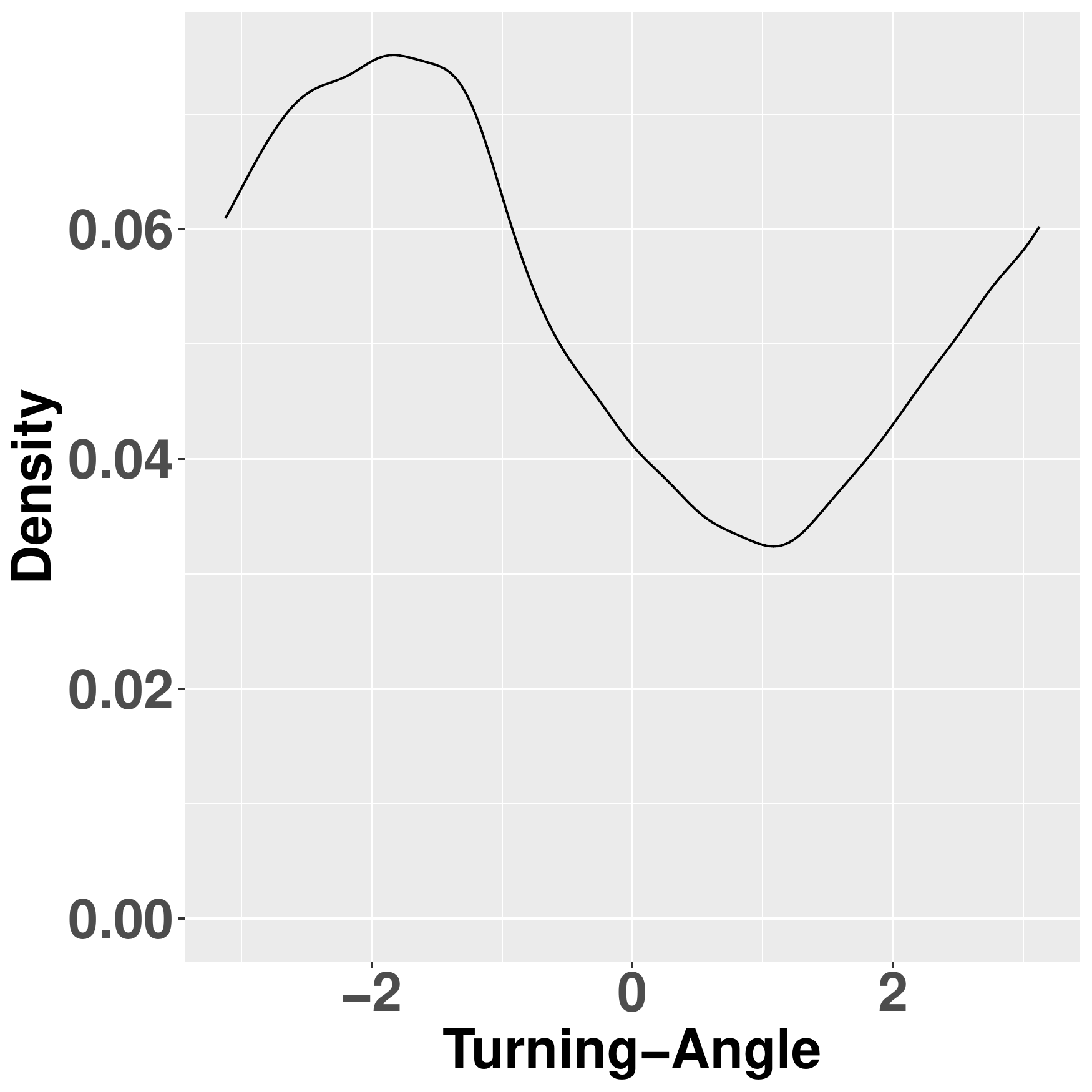}}}
    \caption{{The wrapped-Cauchy densities used to simulate the paths $\mathbf{s}^*$ (first row), and the kernel density estimates of the turning-angle distributions of $\mathbf{s}$ (second row). }} \label{fig:pathsim}
\end{figure}
%

{In this section, we want to show that even though  a path, at a given time-interval, can be described by a unimodal and symmetric circular distribution, the distribution over the turning-angle (or bearing-angle) can be asymmetric and multimodal for a greater time-interval.}

{Let us suppose we have a path $\mathbf{s}^* = (\mathbf{s}_{t_1}^*,\dots, \mathbf{s}_{t_{T^*}}^*)$, where $t_1, \dots , t_{T^*}$ are equally-spaced temporal-indices, i.e., $t_{i}-{t_{i-1}} = t_{i-1}-{t_{i-2}}$ for all $i = 3,4,\dots , T^* $. We assume that the path is a CRW with the following distribution for the step-length and turning-angle:
\begin{align} \label{eq:dd}
    r_{t_i}^* = || \mathbf{y}_{t_i}^*||_2  &\sim \text{Weibull} (a_{\text{sim}},b_{\text{sim}}),\\
\theta_{t_i}^* = \text{atan}^*(y_{t_i,2}^*, y_{t_i,1}^*) &\sim \text{wCauchy}(\lambda_{\text{sim}},\epsilon_{\text{sim}} ),
\end{align}
where $\mathbf{y}_{t_i}^* = (y_{t_i,1}^*,y_{t_i,2}^*)' = R(\phi_{t_{i-1}}^*) (\mathbf{s}_{t_{i+1}}^*-\mathbf{s}_{t_{i}}^*)$, $\phi_{t_{i}}^* = \text{atan}^*(s_{t_{i+1},2}^*-s_{t_{i},2}^*,s_{t_{i+1},1}^*-s_{t_{i},1}^*)$, $\theta_{t_i}^* = \phi_{t_{i}}^*-\phi_{t_{i-1}}^*$,
$\mathbf{s}_{t_0}^*=(-1,0)'$, and $\mathbf{s}_{t_1}^*=(0,0)'$. It should be noted that the wrapped-Cauchy is  symmetric and unimodal.  We assume that the distributions in \eqref{eq:dd} are the ones that rule the animal's path with a time-step $t_{i}-{t_{i-1}}$, but we are only able to observe/record the path at the time-interval  $d(t_{i}-{t_{i-1}}) =  t_{id}-{t_{(i-1)d}}$ with
$d \geq 2$.  Hence,  we define $\mathbf{s} = (\mathbf{s}_{1},\dots, \mathbf{s}_{T})$ as the recorded path with
\begin{equation} \label{eq:path222dd}
\mathbf{s}_i = \mathbf{s}_{t_{id}}^*.
\end{equation}
From \eqref{eq:path222dd}, we can compute the movement metrics, as well as the other sets of coordinates, as explained in Section \ref{sec:Model}, e.g.,  $\phi_{{i}} = \text{atan}^*(s_{{i+1},2}-s_{{i},2},s_{{i+1},1}-s_{{i},1})$ and $\theta_{i} = \phi_{{i}}-\phi_{{i-1}}$.}


{We simulate three paths $\mathbf{s}^* $ (equation \eqref{eq:dd}) with $T^* = 100\,000$  and the following sets of parameters:
\begin{itemize}
\item set1=$\{\lambda_{\text{sim}} = \pi-0.1,\epsilon_{\text{sim}}=0.1,a_{\text{sim}}=1,b_{\text{sim}}=1\}$;
\item set2=$\{\lambda_{\text{sim}} = \pi-0.35,\epsilon_{\text{sim}}=0.1,a_{\text{sim}}=1.7,b_{\text{sim}}=5\}$;
\item set3=$\{\lambda_{\text{sim}} = \pi/4-0.2,\epsilon_{\text{sim}}=0.1,a_{\text{sim}}=15,b_{\text{sim}}=10\}$.
\end{itemize}
 We can see the highly concentrated wrapped-Cauchy distributions used to simulate the    paths, i.e., the distributions of $\theta_{t_i}^*$, in the first row of Figure \ref{fig:pathsim}.  We obtain three datasets of recorded coordinates by assuming     $d=2$ in the first path, while in the second path we assume $d=3$, and in the third $d=9$.
Using
\eqref{eq:disp1} and
\eqref{eq:theta_atan},
we derive the samples of the turning-angles $\theta_{i}$ of the recorded paths $\mathbf{s}$ and  we evaluate the distributions induced over  $\theta_i$ with the given  time-intervals $t_{id}-{t_{(i-1)d}}$  by using a  kernel estimator; these densities are shown in the second row of Figure  \ref{fig:pathsim}. It is clear from the figures  that, even though  the data at a given time-interval come from a  unimodal  and symmetric circular distribution,  the circular distribution can be bimodal for a greater time-interval (Figure  \ref{fig:pathsim} (d) and (e)) or asymmetric (Figure  \ref{fig:pathsim} (f)). For this reason,  a  distribution  over the circular variable that may be  multimodal and asymmetric should  be used  to describe the data at any time-interval.
}


\section{Sensitivity analysis of the hyper-parameter priors} \label{sec:priors}
\begin{figure}[t]
    \centering
    \includegraphics[scale=0.26]{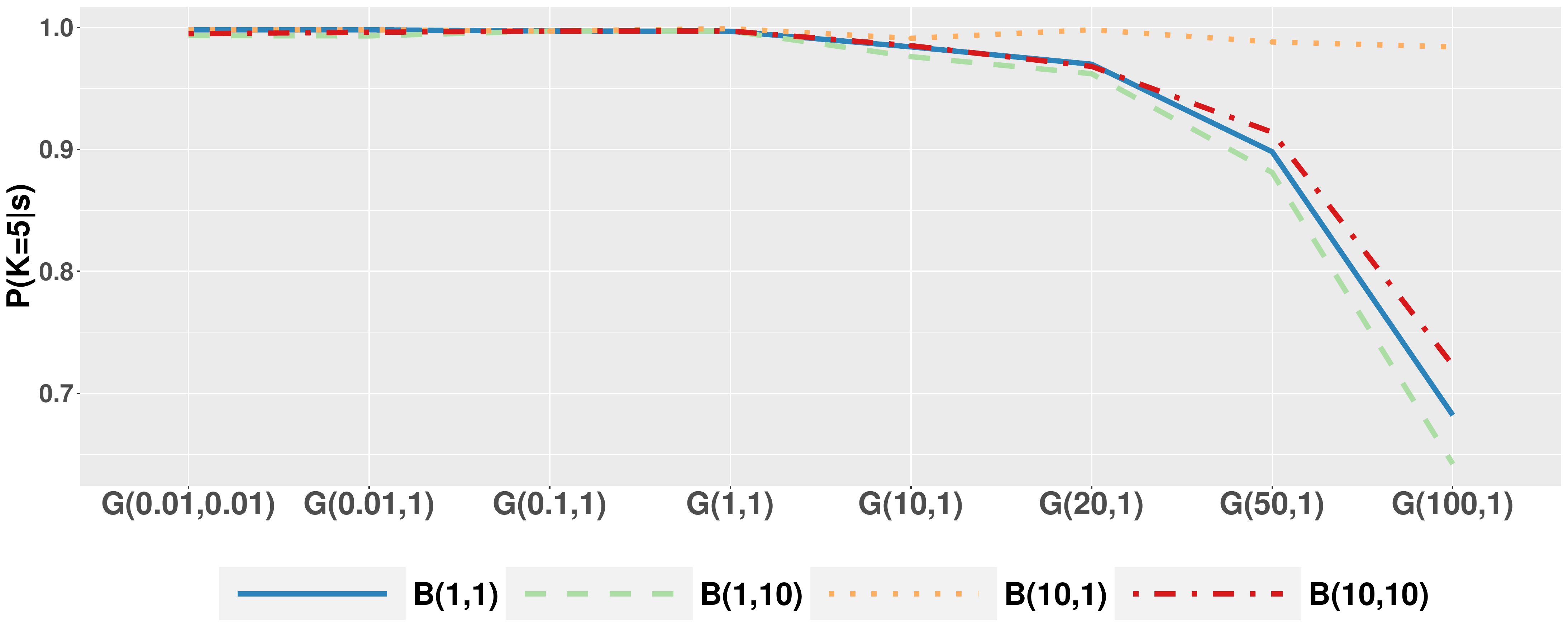}
    \caption{{Values of the posterior probability of the number of behaviors being equal to 5,  under 9 Gamma priors for  $\alpha+\kappa$ and $\gamma$, and 4 Beta priors for $\kappa/(\alpha+\kappa)$.}}\label{fig:Sens}
\end{figure}

{The number of latent behaviors of  the sHDP-HMM depends  on the hyper-parameters $\alpha$, $\kappa$ and $\gamma$, which  are random quantities in our model  and  are assumed to be  distributed according to the following priors   (see Section \ref{sec:realdata}): $\alpha+\kappa\sim G(0.1, 1)$, $\kappa/(\alpha+\kappa) \sim B(10,1)$, $\gamma \sim G(0.1,1)$.
%
In this section, we want to show how the posterior distribution  of $K$  changes if we change  these priors. }

{
Parameters  $\alpha+\kappa$ and $\gamma$ can be interpreted in a similar way, since the expected value of the number of behaviors increases with both parameters,  while $\kappa/(\alpha+\kappa)$
is the proportion of  the extra weight, $\kappa$,  added to the  self transitions,  with respect to the total weight, $\alpha+\kappa$; for more details see \cite{fox2011}.
We estimate  different models on the real data, using  the same iterations and priors as in Section \ref{sec:realdata}, with the exception of those for  $\alpha+\kappa$, $\kappa/(\alpha+\kappa)$, and $\gamma$. We  assume that  $\alpha+\kappa$ and $\gamma$  have  the same prior
and we  test all the possible combinations of 8 Gamma priors for  $\alpha+\kappa$ and $\gamma$ and 4 Beta priors for $\kappa/(\alpha+\kappa)$,  thus obtaining  24 models. In detail, the priors are:
\begin{itemize}
\item Gamma priors - $G(0.01,1)$, $G(0.1,1)$, $G(1,1)$, $G(10,1)$, $G(20,1)$, $G(50,1)$, $G(100,1)$, $G(0.01,0.01)$;
\item Beta priors - $B(1,1)$, $B(10,1)$, $B(1,10)$, $B(10,10)$.
\end{itemize}
We have a uniform distribution for $\kappa/(\alpha+\kappa)$ ($B(10,10)$),
while the other three have  expected values equal to  0.09 $(B(10,1))$, 0.91 ($B(1,10)$), 0.5 ($B(10,10)$), and  smaller variances than 0.012.
The mean  and the variance  of  7 out of the 8   Gamma distributions have the same value and  are equal to  0.01 $(G(0.01,1))$, 0.1 ($G(0.1,1)$), 1 ($G(1,1)$),  10 ($G(10,1)$,  20 ($G(20,1)$,  50 ($G(50,1)$) and  100 ($G(100,1)$), while the last one, $G(0.01,0.01)$, has mean 1 and variance 100.
Hence, we have 3  Gamma priors that, a priori,  put a greater probability mass on a small number of behaviors ($G(0.01,1)$,  $G(0.1,1)$, $G(1,1)$),  and 4 priors ($G(10,1)$,  $G(20,1)$,
 $G(50,1)$,  $G(100,1)$) which, a priori, support a high number of behaviors, while $G(0.01,0.01)$ is often used as a weakly informative prior.}

 {In Figure \ref{fig:Sens}, we show the posterior probability that the number of latent behaviors is equal to $5$  in the 24  models, which is the number found when using the model estimated  in Section \ref{sec:realdata}. Since    $\mathbb{P}(K=5 | \mathbf{s})$ is always greater than 0.5, $K=5$ is   the number of   behaviors with the highest posterior probability for any possible combination of priors, and this is $>0.95$ for  all the priors, with the exception of  $G(50,1)$ and $G(100,1)$, where   $\mathbb{P}(K=5 | \mathbf{s})>0.64$. It is interesting to note that  when using  the two weakly informative priors $G(0.01,0.01)$ and $B(1,1)$,
  we find that $\mathbb{P}(K=5|\mathbf{s})  $ is approximatively 1. The posterior samples of $K$ have values in
 $\{5,6\}$ for  the
priors $G(0.01,0.01)$, $G(0.01,1)$,  $G(0.1,1)$, and  $G(1,1)$, values in $\{5,6,7\}$ for the priors
$G(10,1)$ and $G(20,1)$, and values in $\{5,6,7,8\}$ for the priors
$G(50,1)$ and $G(100,1)$.}

 {As expected,  $\mathbb{P}(K=5|\mathbf{s})  $    tends to decrease   if we have a greater prior probability mass on higher values of  $\alpha+\kappa$ and $\gamma$, and it  also increases the number of elements of $\mathbb{N}$ that have a significant posterior probability,  that is a measure of the variability of $K$.   The prior on $\kappa/(\alpha+\kappa)$ has a limited impact on $\mathbb{P}(K=5|\mathbf{s})$ if the distribution over   $\alpha+\kappa$ and $\gamma$  is $G(0.01,0.01)$, $G(0.01,1)$,  $G(0.1,1)$, or  $G(1,1)$, and
the model  with a $B(10,1)$  prior  tends to have the  highest   $\mathbb{P}(K=5|\mathbf{s})$  for a given Gamma prior, while a $B(1,10)$  prior has the smallest. Nonetheless, since $\mathbb{P}(K=5|\mathbf{s})$ is very large in all of the 24 models, the number of estimated behaviors is always 5, even though we use   weakly informative priors ($G(0.01,0.01)$ and $B(1,1)$).
 }

%
%
%
%

\section{Real data applications - Simulated trajectories}\label{sec:simTraj}
\begin{figure}[t]
    \centering
    {\subfloat[]{\includegraphics[scale=0.23]{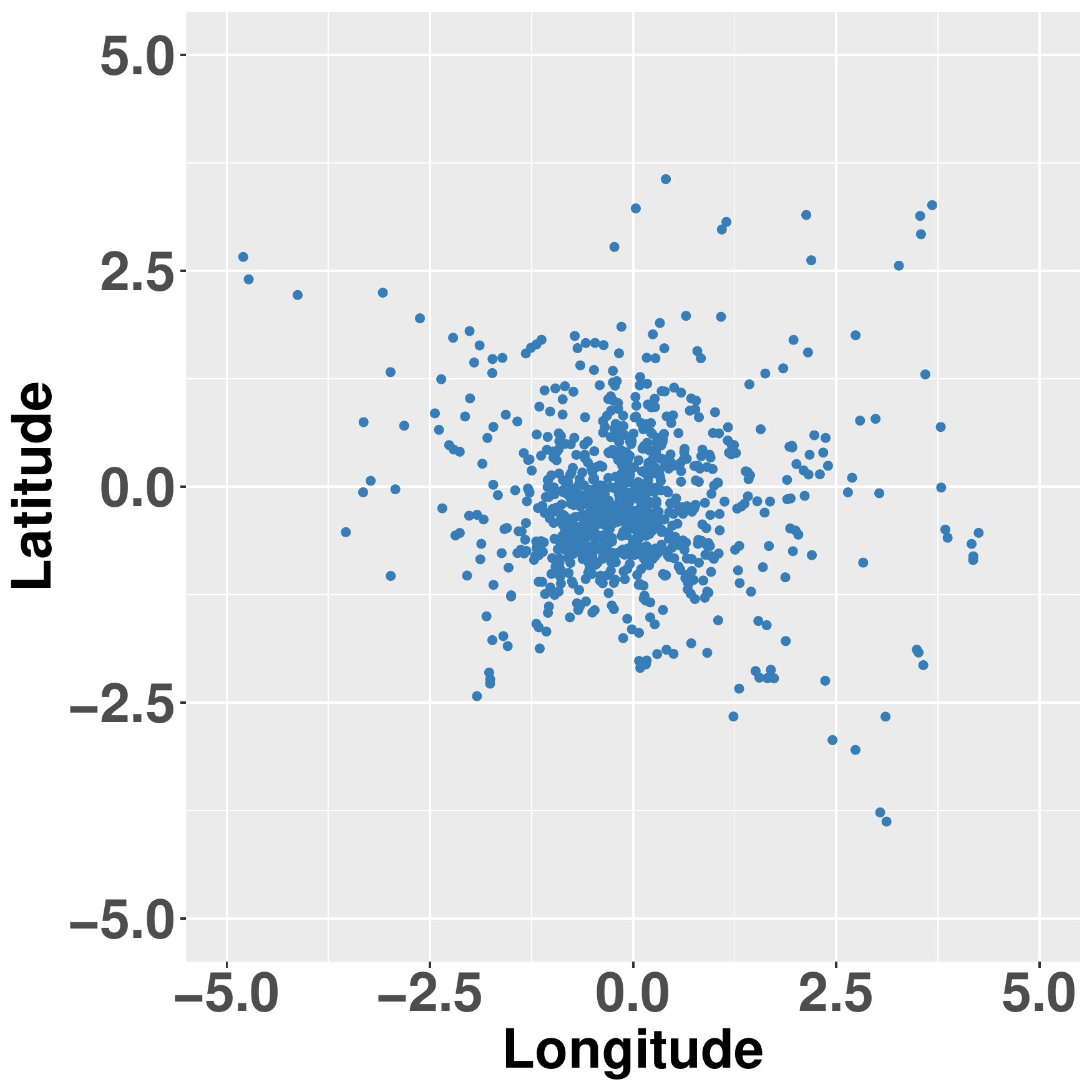}}}
    {\subfloat[]{\includegraphics[scale=0.23]{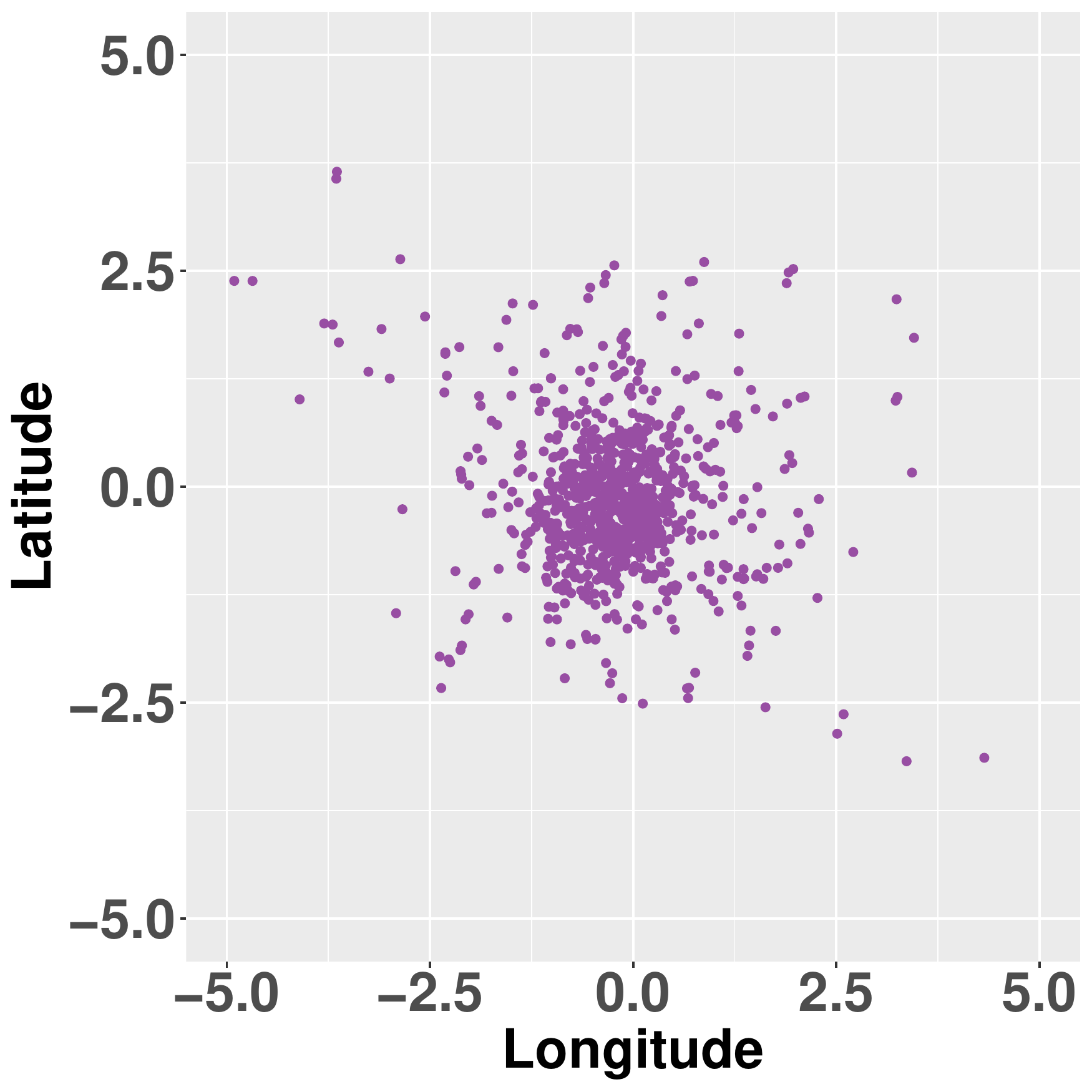}}}
    {\subfloat[]{\includegraphics[scale=0.23]{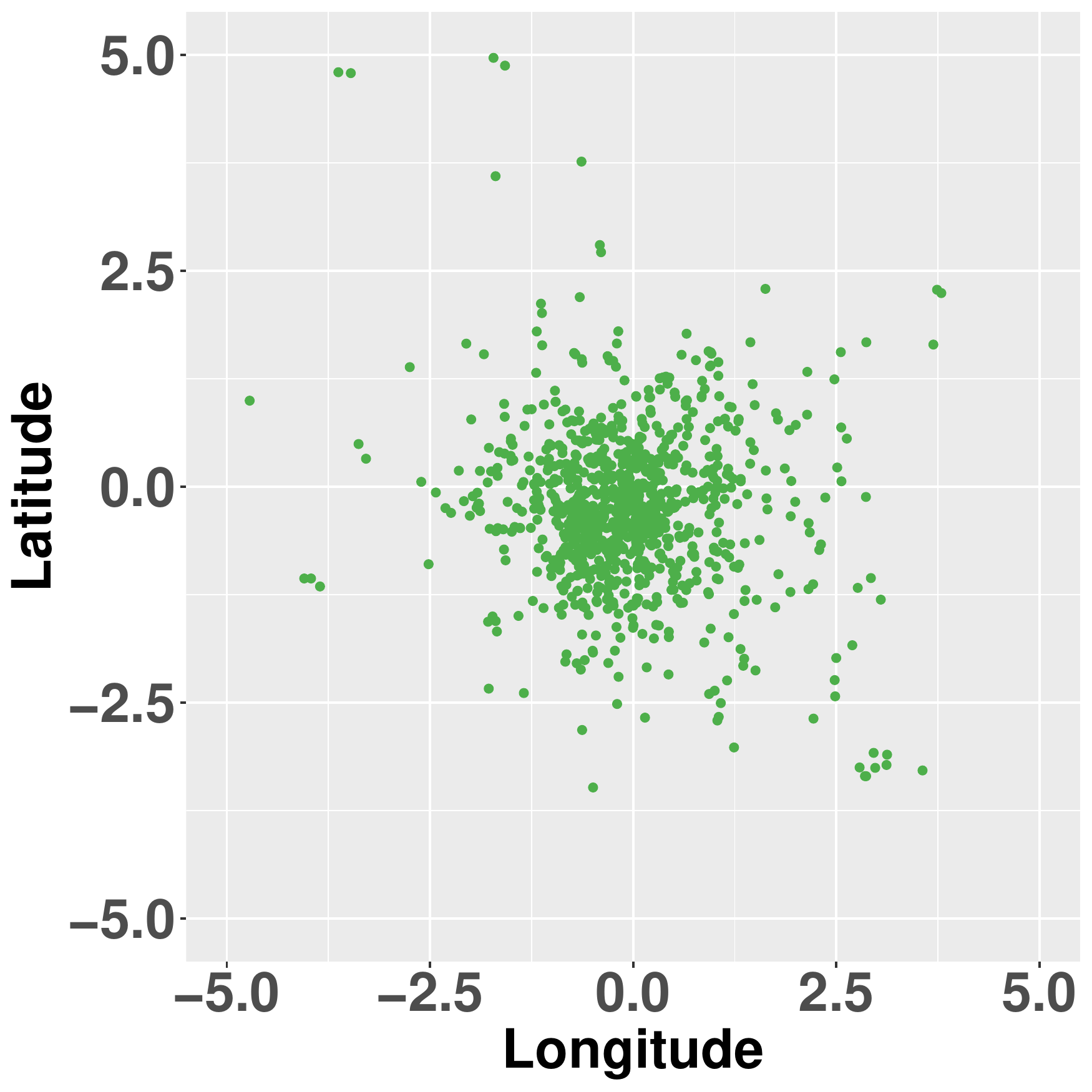}}}
    \caption{Simulated spatial locations using  the posterior means of the STAP-HMM as parameters.}\label{fig:simTRaj}
\end{figure}

In order to show that our model can describe the data in Section \ref{sec:realdata},  we show, in  Figure \ref{fig:simTRaj},  three  sets of coordinates obtained by simulating from the STAP-HMM, using  the posterior means in Table \ref{tab:resSTAP} as  parameters. We can clearly see that the simulated data have  the central bulk of observations, as well as the extension of the ``explored space'', as the real data.

\section{Real data applications - CRW-HMM and BRW-HMM}\label{sec:realOtherRes}

%
%

\begin{table}[t]
    \scriptsize
      \centering
    \begin{tabular}{cr|ccccc|ccc}
        \hline
\multicolumn{2}{c|}{}        &\multicolumn{5}{|c|}{BRW-HMM} & \multicolumn{3}{c}{CRW-HMM} \\
&  &	First&	Second&	Third	&Fourth&	Fifth	&First	&Second&	Third\\ \hline
 &First&	4259	&13&	1&	1&	3&	4266&	9&	2\\
 &Second&	10&	773&	7&	41&	1&	8&	808&	16\\
 STAP-HMM&Third&	0&	27	&349	&5&	0	&0&	4&	377\\
 &Forth	&1&	31&	8&	521	&0&	2&	72	&487\\
 &Fifth&	8&	1&	1&	0&	273	&2&	123&	158\\
     \hline \hline
    \end{tabular}
    \caption{{Two-way table between the  MAP behaviors  of the  STAP-HMM (rows), and the  BRW-HMM and   CRW-HMM (columns).} }\label{fig:beavS}
\end{table}

\begin{figure}[t]
    \centering
    \includegraphics[trim = 0 10 0 40,scale=0.28]{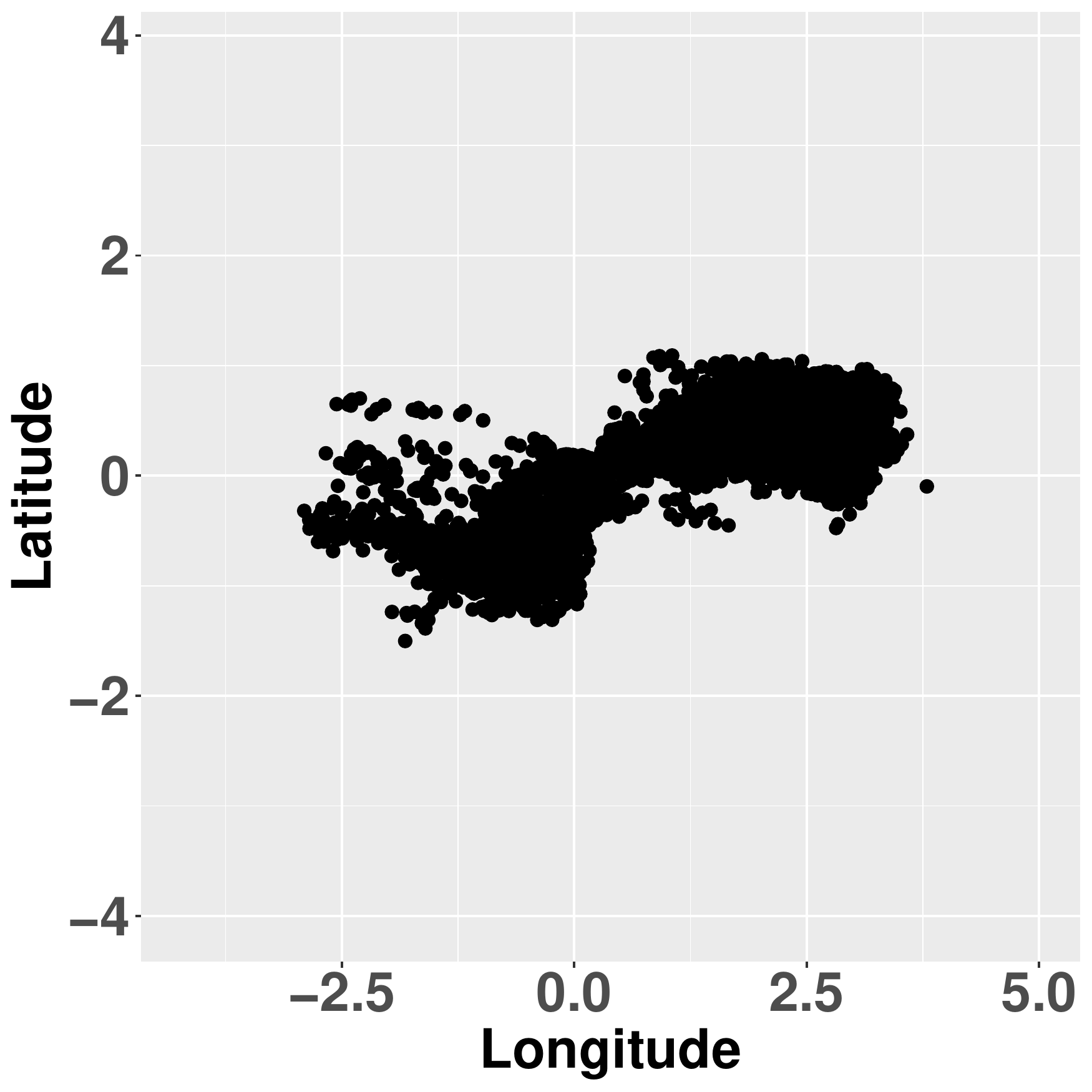}
    \caption{Observed  locations of the sheep.}\label{fig:Sheep}
\end{figure}

\begin{table}[t]
\scriptsize
  \centering
\begin{tabular}{c|ccccc}
  \hline
 & j=1& j=2 & j=3  & j=4& j=5\\
 \hline \hline
 $\hat{\mu}_{j,1}$   &  -3.168  &  -0.105  &  17.062  &  -0.356  &  0.18  \\
 (CI)  & (-52.457 43.823) & (-22.525 23.693) & (-15.255 59.094) & (-0.423 -0.295) & (0.17 0.19)  \\
 $\hat{\mu}_{j,2}$   &  -0.294  &  3.153  &  5.021  &  -0.314  &  -0.404  \\
 (CI)  & (-46.638 45.392) & (-7.564 31.502) & (-25.564 41.621) & (-0.389 -0.244) & (-0.412 -0.395)  \\
 $\hat{\tau}_{j}$   &  0  &  0.008  &  0.008  &  0.624  &  0.985  \\
 (CI)  & (0 0) & (0 0.02) & (0 0.038) & (0.572 0.681) & (0.974 0.996)  \\
 $\hat{\boldsymbol{\Sigma}}_{j,1,1}$   &  0.001  &  0.035  &  0.854  &  0.12  &  0.006  \\
 (CI)  & (0.001 0.001) & (0.025 0.05) & (0.709 1.062) & (0.097 0.15) & (0.005 0.007)  \\
 $\hat{\boldsymbol{\Sigma}}_{j,1,2}$   &  0  &  -0.002  &  -0.083  &  0.031  &  0.001  \\
 (CI)  & (0 0) & (-0.005 0.001) & (-0.155 -0.023) & (0.015 0.047) & (0 0.002)  \\
 $\hat{\boldsymbol{\Sigma}}_{j,2,2}$   &  0.001  &  0.019  &  0.427  &  0.164  &  0.005  \\
 (CI)  & (0.001 0.001) & (0.014 0.027) & (0.357 0.522) & (0.134 0.204) & (0.004 0.006)  \\
 $\hat{{\pi}}_{1,j}$   &  0.847  &  0.077  &  0.023  &  0.042  &  0.011  \\
 (CI)  & (0.834 0.86) & (0.066 0.089) & (0.015 0.031) & (0.032 0.052) & (0.005 0.019)  \\
 $\hat{{\pi}}_{2,j}$   &  0.28  &  0.396  &  0.143  &  0.085  &  0.096  \\
 (CI)  & (0.236 0.33) & (0.338 0.458) & (0.102 0.185) & (0.048 0.125) & (0.067 0.126)  \\
 $\hat{{\pi}}_{3,j}$   &  0.07  &  0.209  &  0.504  &  0.161  &  0.056  \\
 (CI)  & (0.039 0.107) & (0.148 0.276) & (0.429 0.571) & (0.11 0.217) & (0.026 0.093)  \\
 $\hat{{\pi}}_{4,j}$   &  0.17  &  0.19  &  0  &  0.447  &  0.193  \\
 (CI)  & (0.125 0.22) & (0.123 0.262) & (0 0.006) & (0.359 0.526) & (0.151 0.24)  \\
 $\hat{{\pi}}_{5,j}$   &  0.883  &  0  &  0.004  &  0.003  &  0.11  \\
 (CI)  & (0.777 0.979) & (0 0.003) & (0 0.029) & (0 0.026) & (0.017 0.212)  \\
 $\hat{\beta}_{j}$   &  0.25  &  0.174  &  0.122  &  0.178  &  0.223  \\
 (CI)  & (0.074 0.48) & (0.04 0.373) & (0.016 0.308) & (0.042 0.386) & (0.068 0.44)  \\
 \hline \hline
 & $\alpha$ & $\kappa$  & $\gamma$   \\
 \hline
 $\hat{}$   &  0.266  &  2.613  &  1.204  \\
 (CI)  & (0.007 1.008) & (1.338 4.357) & (0.296 2.904)  \\
 \hline \hline
\end{tabular} \caption{Posterior means and 95\% CIs of the BRW-HMM parameters for  K=5.}\label{tab:resBRW}
\end{table}

\begin{table}[!t]
\scriptsize
  \centering
\begin{tabular}{c|ccc}
  \hline
 & j=1& j=2 & j=3  \\
 \hline \hline
 $\hat{\eta}_{j,1}$   &  -0.005  &  -0.044  &  0.026  \\
 (CI)  & (-0.006 -0.004) & (-0.059 -0.029) & (-0.021 0.073)  \\
 $\hat{\eta}_{j,2}$   &  0  &  0.001  &  -0.008  \\
 (CI)  & (-0.001 0.001) & (-0.009 0.011) & (-0.046 0.029)  \\
 $\hat{\boldsymbol{\Sigma}}_{j,1,1}$   &  0.001  &  0.036  &  0.597  \\
 (CI)  & (0.001 0.001) & (0.029 0.045) & (0.541 0.657)  \\
 $\hat{\boldsymbol{\Sigma}}_{j,1,2}$   &  0  &  0  &  0.013  \\
 (CI)  & (0 0) & (-0.002 0.002) & (-0.017 0.042)  \\
 $\hat{\boldsymbol{\Sigma}}_{j,2,2}$   &  0.001  &  0.018  &  0.394  \\
 (CI)  & (0.001 0.001) & (0.015 0.022) & (0.357 0.435)  \\
 $\hat{{\pi}}_{1,j}$   &  0.849  &  0.086  &  0.064  \\
 (CI)  & (0.837 0.861) & (0.074 0.099) & (0.055 0.075)  \\
 $\hat{{\pi}}_{2,j}$   &  0.349  &  0.415  &  0.236  \\
 (CI)  & (0.308 0.394) & (0.364 0.463) & (0.198 0.274)  \\
 $\hat{{\pi}}_{3v}$   &  0.245  &  0.213  &  0.542  \\
 (CI)  & (0.213 0.279) & (0.168 0.256) & (0.501 0.583)  \\
 $\hat{\beta}_{j}$   &  0.321  &  0.3  &  0.303  \\
 (CI)  & (0.047 0.696) & (0.047 0.675) & (0.046 0.671)  \\
 \hline \hline
 & $\alpha$ & $\kappa$  & $\gamma$   \\
 \hline
 $\hat{}$   &  0.229  &  2.175  &  0.765  \\
 (CI)  & (0.005 0.907) & (0.778 4.439) & (0.088 2.343)  \\
 \hline \hline
\end{tabular} \caption{Posterior means and 95\% CIs of the CRW-HMM parameters for K=3.}\label{tab:resCRW}
\end{table}

\begin{figure}[t]
    \centering
    \includegraphics[scale=0.4]{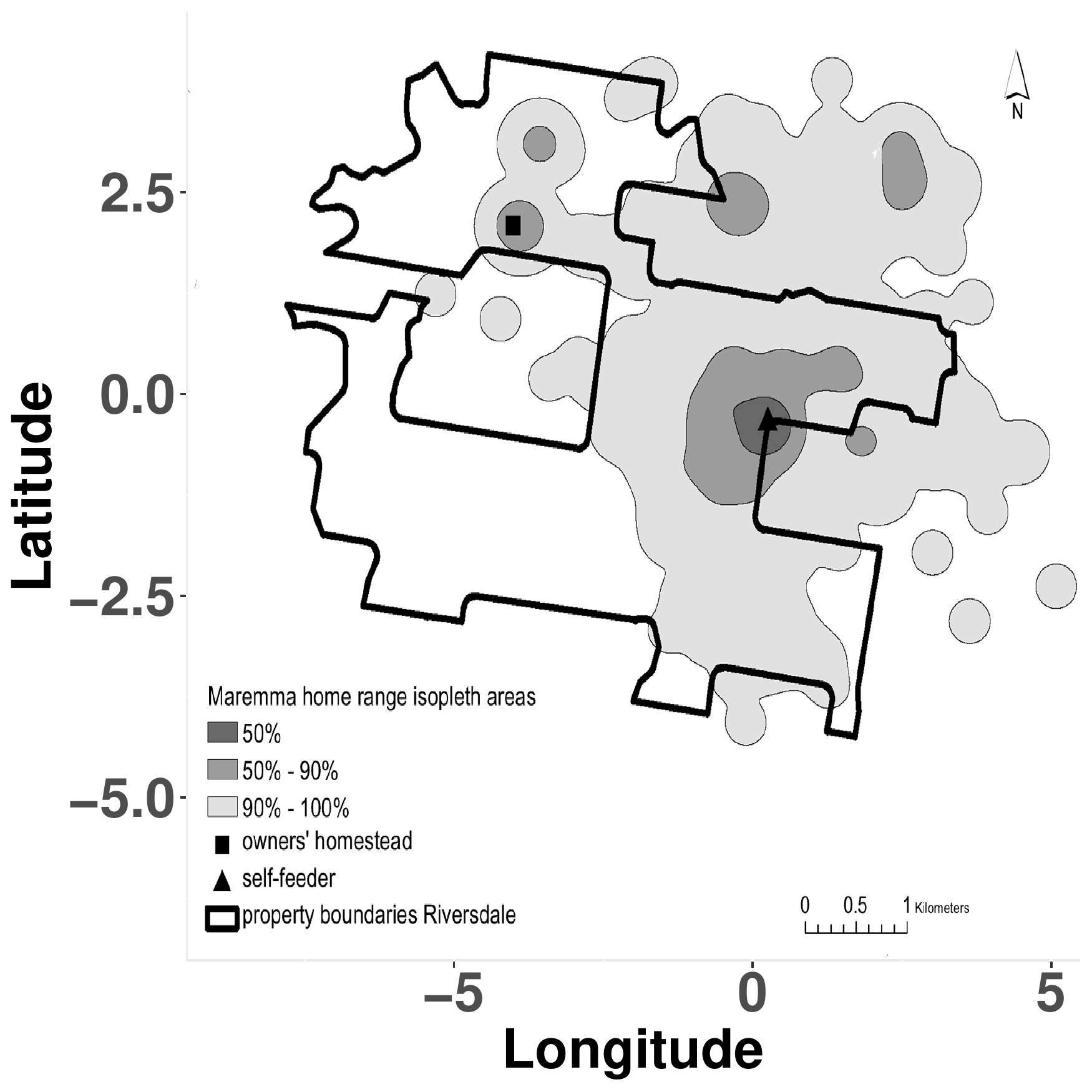}
    \caption{ {A copy of the second row of Figure  1 from \cite{Sheepdog}, which shows the location of the self-feeder (triangle). The coordinates are not displayed in the original figure  and have been added by matching the observed coordinates of all the dogs in the Riversdale property with the home range  isopleth.}}\label{fig:PlosOne}
\end{figure}

\begin{figure}[t]
    \centering
    {\subfloat[]{\includegraphics[trim = 0 10 0 40,scale=0.25]{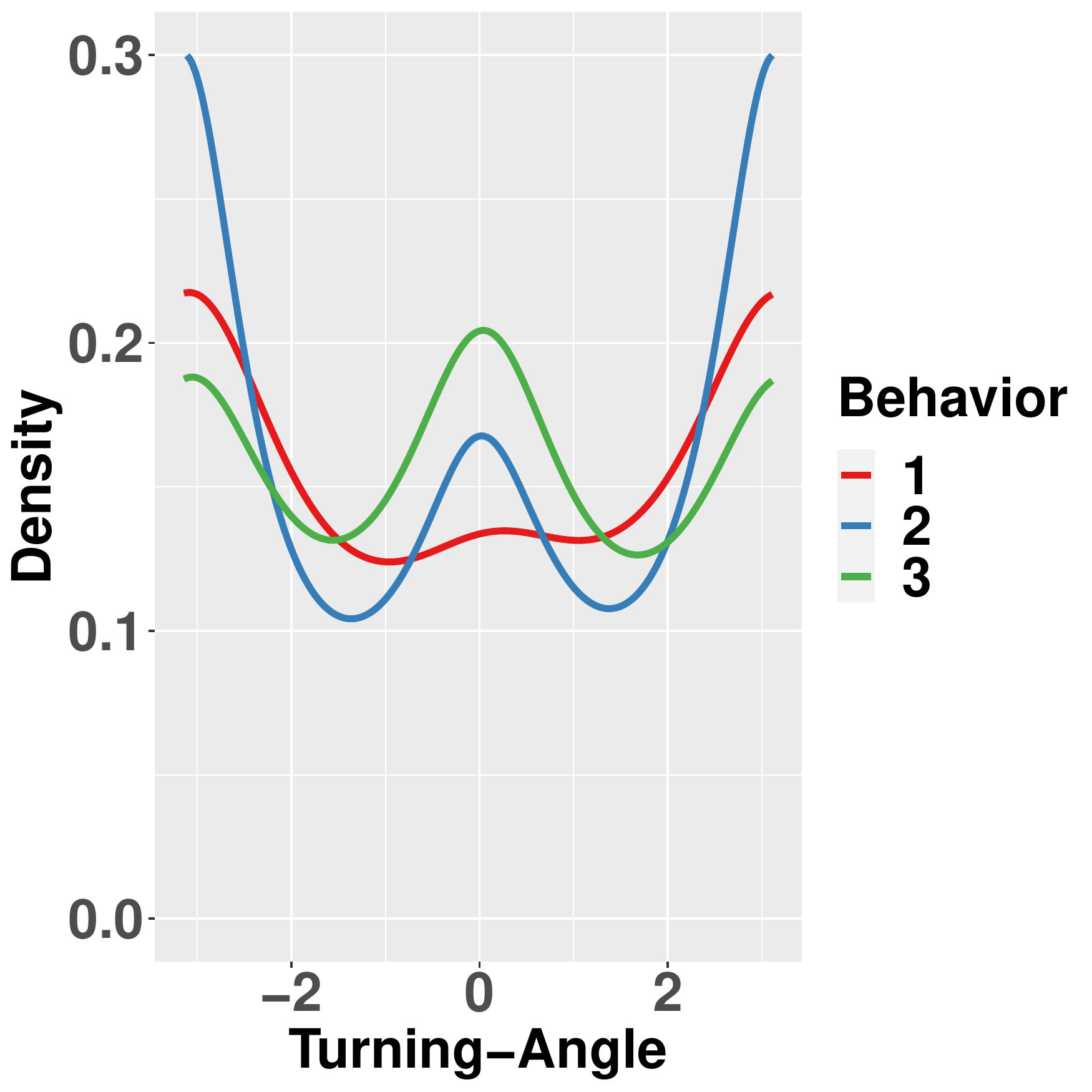}}}
    {\subfloat[]{\includegraphics[trim = 0 10 0 40,scale=0.25]{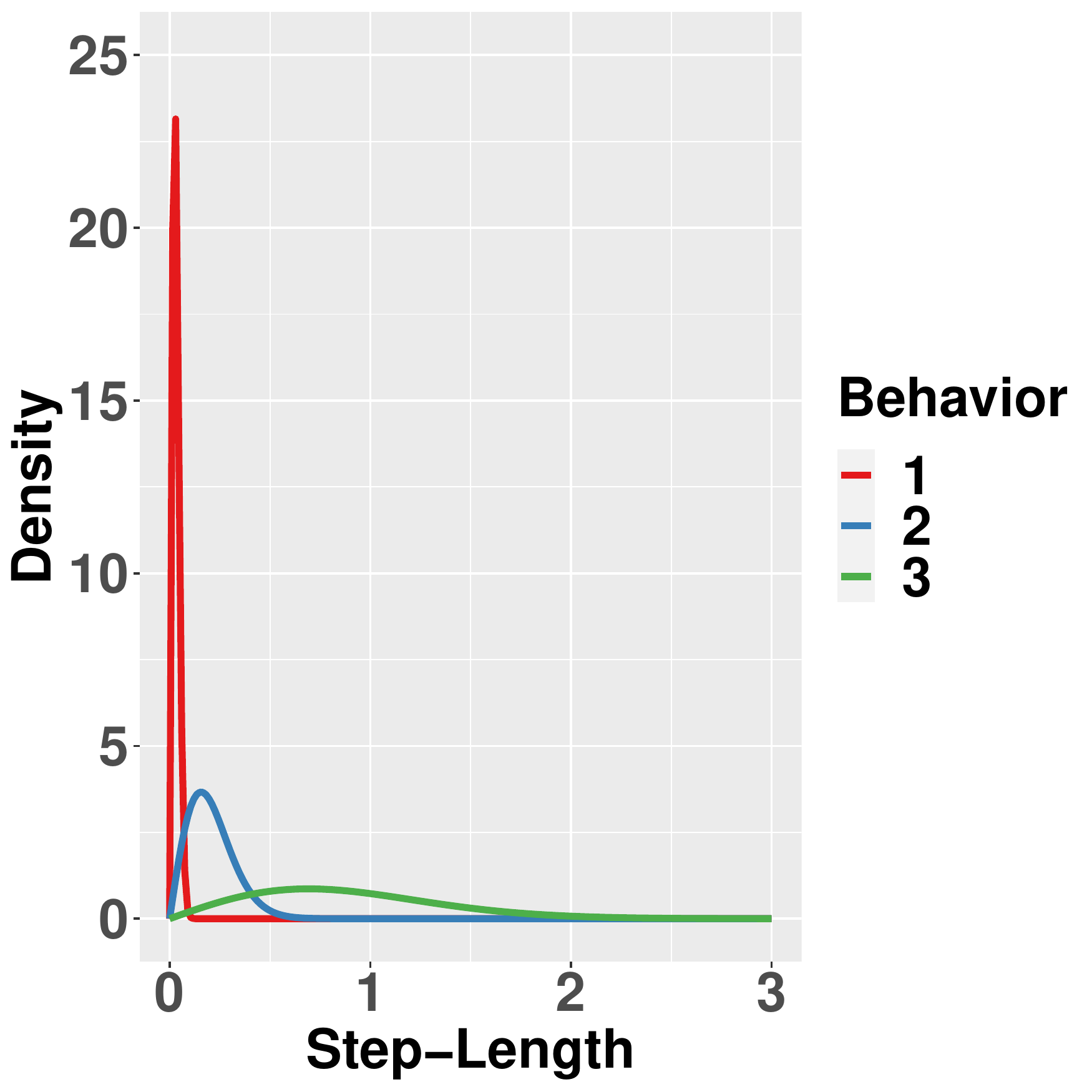}}}
    \caption{Turning-angle (a) and log step-length (b)  predictive distributions for the CRW-HMM   behaviors. }\label{fig:steptunrCWR}
\end{figure}


In this Appendix, we show the results obtained when an HMM that can only estimate CRWs (CRW-HMM) or BRWs (BRW-HMM)  is considered. The priors, iterations, burnin and thin are the same  as those used to obtain posterior samples of the STAP-HMM.
After fitting the model, we found that the number of estimated behaviors  is 5 (with probability 0.99) for the BRW-HMM, and   3 (with probability 0.98)  for  the CRW-HMM. The posterior means  and CIs of all the  parameters are shown in Tables  \ref{tab:resBRW} and  \ref{tab:resCRW}.  We indicate the $j$-th behavior for the BRW-HMM and CRW-HMM, respectively,  as   $\text{LB}_{j, \text{BRW}}$ and $\text{LB}_{j, \text{CRW}}$.

\paragraph*{BRW-HMM}
 It is clear, from  Table \ref{fig:beavS},
 that the MAP behaviors of the STAP-HMM and BRW-HMM have the same temporal dynamics, i.e., the animal exhibits behavior $\text{LB}_{j}$ and $\text{LB}_{j, \text{BRW}}$   at the same temporal points.
This almost one-to-one relation explains why the parameters estimated for    $\text{LB}_{4}$ and $\text{LB}_{5}$, i.e., the BRW-type behaviors of our proposal, are very similar to   $\text{LB}_{4, \text{BRW}}$ and $\text{LB}_{5, \text{BRW}}$, respectively (see Tables \ref{tab:resSTAP} and \ref{tab:resBRW}).
As the behaviors have  the same temporal evolution as  the STAP-HMM, they have the same spatial locations and time of the day when they are more likely to be observed.  This means that the interpretation is similar, but  the BRW-HMM fails  to recognize the directional persistence that is present in the  first and second behaviors, and estimates   RWs, i.e.,   $\hat{\tau}_{1} = 0$ and $ \hat{\tau}_2 = 0.008$, respectively.
%

\paragraph*{CRW-HMM}
\begin{figure}[t]
    \centering
    {\subfloat[First behavior]{\includegraphics[trim = 0 10 0 40,scale=0.25]{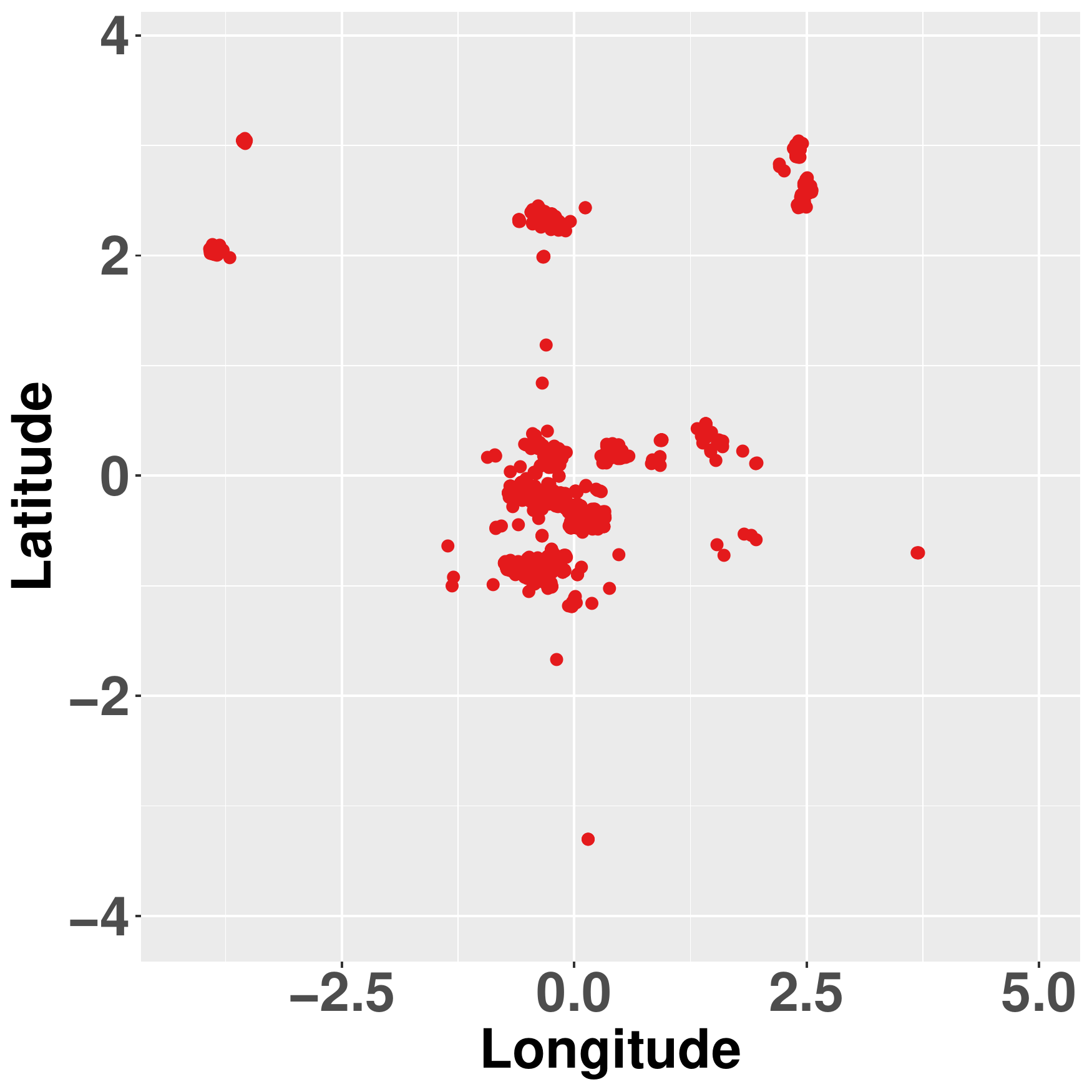}}}
    {\subfloat[Second behavior]{\includegraphics[trim = 0 10 0 40,scale=0.25]{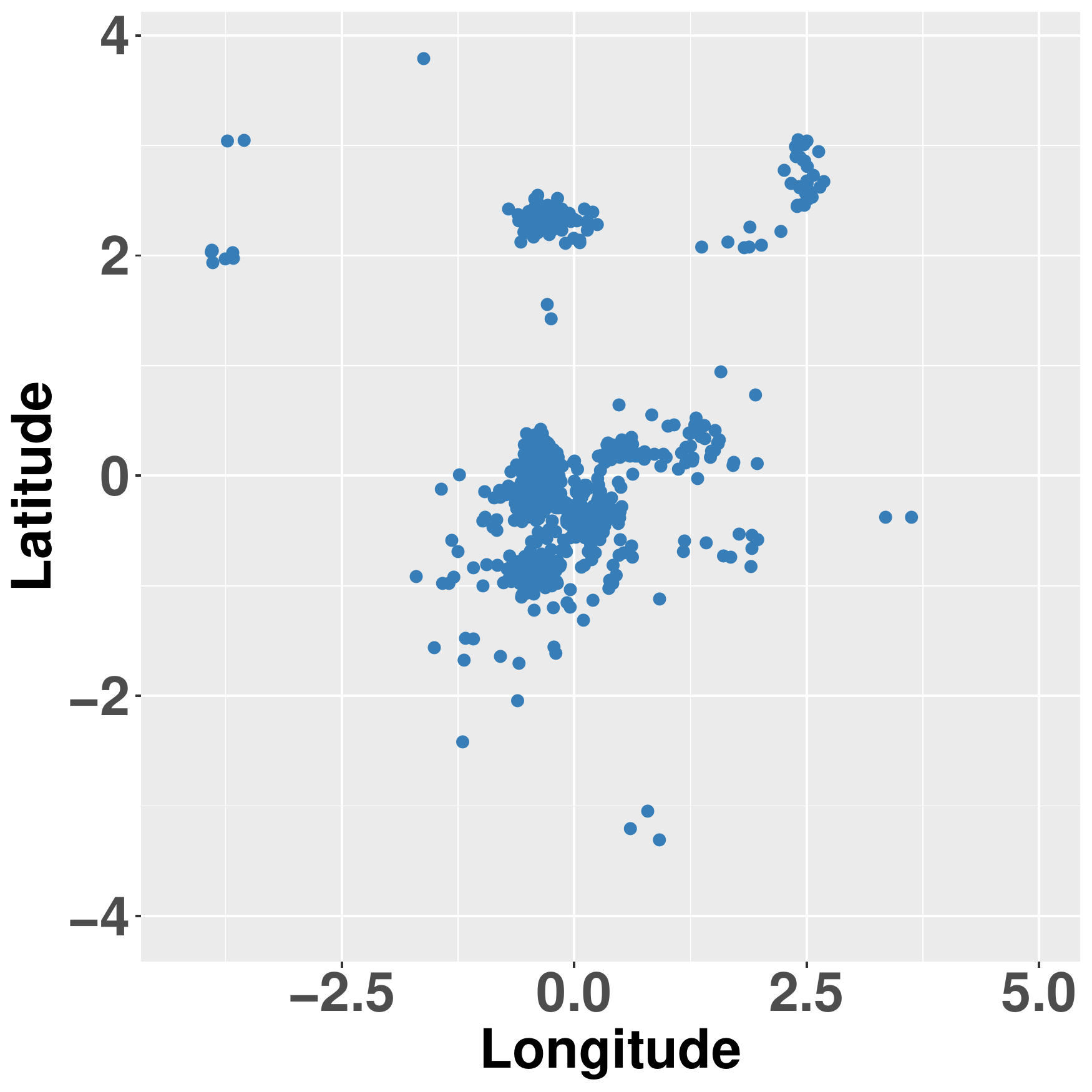}}}
    {\subfloat[Third behavior]{\includegraphics[trim = 0 10 0 40,scale=0.25]{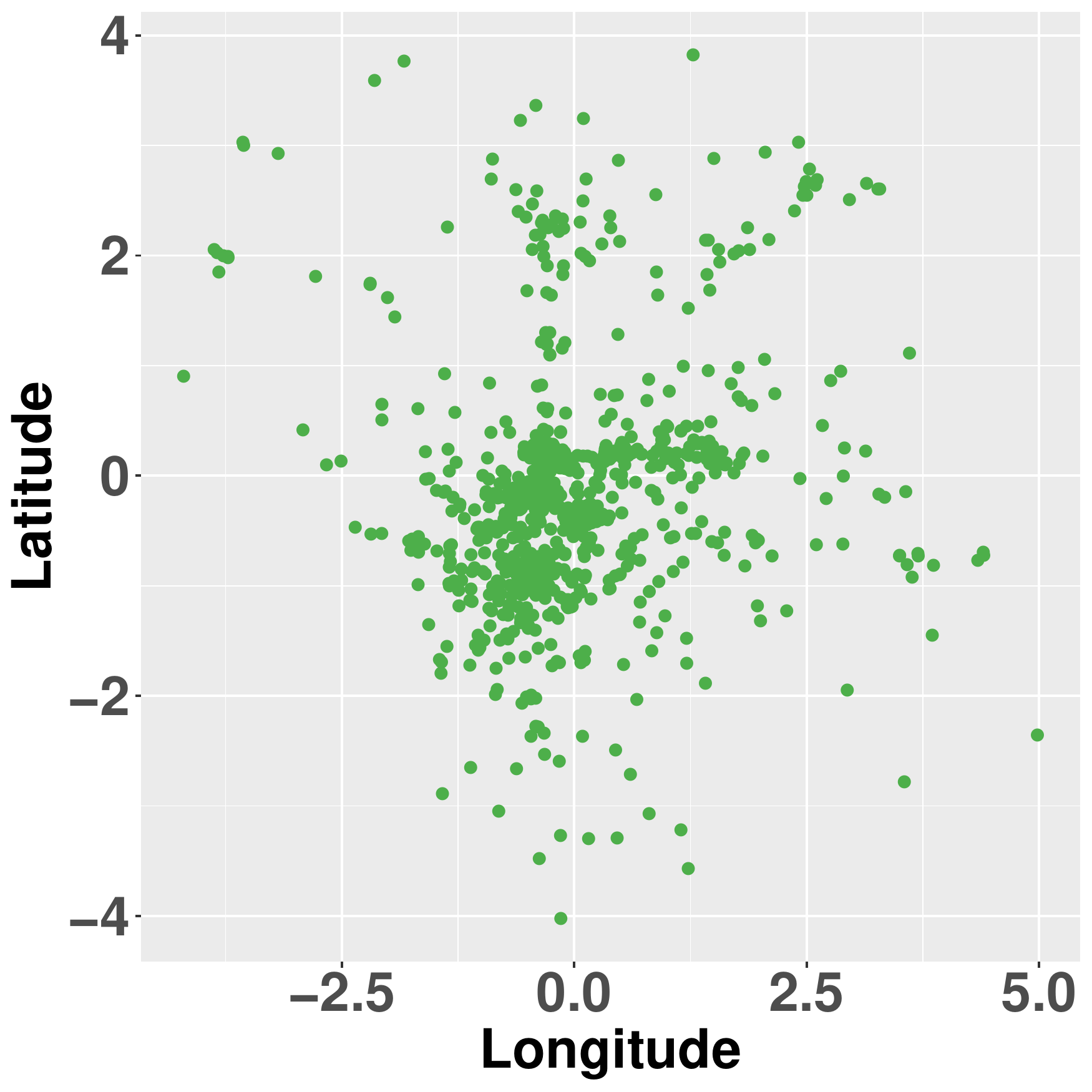}}}
    \caption{Spatial locations of the MAP behavior for the CRW-HMM model.  }\label{fig:locpost_CRW}
\end{figure}
\begin{figure}[t]
    \centering
    {\subfloat[First behavior]{\includegraphics[scale=0.2]{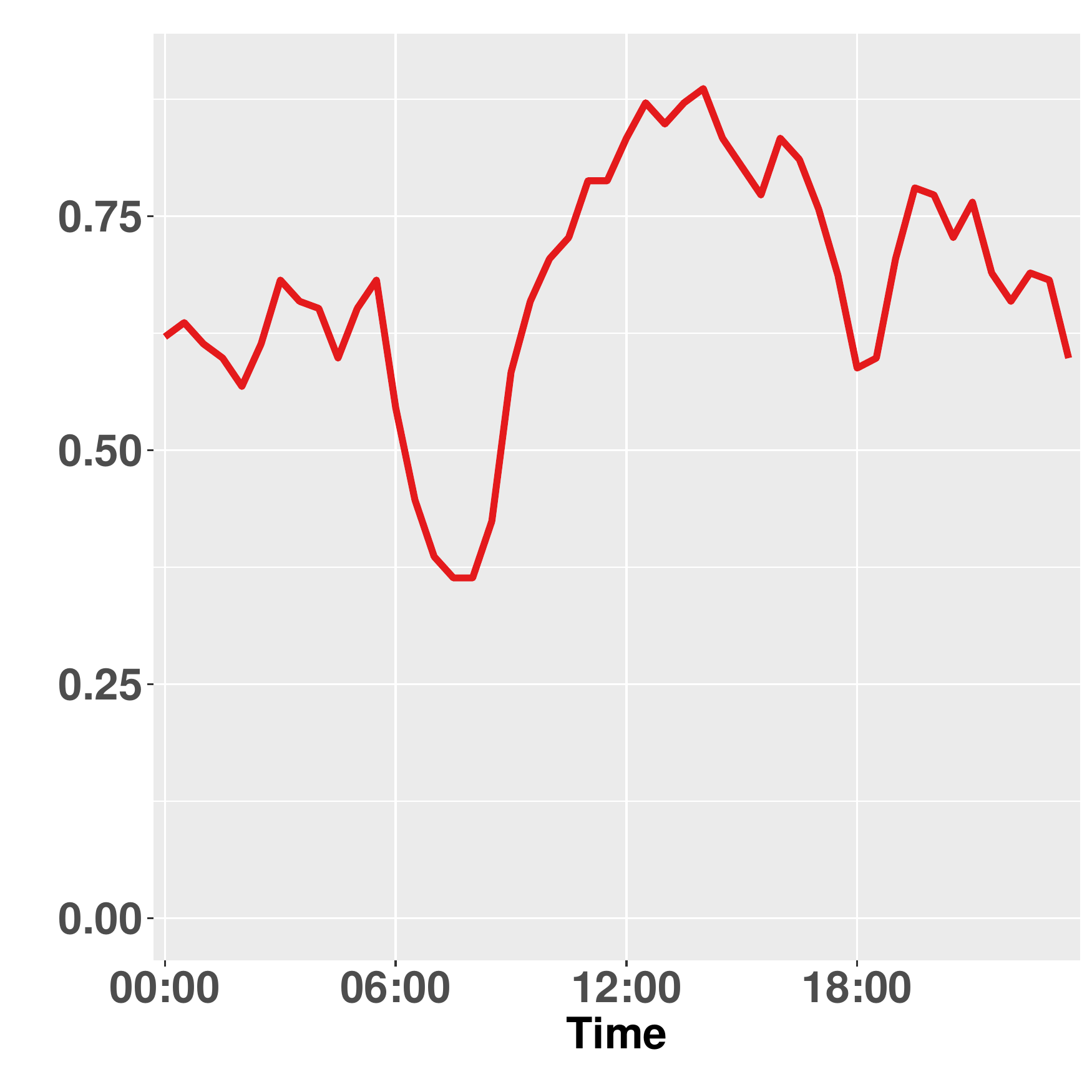}}}
    {\subfloat[Second behavior]{\includegraphics[scale=0.2]{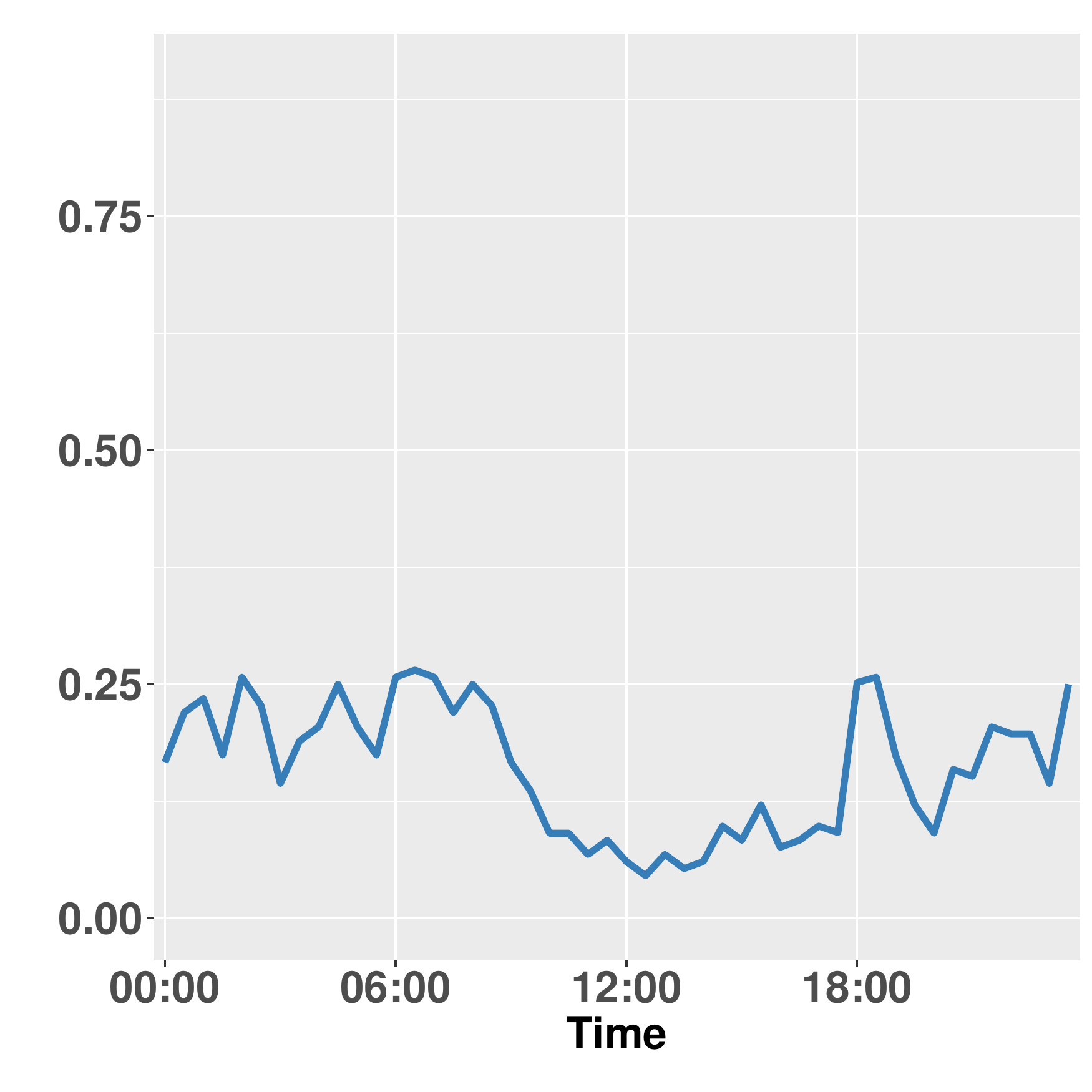}}}
    {\subfloat[Third behavior]{\includegraphics[scale=0.2]{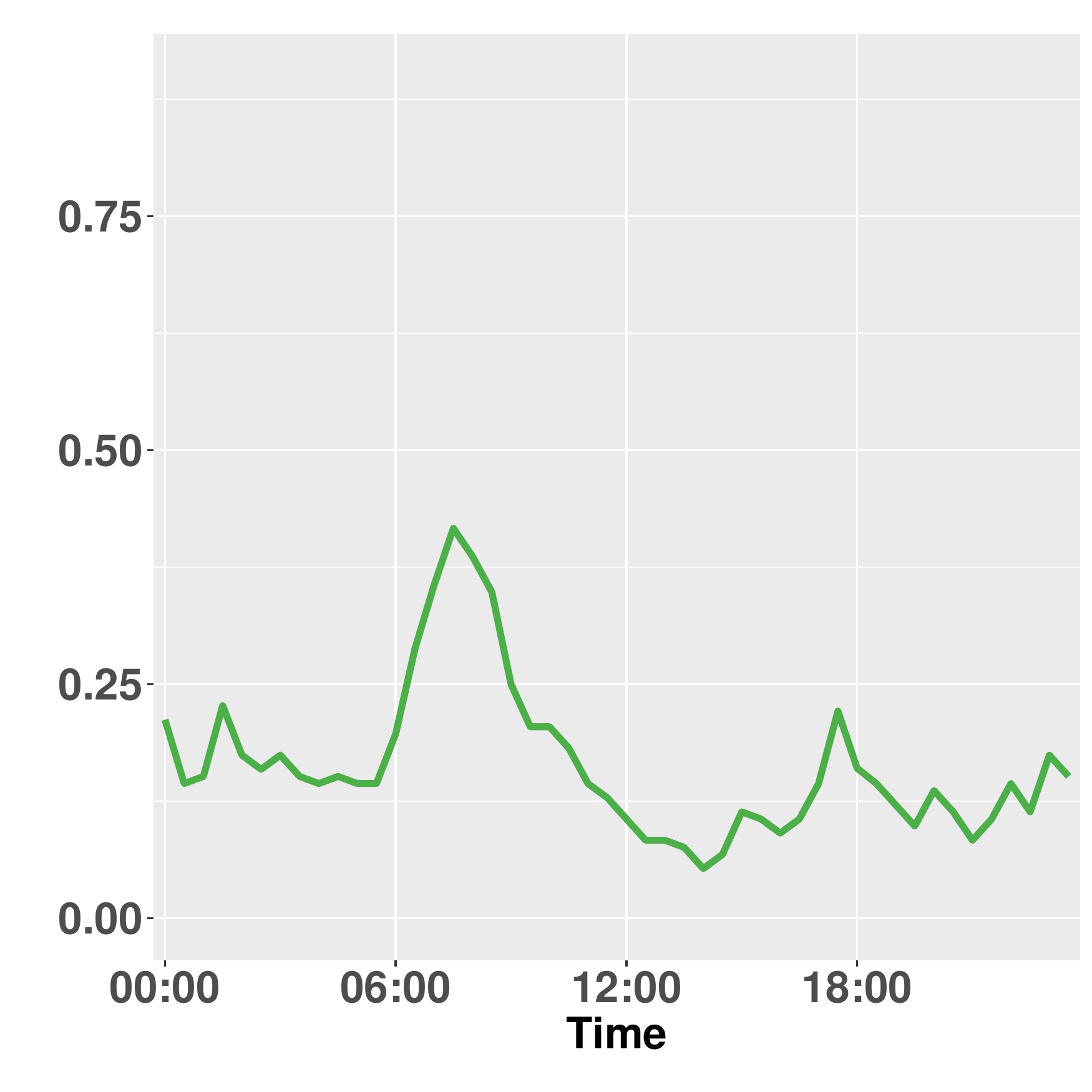}}}
    \caption{ The values of the  lines at each time (x-axis) represent the proportion of times that   the CRW-HMM MAP behaviors are observed (y-axis).}\label{fig:btimeCRW}
\end{figure}
%


As can be seen from Table  \ref{fig:beavS},
the dog follows $\text{LB}_{1}$ and $\text{LB}_{1, \text{CRW}}$ at the same temporal points, which is also shown by the spatial locations of $\text{LB}_{1, \text{CRW}}$  that can be seen in Figure \ref{fig:locpost_CRW} (a). The posterior estimates of the likelihood parameters of  $\text{LB}_{1}$ and $\text{LB}_{1, \text{CRW}}$
are almost identical, and this  can also be  verified from the posterior densities in Figure \ref{fig:steptunrCWR}.
$\text{LB}_{2, \text{CRW}}$ generally has  the same temporal indices as $\text{LB}_{2}$, similar spatial locations (Figure \ref{fig:locpost_CRW} (a)),  similar frequencies during the day  (Figure \ref{fig:btimeCRW} (b)), a similar step-length distribution (Figure \ref{fig:steptunrCWR} (b)), but a different circular one (Figure \ref{fig:steptunrCWR} (d)).
 This behavior has a higher speed than $\text{LB}_{1, CRW}$, and  a bimodal circular distribution with  the major mode at $-\pi$,  and a smaller one at 0.
$\text{LB}_{3, CRW}$  has the highest speed and a circular distribution with two modes, at 0 and  $-\pi$, which have  approximatively  the same density (see Figure \ref{fig:steptunrCWR}), and are more likely  during the night or  in the early hours of the morning.

\paragraph*{General comments}
Our proposal is able to estimate behaviors that are similar, in interpretation and posterior inference,  to   $\text{LB}_{4, \text{BRW}}$, $\text{LB}_{5, \text{BRW}}$ and $\text{LB}_{1, \text{CRW}}$,  as well as to find two behaviors,  $\text{LB}_{2}$ and $\text{LB}_{3}$, that are not estimated by the other two models.
The  CRW-HMM finds 3 behaviors
but, since none of them has a spatial distribution concentrated on the livestock paddock (Figure \ref{fig:Sheep}) as $\text{LB}_{4}$, it does not identify when the dog is attending the livestock. The  BRW-HMM  fails to find any structure in  $\text{LB}_{1, \text{BRW}}$ or in $\text{LB}_{2, \text{BRW}}$, where it estimates a simple RW.

\end{appendices}

\bibliographystyle{natbib}
\bibliography{all}       

\begin{thebibliography}{}

\bibitem[Abe and Ley(2017)Abe and Ley]{ABE201791}
Abe, T. and Ley, C. (2017).
\newblock A tractable, parsimonious and flexible model for cylindrical data,
  with applications.
\newblock {\em Econometrics and Statistics\/}, {\bf 4}, 91--104.

\bibitem[Anderson and Lindzey(2003)Anderson and Lindzey]{Anderson2003}
Anderson, C.~R. and Lindzey, F.~G. (2003).
\newblock Estimating cougar predation rates from gps location clusters.
\newblock {\em The Journal of Wildlife Management\/}, {\bf 67}(2), 307--316.

\bibitem[Barton {\em et~al.}(2009)Barton, Phillips, Morales, and
  Travis]{Barton2009}
Barton, K.~A., Phillips, B.~L., Morales, J.~M., and Travis, J. M.~J. (2009).
\newblock The evolution of an `intelligent' dispersal strategy: biased,
  correlated random walks in patchy landscapes.
\newblock {\em Oikos\/}, {\bf 118}(2), 309--319.

\bibitem[Bezanson {\em et~al.}(2017)Bezanson, Edelman, Karpinski, and
  Shah]{bezanson2017julia}
Bezanson, J., Edelman, A., Karpinski, S., and Shah, V.~B. (2017).
\newblock Julia: A fresh approach to numerical computing.
\newblock {\em SIAM review\/}, {\bf 59}(1), 65--98.

\bibitem[Biernacki {\em et~al.}(2000)Biernacki, Celeux, and
  Govaert]{Biernacki:2000}
Biernacki, C., Celeux, G., and Govaert, G. (2000).
\newblock Assessing a mixture model for clustering with the integrated
  completed likelihood.
\newblock {\em IEEE Trans. Pattern Anal. Mach. Intell.}, {\bf 22}(7), 719--725.

\bibitem[Blackwell(1997)Blackwell]{BLACKWELL199787}
Blackwell, P. (1997).
\newblock Random diffusion models for animal movement.
\newblock {\em Ecological Modelling\/}, {\bf 100}(1), 87 -- 102.

\bibitem[Brook {\em et~al.}(2012)Brook, Johnson, and Ritchie]{Brook2012}
Brook, L.~A., Johnson, C.~N., and Ritchie, E.~G. (2012).
\newblock Effects of predator control on behaviour of an apex predator and
  indirect consequences for mesopredator suppression.
\newblock {\em Journal of Applied Ecology\/}, {\bf 49}(6), 1278--1286.

\bibitem[Brost {\em et~al.}(2015)Brost, Hooten, Hanks, and Small]{Brost2015}
Brost, B.~M., Hooten, M.~B., Hanks, E.~M., and Small, R.~J. (2015).
\newblock Animal movement constraints improve resource selection inference in
  the presence of telemetry error.
\newblock {\em Ecology\/}, {\bf 96}(10), 2590--2597.

\bibitem[Buderman {\em et~al.}(2018a)Buderman, Hooten, Ivan, and
  Shenk]{Buderman2018a}
Buderman, F.~E., Hooten, M.~B., Ivan, J.~S., and Shenk, T.~M. (2018a).
\newblock Large-scale movement behavior in a reintroduced predator population.
\newblock {\em Ecography\/}, {\bf 41}(1), 126--139.

\bibitem[Buderman {\em et~al.}(2018b)Buderman, Hooten, Alldredge, Hanks, and
  Ivan]{Buderman2018}
Buderman, F.~E., Hooten, M.~B., Alldredge, M.~W., Hanks, E.~M., and Ivan, J.~S.
  (2018b).
\newblock Time-varying predatory behavior is primary predictor of fine-scale
  movement of wildland-urban cougars.
\newblock {\em Movement Ecology\/}, {\bf 6}(1), 22.

\bibitem[Cagnacci {\em et~al.}(2010)Cagnacci, Boitani, Powell, and
  Boyce]{Cagnacci2010}
Cagnacci, F., Boitani, L., Powell, R.~A., and Boyce, M.~S. (2010).
\newblock Animal ecology meets gps-based radiotelemetry: a perfect storm of
  opportunities and challenges.
\newblock {\em Philosophical Transactions of the Royal Society of London B:
  Biological Sciences\/}, {\bf 365}(1550), 2157--2162.

\bibitem[Celeux {\em et~al.}(2006)Celeux, Forbes, Robert, Titterington, Futurs,
  and Rh\^{o}ne-alpes]{celeux2006deviance}
Celeux, G., Forbes, F., Robert, C.~P., Titterington, D.~M., Futurs, I., and
  Rh\^{o}ne-alpes, I. (2006).
\newblock {Deviance information criteria for missing data models}.
\newblock {\em {B}ayesian Analysis\/}, {\bf 4}, 651--674.

\bibitem[Christ {\em et~al.}(2008)Christ, Hoef, and Zimmerman]{Christ2008}
Christ, A., Hoef, J.~V., and Zimmerman, D.~L. (2008).
\newblock An animal movement model incorporating home range and habitat
  selection.
\newblock {\em Environmental and Ecological Statistics\/}, {\bf 15}(1), 27--38.

\bibitem[Codling and Hill(2005)Codling and Hill]{CODLING2005573}
Codling, E. and Hill, N. (2005).
\newblock Sampling rate effects on measurements of correlated and biased random
  walks.
\newblock {\em Journal of Theoretical Biology\/}, {\bf 233}(4), 573 -- 588.

\bibitem[Codling {\em et~al.}(2008)Codling, Plank, and Benhamou]{codling2008}
Codling, E.~A., Plank, M.~J., and Benhamou, S. (2008).
\newblock Random walk models in biology.
\newblock {\em Journal of The Royal Society Interface\/}, {\bf 5}(25),
  813--834.

\bibitem[Dunn and Gipson(1977)Dunn and Gipson]{dunn77}
Dunn, J.~E. and Gipson, P.~S. (1977).
\newblock Analysis of radiotelemetry data in studies of home range.
\newblock {\em Biometrics\/}, {\bf 33}(1).

\bibitem[Fleming {\em et~al.}(2014)Fleming, Calabrese, Mueller, Olson,
  Leimgruber, and Fagan]{fleming2014non}
Fleming, C.~H., Calabrese, J.~M., Mueller, T., Olson, K.~A., Leimgruber, P.,
  and Fagan, W.~F. (2014).
\newblock Non-markovian maximum likelihood estimation of autocorrelated
  movement processes.
\newblock {\em Methods in Ecology and Evolution\/}, {\bf 5}(5), 462--472.

\bibitem[Fortin {\em et~al.}(2005)Fortin, Morales, and Boyce]{Fortin2005}
Fortin, D., Morales, J.~M., and Boyce, M.~S. (2005).
\newblock Elk winter foraging at fine scale in yellowstone national park.
\newblock {\em Oecologia\/}, {\bf 145}(2), 334--342.

\bibitem[Fox {\em et~al.}(2011)Fox, Sudderth, Jordan, and Willsky]{fox2011}
Fox, E.~B., Sudderth, E.~B., Jordan, M.~I., and Willsky, A.~S. (2011).
\newblock A sticky hdp-hmm with application to speaker diarization.
\newblock {\em The Annals of Applied Statistics\/}, {\bf 5}(2A), 1020--1056.

\bibitem[Frair {\em et~al.}(2010)Frair, Fieberg, Hebblewhite, Cagnacci,
  DeCesare, and Pedrotti]{Frair2010}
Frair, J.~L., Fieberg, J., Hebblewhite, M., Cagnacci, F., DeCesare, N.~J., and
  Pedrotti, L. (2010).
\newblock Resolving issues of imprecise and habitat-biased locations in
  ecological analyses using gps telemetry data.
\newblock {\em Philosophical Transactions of the Royal Society of London B:
  Biological Sciences\/}, {\bf 365}(1550), 2187--2200.

\bibitem[Friendly {\em et~al.}(2013)Friendly, Monette, and Fox]{Friendly2012}
Friendly, M., Monette, G., and Fox, J. (2013).
\newblock Elliptical insights: Understanding statistical methods through
  elliptical geometry.
\newblock {\em Statistical Science\/}, {\bf 28}(1), 1–39.

\bibitem[Fr{\"u}hwirth-Schnatter and
  Malsiner-Walli(2019)Fr{\"u}hwirth-Schnatter and Malsiner-Walli]{Schnatte2019}
Fr{\"u}hwirth-Schnatter, S. and Malsiner-Walli, G. (2019).
\newblock From here to infinity: sparse finite versus dirichlet process
  mixtures in model-based clustering.
\newblock {\em Advances in Data Analysis and Classification\/}, {\bf 13}(1),
  33--64.

\bibitem[Fryxell {\em et~al.}(2008)Fryxell, Hazell, B{\"o}rger, Dalziel,
  Haydon, Morales, McIntosh, and Rosatte]{Fryxell2008}
Fryxell, J.~M., Hazell, M., B{\"o}rger, L., Dalziel, B.~D., Haydon, D.~T.,
  Morales, J.~M., McIntosh, T., and Rosatte, R.~C. (2008).
\newblock Multiple movement modes by large herbivores at multiple
  spatiotemporal scales.
\newblock {\em Proceedings of the National Academy of Sciences\/}, {\bf
  105}(49), 19114--19119.

\bibitem[Gehring {\em et~al.}(2017)Gehring, VerCauteren, and Cellar]{Gehring}
Gehring, T.~M., VerCauteren, K.~C., and Cellar, A.~C. (2017).
\newblock Good fences make good neighbors: Implementation of electric fencing
  for establishing effective livestock-protection dogs.
\newblock {\em Human-Wildlife Interactions\/}, {\bf 5}(1), 106--111.

\bibitem[Gelman {\em et~al.}(1995)Gelman, Robert, Chopin, and
  Rousseau]{Gelman95bayesiandata}
Gelman, A., Robert, C., Chopin, N., and Rousseau, J. (1995).
\newblock Bayesian data analysis.

\bibitem[Hanks {\em et~al.}(2015)Hanks, Hooten, and Alldredge]{hanks2015}
Hanks, E.~M., Hooten, M.~B., and Alldredge, M.~W. (2015).
\newblock Continuous-time discrete-space models for animal movement.
\newblock {\em Annals of Applied Statistics\/}, {\bf 9}(1), 145--165.

\bibitem[Harris and Blackwell(2013)Harris and Blackwell]{Harris201329}
Harris, K.~J. and Blackwell, P.~G. (2013).
\newblock Flexible continuous-time modelling for heterogeneous animal movement.
\newblock {\em Ecological Modelling\/}, {\bf 255}, 29 -- 37.

\bibitem[Hastie and Green(2012)Hastie and Green]{Hastie2012}
Hastie, D.~I. and Green, P.~J. (2012).
\newblock Model choice using reversible jump markov chain monte carlo.
\newblock {\em Statistica Neerlandica\/}, {\bf 66}(3), 309--338.

\bibitem[Hebblewhite and Merrill(2008)Hebblewhite and Merrill]{Hebblewhite}
Hebblewhite, M. and Merrill, E. (2008).
\newblock Modelling wildlife and uman relationships for social species with
  mixed-effects resource selection models.
\newblock {\em Journal of Applied Ecology\/}, {\bf 45}(3), 834--844.

\bibitem[Hooten {\em et~al.}(2017)Hooten, Johnson, McClintock, and
  Morales]{hooten2017animal}
Hooten, M., Johnson, D., McClintock, B., and Morales, J. (2017).
\newblock {\em Animal Movement: Statistical Models for Telemetry Data\/}.
\newblock CRC Press.

\bibitem[Ishwaran and Zarepour(2002)Ishwaran and Zarepour]{Ishwaran2002}
Ishwaran, H. and Zarepour, M. (2002).
\newblock Exact and approximate sum representations for the {D}irichlet
  process.
\newblock {\em Canadian Journal of Statistics\/}, {\bf 30}(2), 269--283.

\bibitem[Jammalamadaka and Kozubowski(2004)Jammalamadaka and
  Kozubowski]{Jammalamadaka2004}
Jammalamadaka, S.~R. and Kozubowski, T.~J. (2004).
\newblock New families of wrapped distributions for modeling skew circular
  data.
\newblock {\em Communications in Statistics - Theory and Methods\/}, {\bf
  33}(9), 2059--2074.

\bibitem[Johnson {\em et~al.}(2008)Johnson, London, Lea, and
  Durban]{Johnson2008a}
Johnson, D.~S., London, J.~M., Lea, M.-A., and Durban, J.~W. (2008).
\newblock Continuous-time correlated random walk model for animal telemetry
  data.
\newblock {\em Ecology\/}, {\bf 89}(5), 1208--1215.

\bibitem[Johnson {\em et~al.}(2013)Johnson, Hooten, and Kuhn]{Johnson2013}
Johnson, D.~S., Hooten, M.~B., and Kuhn, C.~E. (2013).
\newblock Estimating animal resource selection from telemetry data using point
  process models.
\newblock {\em Journal of Animal Ecology\/}, {\bf 82}(6), 1155--1164.

\bibitem[Jonsen {\em et~al.}(2005)Jonsen, Flemming, and Myers]{Jonsen2005}
Jonsen, I.~D., Flemming, J.~M., and Myers, R.~A. (2005).
\newblock Robust state-space modeling of animal movement data.
\newblock {\em Ecology\/}, {\bf 86}(11), 2874--2880.

\bibitem[Langrock {\em et~al.}(2012)Langrock, King, Matthiopoulos, Thomas,
  Fortin, and Morales]{Langrock2012}
Langrock, R., King, R., Matthiopoulos, J., Thomas, L., Fortin, D., and Morales,
  J.~M. (2012).
\newblock Flexible and practical modeling of animal telemetry data: hidden
  {M}arkov models and extensions.
\newblock {\em Ecology\/}, {\bf 93}(11), 2336--2342.

\bibitem[Langrock {\em et~al.}(2014)Langrock, Hopcraft, Blackwell, Goodall,
  King, Niu, Patterson, Pedersen, Skarin, and Schick]{langrock2014b}
Langrock, R., Hopcraft, G., Blackwell, P., Goodall, V., King, R., Niu, M.,
  Patterson, T., Pedersen, M., Skarin, A., and Schick, R. (2014).
\newblock Modelling group dynamic animal movement.
\newblock {\em Methods in Ecology and Evolution\/}, {\bf 5}(2), 190--199.

\bibitem[Mastrantonio(2018)Mastrantonio]{MASTRANTONIO2018}
Mastrantonio, G. (2018).
\newblock The joint projected normal and skew-normal: A distribution for
  poly-cylindrical data.
\newblock {\em Journal of Multivariate Analysis\/}, {\bf 165}, 14 -- 26.

\bibitem[Mastrantonio {\em et~al.}(2015)Mastrantonio, Maruotti, and
  Jona~Lasinio]{mastrantonio2015}
Mastrantonio, G., Maruotti, A., and Jona~Lasinio, G. (2015).
\newblock {B}ayesian hidden {M}arkov modelling using circular-linear general
  projected normal distribution.
\newblock {\em Environmetrics\/}, {\bf 26}, 145--158.

\bibitem[Mastrantonio {\em et~al.}(2016)Mastrantonio, Jona~Lasinio, and
  Gelfand]{mastrantonio2015b}
Mastrantonio, G., Jona~Lasinio, G., and Gelfand, A.~E. (2016).
\newblock Spatio-temporal circular models with non-separable covariance
  structure.
\newblock {\em TEST\/}, {\bf 23}, 331--350.

\bibitem[Mastrantonio {\em et~al.}(2019)Mastrantonio, Grazian, Mancinelli, and
  Bibbona]{mastrantonio2019}
Mastrantonio, G., Grazian, C., Mancinelli, S., and Bibbona, E. (2019).
\newblock New formulation of the logistic-gaussian process to analyze
  trajectory tracking data.
\newblock {\em Ann. Appl. Stat.}, {\bf 13}(4), 2483--2508.

\bibitem[McClintock and Michelot(2018)McClintock and Michelot]{momentuHMM2018}
McClintock, B.~T. and Michelot, T. (2018).
\newblock momentuhmm: R package for generalized hidden markov models of animal
  movement.
\newblock {\em Methods in Ecology and Evolution\/}, {\bf 9}(6), 1518--1530.

\bibitem[McClintock {\em et~al.}(2012)McClintock, King, Thomas, Matthiopoulos,
  McConnell, and Morales]{McClintock2012}
McClintock, B.~T., King, R., Thomas, L., Matthiopoulos, J., McConnell, B.~J.,
  and Morales, J.~M. (2012).
\newblock A general discrete-time modeling framework for animal movement using
  multistate random walks.
\newblock {\em Ecological Monographs\/}, {\bf 82}(3), 335--349.

\bibitem[McClintock {\em et~al.}(2014)McClintock, Johnson, Hooten, Ver~Hoef,
  and Morales]{McClintock2014}
McClintock, B.~T., Johnson, D.~S., Hooten, M.~B., Ver~Hoef, J.~M., and Morales,
  J.~M. (2014).
\newblock When to be discrete: the importance of time formulation in
  understanding animal movement.
\newblock {\em Movement Ecology\/}, {\bf 2}(1), 21.

\bibitem[McGrew and Blakesley(1982)McGrew and Blakesley]{MCgrew}
McGrew, J.~C. and Blakesley, C.~S. (1982).
\newblock How komondor dogs reduce sheep losses to coyotes.
\newblock {\em Journal of Range Management\/}, {\bf 6}(35), 693--696.

\bibitem[Merrill and David~Mech(2000)Merrill and David~Mech]{MERRILL2000}
Merrill, S.~B. and David~Mech, L. (2000).
\newblock Details of extensive movements by minnesota wolves (canis lupus).
\newblock {\em The American Midland Naturalist\/}, {\bf 144}(2), 428--433.

\bibitem[Michelot and Blackwell(2019)Michelot and Blackwell]{Michelot2019}
Michelot, T. and Blackwell, P.~G. (2019).
\newblock State-switching continuous-time correlated random walks.
\newblock {\em Methods in Ecology and Evolution\/}, {\bf 10}(5), 637--649.

\bibitem[Michelot {\em et~al.}(2016)Michelot, Langrock, and
  Patterson]{Michelot2016}
Michelot, T., Langrock, R., and Patterson, T.~A. (2016).
\newblock movehmm: an r package for the statistical modelling of animal
  movement data using hidden markov models.
\newblock {\em Methods in Ecology and Evolution\/}, {\bf 7}(11), 1308--1315.

\bibitem[Morales and Ellner(2002)Morales and Ellner]{morales2002}
Morales, J.~M. and Ellner, S.~P. (2002).
\newblock Scaling up animal movements in heterogeneous landscapes: the
  importance of behavior.
\newblock {\em Ecology\/}, {\bf 83}(8), 2240--2247.

\bibitem[Morales {\em et~al.}(2004)Morales, Haydon, Frair, Holsinger, and
  Fryxell]{morales2004}
Morales, J.~M., Haydon, D.~T., Frair, J., Holsinger, K.~E., and Fryxell, J.~M.
  (2004).
\newblock Extracting more out of relocation data: building movement models as
  mixtures of random walks.
\newblock {\em Ecology\/}, {\bf 85}(9), 2436--2445.

\bibitem[Nathan {\em et~al.}(2008)Nathan, Getz, Revilla, Holyoak, Kadmon,
  Saltz, and Smouse]{Nathan2008}
Nathan, R., Getz, W.~M., Revilla, E., Holyoak, M., Kadmon, R., Saltz, D., and
  Smouse, P.~E. (2008).
\newblock A movement ecology paradigm for unifying organismal movement
  research.
\newblock {\em Proceedings of the National Academy of Sciences\/}, {\bf
  105}(49), 19052--19059.

\bibitem[Parton and Blackwell(2017)Parton and Blackwell]{Parton2017}
Parton, A. and Blackwell, P.~G. (2017).
\newblock Bayesian inference for multistate `step and turn' animal movement in
  continuous time.
\newblock {\em Journal of Agricultural, Biological and Environmental
  Statistics\/}, {\bf 22}(3), 373--392.

\bibitem[Patterson {\em et~al.}(2008)Patterson, Thomas, Wilcox, Ovaskainen, and
  Matthiopoulos]{Patterson2222222}
Patterson, T., Thomas, L., Wilcox, C., Ovaskainen, O., and Matthiopoulos, J.
  (2008).
\newblock State-space models of individual animal movement.
\newblock {\em Trends in Ecology \& Evolution\/}, {\bf 23}(2), 87--94.

\bibitem[Patterson {\em et~al.}(2017)Patterson, Parton, Langrock, Blackwell,
  Thomas, and King]{Patterson2017}
Patterson, T.~A., Parton, A., Langrock, R., Blackwell, P.~G., Thomas, L., and
  King, R. (2017).
\newblock Statistical modelling of individual animal movement: an overview of
  key methods and a discussion of practical challenges.
\newblock {\em AStA Advances in Statistical Analysis\/}, {\bf 101}(4),
  399--438.

\bibitem[Pohle {\em et~al.}(2017)Pohle, Langrock, van Beest, and
  Schmidt]{Pohle2017}
Pohle, J., Langrock, R., van Beest, F.~M., and Schmidt, N.~M. (2017).
\newblock Selecting the number of states in hidden markov models: Pragmatic
  solutions illustrated using animal movement.
\newblock {\em Journal of Agricultural, Biological and Environmental
  Statistics\/}, {\bf 22}(3), 270--293.

\bibitem[Rivest {\em et~al.}(2016)Rivest, Duchesne, Nicosia, and
  Fortin]{rivest2016}
Rivest, L.-P., Duchesne, T., Nicosia, A., and Fortin, D. (2016).
\newblock A general angular regression model for the analysis of data on animal
  movement in ecology.
\newblock {\em Journal of the Royal Statistical Society: Series C (Applied
  Statistics)\/}, {\bf 65}(3), 445--463.

\bibitem[Schultz and Crone(2001)Schultz and Crone]{Schultz2001}
Schultz, C.~B. and Crone, E.~E. (2001).
\newblock Edge-mediated dispersal behavior in a prairie butterfly.
\newblock {\em Ecology\/}, {\bf 82}(7), 1879--1892.

\bibitem[van and Johnson(2014)van and Johnson]{SheepdogRep}
van, Bommel, L. and Johnson, C. (2014).
\newblock Data from: Where do livestock guardian dogs go? movement patterns of
  free-ranging maremma sheepdogs, doi:10.5441/001/1.pv048q7v.

\bibitem[van Bommel and {Invasive Animals Cooperative Research Centre}(2010)van
  Bommel and {Invasive Animals Cooperative Research Centre}]{van2010guardian}
van Bommel, L. and {Invasive Animals Cooperative Research Centre} (2010).
\newblock {\em Guardian Dogs: Best Practice Manual for the Use of Livestock
  Guardian Dogs\/}.
\newblock Invasive Animals Cooperative Research Centre.

\bibitem[van Bommel and Johnson(2012)van Bommel and Johnson]{Bommel3}
van Bommel, L. and Johnson, C.~N. (2012).
\newblock Good dog! using livestock guardian dogs to protect livestock from
  predators in australia's extensive grazing systems.
\newblock {\em Wildlife Research\/}, {\bf 39}(3), 220--229.

\bibitem[van Bommel and Johnson(2014)van Bommel and Johnson]{Sheepdog}
van Bommel, L. and Johnson, C.~N. (2014).
\newblock Where do livestock guardian dogs go? movement patterns of
  free-ranging maremma sheepdogs.
\newblock {\em PLOS ONE\/}, {\bf 9}(10), 1--12.

\bibitem[van Bommel and Johnson(2016)van Bommel and Johnson]{bommel2}
van Bommel, L. and Johnson, C.~N. (2016).
\newblock Livestock guardian dogs as surrogate top predators? how maremma
  sheepdogs affect a wildlife community.
\newblock {\em Ecology and Evolution\/}, {\bf 6}(18), 6702--6711.

\bibitem[Volant {\em et~al.}(2014)Volant, B{\'e}rard, Martin-Magniette, and
  Robin]{Volant2014}
Volant, S., B{\'e}rard, C., Martin-Magniette, M.-L., and Robin, S. (2014).
\newblock Hidden markov models with mixtures as emission distributions.
\newblock {\em Statistics and Computing\/}, {\bf 24}(4), 493--504.

\bibitem[Walton {\em et~al.}(2017)Walton, Samelius, Odden, and
  Willebrand]{Walton2017}
Walton, Z., Samelius, G., Odden, M., and Willebrand, T. (2017).
\newblock Variation in home range size of red foxes vulpes vulpes along a
  gradient of productivity and human landscape alteration.
\newblock {\em PLOS ONE\/}, {\bf 12}(4), 1--14.

\bibitem[Wang and Gelfand(2013)Wang and Gelfand]{Wang2013}
Wang, F. and Gelfand, A.~E. (2013).
\newblock Directional data analysis under the general projected normal
  distribution.
\newblock {\em Statistical Methodology\/}, {\bf 10}(1), 113--127.

\end{thebibliography}

%
\end{document}